\documentclass[longauth]{aa}  
\usepackage{ulem}
\usepackage{multirow}
\usepackage{graphicx}
\usepackage{txfonts}
\usepackage[colorlinks=true,linkcolor=red,anchorcolor=blue,citecolor=blue,filecolor=black,menucolor=black,runcolor=black,urlcolor=blue]{hyperref}

\usepackage[dvipsnames]{xcolor}

\newcommand{\arcs}{$\arcsec$\xspace}

\begin{document}

   \title{Insight into the Starburst Nature of Galaxy GN-z11 with JWST MIRI Spectroscopy}

   \author{J. \'Alvarez-M\'arquez\inst{\ref{inst:CAB}} \and A. Crespo G\'omez\inst{\ref{inst:CAB},\ref{inst:STScI}} \and L. Colina\inst{\ref{inst:CAB}}  \and D. Langeroodi\inst{\ref{inst:DARK}} \and R. Marques-Chaves\inst{\ref{inst:geneve}} \and C. Prieto-Jim\'enez\inst{\ref{inst:CAB}, \ref{inst:UCM}} \and A. Bik\inst{\ref{inst:Stockholm}}
   \and A. Alonso-Herrero\inst{\ref{inst:CAB-ESAC}} \and L. Boogaard\inst{\ref{inst:MPIA}} \and L. Costantin\inst{\ref{inst:CAB}} \and M. Garc\'ia-Mar\'in\inst{\ref{inst:ESA}} \and S. Gillman\inst{\ref{inst:DAWN},\ref{inst:DTU}} \and J. Hjorth\inst{\ref{inst:DARK}} \and E. Iani\inst{\ref{inst:Groningen}} \and I. Jermann\inst{\ref{inst:DAWN},\ref{inst:DTU}} \and A. Labiano\inst{\ref{inst:Telespazio}} \and J. Melinder\inst{\ref{inst:Stockholm}} \and R. Meyer\inst{\ref{inst:geneve}} \and G. {\"O}stlin\inst{\ref{inst:Stockholm}} \and P. G. P\'erez-Gonz\'alez\inst{\ref{inst:CAB}} \and P. Rinaldi\inst{\ref{inst:Groningen},\ref{inst:tucson}} \and F. Walter\inst{\ref{inst:MPIA}} \and P. van der Werf\inst{\ref{inst:Leiden}} \and G. Wright\inst{\ref{inst:UKATC}} } 

   \institute{Centro de Astrobiolog\'{\i}a (CAB), CSIC-INTA, Ctra. de Ajalvir km 4, Torrej\'on de Ardoz, E-28850, Madrid, Spain\\  \email{jalvarez@cab.inta-csic.es} \label{inst:CAB}
   \and Space Telescope Science Institute (STScI), 3700 San Martin Drive, Baltimore, MD 21218, USA \label{inst:STScI}
    \and DARK, Niels Bohr Institute, University of Copenhagen, Jagtvej 155A, 2200 Copenhagen, Denmark \label{inst:DARK} 
    \and Department of Astronomy, University of Geneva, Chemin Pegasi 51, 1290 Versoix, Switzerland \label{inst:geneve} 
    \and Departamento de F\'{i}sica de la Tierra y Astrof\'{i}sica, Facultad de Ciencias F\'{i}sicas, Universidad Complutense de Madrid, E-28040, Madrid, Spain \label{inst:UCM} 
    \and Department of Astronomy, Stockholm University, Oscar Klein Centre, AlbaNova University Centre, 106 91 Stockholm, Sweden \label{inst:Stockholm}
    \and Centro de Astrobiolog\'ia (CAB), CSIC-INTA, Camino Viejo del Castillo s/n, 28692 Villanueva de la Ca\~{n}ada, Madrid, Spain \label{inst:CAB-ESAC}
    \and Max-Planck-Institut f\"ur Astronomie, K\"onigstuhl 17, 69117 Heidelberg, Germany\label{inst:MPIA}
    \and Cosmic Dawn Centre (DAWN), Copenhagen, Denmark\label{inst:DAWN}
    \and European Space Agency, Space Telescope Science Institute, Baltimore, Maryland, USA \label{inst:ESA} 
    \and DTU Space, Technical University of Denmark, Elektrovej 327, 2800 Kgs. Lyngby, Denmark \label{inst:DTU}
    \and Kapteyn Astronomical Institute, University of Groningen, P.O. Box 800, 9700 AV Groningen, The Netherlands \label{inst:Groningen}
    \and Telespazio UK for the European Space Agency, ESAC, Camino Bajo del Castillo s/n, 28692 Villanueva de la Ca\~{n}ada, Spain \label{inst:Telespazio}
    \and Steward Observatory, University of Arizona, 933 North Cherry Avenue, Tucson, AZ 85721, USA \label{inst:tucson}
    \and Leiden Observatory, Leiden University, PO Box 9513, 2300 RA Leiden, The Netherlands \label{inst:Leiden}
    \and UK Astronomy Technology Centre, Royal Observatory Edinburgh, Blackford Hill, Edinburgh EH9 3HJ, UK \label{inst:UKATC}  
   }

   \date{Received ; accepted}

\abstract
{This paper presents a deep MIRI/JWST medium resolution spectroscopy (MRS) covering the rest-frame optical spectrum of the GN-z11 galaxy. The [O\,III]\,5008$\AA$ and H$\alpha$ emission lines are detected and spectroscopically resolved. The line profiles are well-modeled by a narrow Gaussian component with intrinsic FWHMs of 189\,$\pm$\,25 and 231\,$\pm$\,52\,km\,s$^{-1}$, respectively. We do not find any evidence of a dominant broad H$\alpha$ emission line component tracing a Broad Line Region in a type 1 active galactic nuclei (AGN). The existence of an accreting black hole dominating the optical continuum and emission lines of GN-z11 is  not compatible with the measured H$\alpha$ and [O\,III]\,5008$\AA$ luminosities. If the well established relations for low-$z$ AGNs apply in GN-z11, the [O\,III]\,5008$\AA$ and H$\alpha$ luminosities would imply extremely large Super-Eddington ratios ($\lambda_{\mathrm{E}}$\,$>$\,290), and bolometric luminosities $\sim$\,20 times those derived from the UV/optical continuum. However, a broad ($\sim$\,430$-$470\,km\,s$^{-1}$) and weak ($<$\,20-30\%) H$\alpha$ line component, tracing a minor AGN contribution in the optical, cannot be ruled out completely with the sensitivity of the present data. The physical and excitation properties of the ionized gas are consistent with a low-metallicity starburst forming stars at a rate of SFR(H$\alpha$)\,$=$\,24\,$\pm$\,3\,$M_{\odot}$\,yr$^{-1}$. The electron temperature of the ionized gas is $T_{\mathrm{e}}$\,(O$^{++}$)\,$=$\,14000\,$\pm$\,2100\,K, while the direct-$T_{\mathrm{e}}$ gas-phase metallicity is 12\,$+$\,$\log$(O/H)\,$=$\,7.91\,$\pm$\,0.07 (Z\,=\,0.17\,$\pm$\,0.03\,Z$_{\odot}$). The optical line ratios locate GN-z11 in the starburst or AGN region but more consistent with those of local low-metallicity starbursts and high-$z$ luminous galaxies detected at redshifts similar to GN-z11. We conclude that the MRS optical spectrum of GN-z11 is consistent with that of a massive, compact, and low-metallicity starburst galaxy. Due to its high SFR and stellar mass surface densities, close to that of the densest stellar clusters, we speculate that GN-z11 could be undergoing a feedback-free, highly efficient starburst phase. Additional JWST data are needed to validate this scenario, and other recently proposed alternatives, to explain the existence of bright compact galaxies in the early Universe.}

\titlerunning{Insights into the Nature of GN-z11}
\maketitle

\section{Introduction}\label{Sect:intro}

JWST \citep{Gardner+23} is revolutionizing our knowledge of the formation of galaxies in the Epoch of Reionization (EoR) and beyond, currently placing the spectroscopically detection frontier of galaxies up to a redshift of about 14.2 \citep{Carniani+24,Carniani+24_ALMA}. The detection of strong optical emission lines, such as H$\beta$, [O\,III]\,4960,5008$\AA$, and H$\alpha$, together with weaker ultraviolet (UV) and optical lines, is starting to reveal the physical properties of galaxies at a redshift beyond 6 (e.g. \citealt{Cameron2023}). Star-forming galaxies (SFGs) are commonly characterized by emission lines with high equivalent widths (EWs), with rest-frame values up to 3000\,$\AA$ (e.g. \citealt{Boyett+24}). Their spectral energy distribution (SED) and high EW lines are indicative of young stellar populations, $<$\,10\,Myr, and intermediate stellar masses, $\log$(M$_{*}$\,[$M_{\odot}$])\,$<$\,9 (e.g.,  \citealt{Matthee+23,Rinaldi+23,Tang+23}). These SFGs have metallicities (Z) ranging from metal poor (0.02\,Z\,$_{\odot}$; \citealt{Vanzella2023}) up to one-third solar (e.g., \citealt{Langeroodi2023, Curti+23, Nakajima+23, Heintz2023, Sanders+24,Morishita2024,Hsiao+2024_MIRI}), high electron temperatures ($T_{\mathrm{e}}$\,$>$\,10$^{4}$\,K; e.g., \citealt{Sanders+24,Hsiao+2024_MIRI}) and densities (n$_{\mathrm{e}}$\,$>$\,100\,cm$^{-3}$; e.g., \citealt{Isobe+23, Abdurrouf+24}), and high photon ionization efficiencies (log$(\zeta_\mathrm{ion})$ ($\mathrm{Hz\,erg^{-1}}$) $\geq$ 25.5; e.g., \citealt{Tang+23, Alvarez-Marquez+23-MACS, Rinaldi+2024}). Also, active galactic nuclei (AGNs) associated with low-mass black holes (BH, 6\,$<$\,$\log$(M$_{\mathrm{BH}}$\,[$M_{\odot}$])\,$<$\,7) have been inferred in the EoR and in numbers larger than previously thought (e.g. \citealt{Greene+24,Harikane+23b,Kocevski+23,Maiolino+23b,Kokorev+23,Juodvzbalis+24}). Some of the properties exhibited by these AGNs suggest accretion properties different from those known in low-$z$ AGNs \citep{Maiolino-Xray2024,Kokubo-Harikane2024}, while the nature and contribution of AGNs in the so-called Little Red Dots (LRDs) identified by JWST are also under discussion \citep{PerezGonzalez-LRD2024, Iani-LRD2024}.

JWST has photometrically detected a large number of galaxies at redshift above 10 (e.g. \citealt{Finkelstein+23, Perez-Gonzalez+23b,Robertson+23}), but only a handful of them have been spectroscopically confirmed with the detection of the Ly$\alpha$ break or UV and optical emission lines \citep{Curtis-Lake+23,Harikane+23a, Arrabal-Haro+23, Bunker+23, Hsiao+23-NIRSpec, Zavala+2024, Castellano2024, Calabro2024,Carniani+24}. The physical properties of these primordial galaxies are still poorly known due to the limited rest-frame spectral range ($<$\,0.4\,$\mu$m) covered by NIRCam and NIRSpec. At these redshifts, the rest-frame optical emission lines and continuum moves into the mid-infrared \citep{Alvarez-Marquez+19_mrs}, the spectral range of MIRI \citep{Rieke+15,Wright+15,Wright+23}. Recent studies have detected the bright optical lines, [O\,III]\,4960,5008$\AA$ and H$\alpha$ in some of  these primordial galaxies highlighting the crucial role of MIRI in  constraining the nature and physical properties of galaxies during the first 500\,Myr of the Universe \citep{Alvarez-Marquez+23-MACS,Hsiao+2024_MIRI,Zavala+2024}.

GN-z11 was discovered by HST and \textit{Spitzer} based on the spectroscopy and multi-wavelength photometry \citep{Oesch+14,Oesch+16}. JWST NIRSpec has confirmed a spectroscopic redshift of 10.6034\,$\pm$\,0.0013 \citep{Bunker+23}. JWST NIRCam imaging has revealed that GN-z11 is an extremely compact galaxy at UV wavelengths, with an effective half-light radius (R$_{\mathrm{e}}$) of 64\,$\pm$\,20\,pc \citep{Tacchella+23}. It is composed of an unresolved central source and a second component defined by a S\'ersic profile with R$_{\mathrm{e}}$\,=\,200\,pc and index n\,$=$\,0.9. Additionally, it has a low-surface-brightness haze about 0\arcs.4 to the northeast of the galaxy \citep{Tacchella+23}. A tentative rotation has been inferred from the spatial analysis of the C\,III]1907,1909$\AA$ lines in the NIRSpec integral field spectroscopy (IFS, \citealt{Xu2024}). JWST multi-wavelength photometry and spectroscopy provide consistent SFR and stellar masses in the range of 20\,-\,30\,$M_{\odot}$\,yr$^{-1}$ and 10$^{8.7-9.1}$\,$M_{\odot}$ \citep{Bunker+23,Tacchella+23}, respectively. Millimeter NOEMA observations did not lead to a detection of the rest-frame 160\,$\mu$m continuum, nor of [C\,II]\,158$\mu$m emission in GN-z11, placing 3$\sigma$ upper limits  of $\log$(M$_{\mathrm{dust}}$\,[$M_{\odot}$])\,$<$\,6.9 and $\log$(M$_{\mathrm{mol,\,[C\,II]}}$\,[$M_{\odot}$])\,$<$\,9.3 on the dust and gas masses \citep{Fudamoto2024}, respectively. These results are consistent with a negligible dust attenuation derived from SED-fitting and Balmer decrement analyses \citep{Bunker+23,Tacchella+23}. Using medium resolution NIRSpec observations, \cite{Maiolino2024_BH} suggest that GN-z11 hosts an AGN with a black hole mass of $\log$(M$_{\mathrm{BH}}$\,[$M_{\odot}$])\,$=$\,6.2\,$\pm$\,0.3 accreting material at about five times its Eddington rate. This conclusion is based on the presence of high-excitation ($>$\,60\,eV), high critical density ($>$\,10$^{5}$\,cm$^{-3}$), and broad (430\,-\,470\,km\,s$^{-1}$) semi-forbidden and permitted emission lines in the UV spectrum. Further claims of an AGN come from the detection of a continuum excess (blue bump) in the rest-frame 3000\,$-$\,3550\,$\AA$ spectral range that could originate from complex Fe\,II emission \citep{Ji+2024}. However, a super-Eddington type 1 AGN scenario for GN-z11 is not supported by cosmological simulations \citep{Bhatt2024}, nor by the low 3$\sigma$ upper limit from deep X-ray Chandra observations (L$_{\mathrm{X}}$\,(2-10\,keV)\,$<$\,3\,$\times$\,10$^{43}$\,erg\,s$^{-1}$, \citealt{Maiolino2024_BH}) that is a factor of about 5 lower than the expected emission of a type 1 AGN given a BH mass of $\log$(M$_{\mathrm{BH}}$\,[$M_{\odot}$])\,$=$\,6.2\,$\pm$\,0.3. NIRSpec spectroscopy also shows GN-z11 with a super-solar N/O abundance, and a series of papers have searched for possible scenarios that could explain that over-abundance: (i) intense and densely clustered star formation with a high concentration of Wolf-Rayet stars \citep{Senchyna2024}, (ii) proto-globular cluster hosting super massive stars \citep{Charbonnel2023}, (iii) stellar collisions in a dense stellar cluster or a tidal disruption event \citep{Cameron2023}, (iv) intermittent star formation history with a quiescent phase lasting about 100\,Myr separating two strong starbursts \citep{Kobayashi2024}, and (v) formation of second generation stars from pristine gas and asymptotic giant branch ejecta in a massive globular cluster \citep{DAntona2023}. Finally, \citep{Maiolino-nebula2024} have suggested the presence of population III stars in the halo of GN-z11.

\begin{figure}
\centering
   \includegraphics[width=\linewidth]{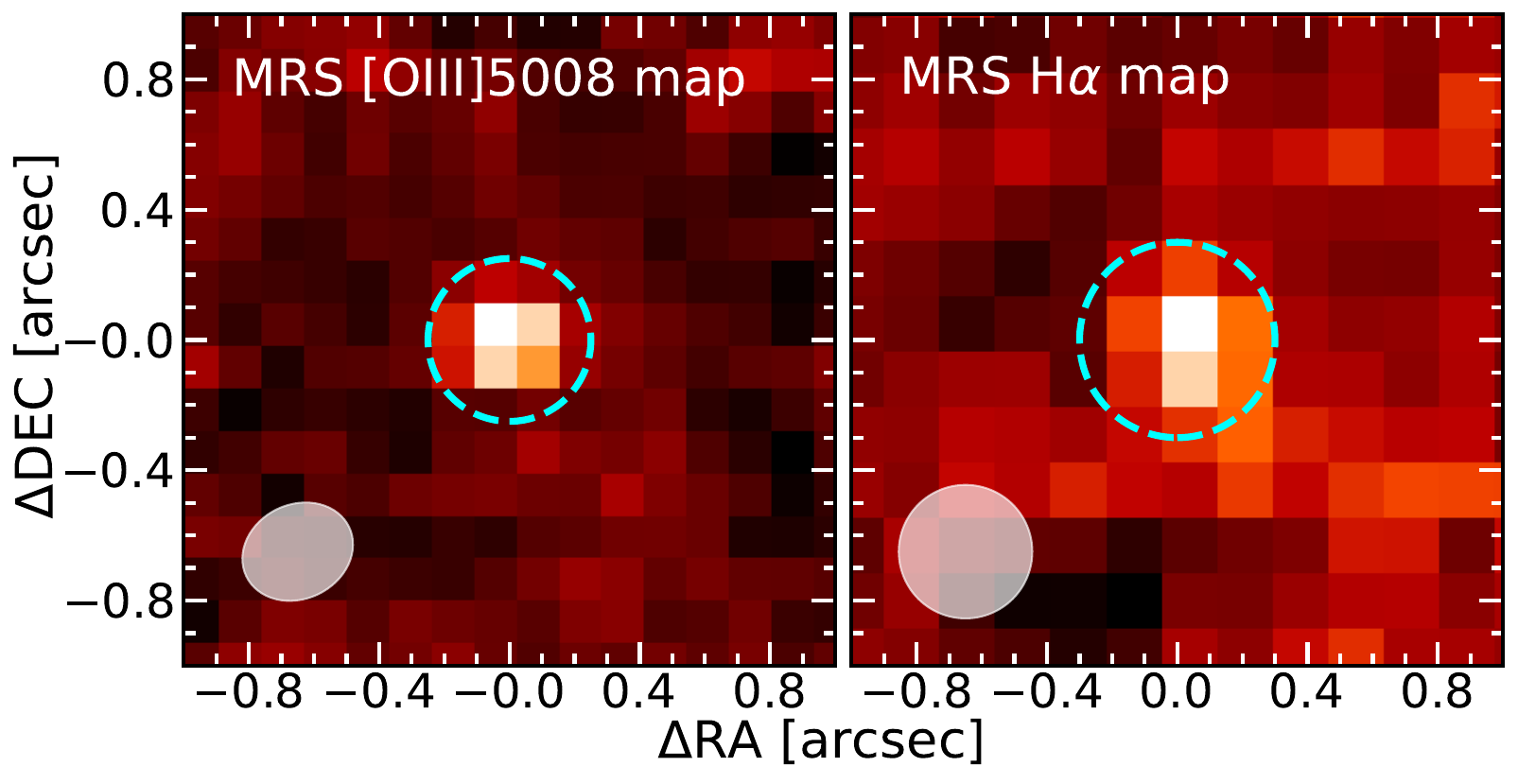}
      \caption{MRS [O\,III]5008$\AA$ (left) and H$\alpha$ (right) emission line maps. The [O\,III]\,5008$\AA$ and H$\alpha$ line maps are generated by integrating a narrow velocity range, $-$150\,<\,$v$\,[km\,s$^{-1}$]\,<\,150, around the peak of each emission line. Cyan dashed circles are the aperture chosen to extract the 1D spectra. The gray area represents the MRS spatial resolution (PSF FWHM) at the observed wavelength of each emission line. These line maps demonstrate that GN-z11 is spatially unresolved in the MRS observations.}
         \label{fig:linemap}
\end{figure}

This paper presents the detection of the brightest optical emission lines, [O\,III]\,4960,5008$\AA$ and H$\alpha$, in the GN-z11 galaxy at a redshift of 10.6. Section \ref{Sect2:data_calibration_lines} introduces the MRS observations and calibrations, together with the [O\,III]\,4960,5008$\AA$ and H$\alpha$ spectra and fluxes. Section \ref{Sect:results_dis} presents the results, which includes subsections discussing: Dust attenuation, SFR, and burstiness ($\S$\,\ref{Sect:SFR}), photon production efficiency and Ly$\alpha$ escape fraction ($\S$\,\ref{Sect:Photon_efficiency}), emission line ratios and diagnostic diagrams ($\S$\,\ref{Sect:line_ratios}), ISM metallicity and physical conditions ($\S$\,\ref{Sect:ISM_conditions}), ionized gas kinematics, gas and dynamical masses ($\S$\,\ref{Sect:kinematic}), and mass-metallicity relation ($\S$\,\ref{Sect:mass-metallicity}). Section \ref{Sect4:disc} discusses the AGN and starburst scenarios for GN-z11. Finally, Section \ref{Sect:conclusion_Summary} gives the summary and conclusion of the paper. Throughout this paper, we assume a Chabrier initial mass function (IMF, \citealt{Chabrier+03}), vacuum emission line wavelengths, and a flat $\Lambda$CDM cosmology with $\Omega_\mathrm{m}$\,=\,0.310, and H$_0$\,=\,67.7\,km\,s$^{-1}$\,Mpc$^{-1}$ \citep{PlanckCollaboration18VI}.

\section{Data, calibration and emission line spectrum}\label{Sect2:data_calibration_lines}

\subsection{MIRI observations and data calibration}\label{Sect:obs_cal}

GN-z11 was observed with the Medium Resolution Spectrograph (MRS, \citealt{Wells+15,Argyriou+23}) on the 13th of March 2024 as part of the cycle 2 JWST program ID\,2926 (PI: L. Colina). The observations were performed with the MRS SHORT and MEDIUM bands covering the wavelength ranges of 4.90$-$6.63\,$\mu$m and 7.51$-$10.13\,$\mu$m for channels 1 and 2, respectively. MRS channels 1 and 2 cover the optical rest-frame spectrum at a redshift of 10.6 including main optical emission lines, such as H$\beta$, [O\,III]\,4960,5008$\AA$, H$\alpha$, [N\,II]\,6550,6585$\AA$, and [S\,II]\,6718,6733$\AA$. The total on-source integration time corresponds to 25248 seconds (7\,hours) per MRS band, distributed in 24 dither positions combining the positive and negative dither patterns optimized for point-like sources. For each dither position, a total of 2 integrations with 22 groups each were obtained using the SLOWR1 readout mode. 

\begin{figure*}
\centering
   \includegraphics[width=\linewidth]{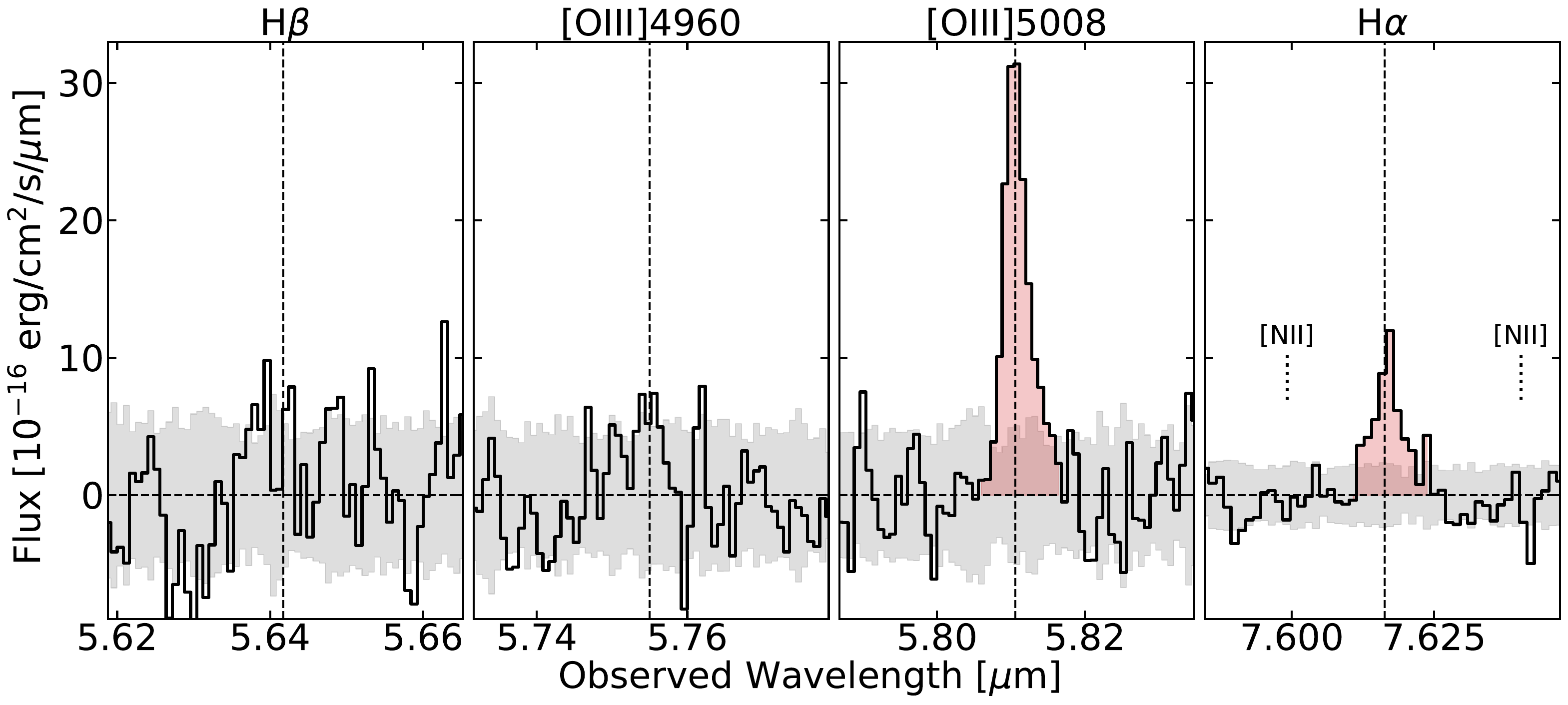}
      \caption{View of the rest-frame optical spectrum of GN-z11 by zooming in the H$\beta$, [O\,III]\,4960,5008$\AA$, and H$\alpha$ emission lines. Back continuous line: 1D extracted MRS spectrum. Gray area: $\pm1\sigma$ noise calculated from the standard deviation of the local background. Red area: spectral range used to calculate the integrated line flux. Black vertical dashed line: wavelength of the peak of each emission line considering a redshift of 10.602.}
         \label{fig:Emission_Lines}
\end{figure*}

The MRS observations are processed with version 1.16.0 of the JWST calibration pipeline and context 1293 of the Calibration Reference Data System (CRDS). We follow the standard MRS pipeline procedure \citep{bushouse_1.14.0}, with additional customized steps to improve the quality of the final MRS calibrated products (see \citealt{Alvarez-Marquez+23-MACS, Alvarez-Marquez+23-SPT} for details). The final channel 1 and 2 cubes have a spatial and spectral sampling of 0.13"\,$\times$\,0.13"\,$\times$\,0.8\,nm and 0.17"\,$\times$\,0.17"\,$\times$\,1.3\,nm \citep{Law+23}, respectively, and a resolving power of about 3500 \citep{Labiano+21, Jones+23}.

\subsection{MRS spectra and emission line fluxes}\label{Sect:MIRI_phot_spec}

There is not an evidence of spatially extended line emission in the MRS cubes (see Figure \ref{fig:linemap}). This is consistent with the small size of GN-z11, effective radius of 0\arcs.016\,$\pm$\,0\arcs.005, derived from NIRCam images \citep{Tacchella+23}. This size is smaller than the point spread function (PSF) full width half maximum (FWHM) of the MRS channels 1 and 2 ($\sim$\,0\arcsec.3\,$-$\,0\arcsec.4, \citealt{Argyriou+23}), and therefore, GN-z11 is considered spatially unresolved in the MRS observations. We extract the 1D integrated spectra of GN-z11 using a circular aperture with radius equal to 0\arcs.25 and 0\arcs.3 for channel 1 and 2 (see cyan dashed circles in Figure \ref{fig:linemap}), respectively. We also extract twenty 1D background spectra using the same apertures at random positions of the MRS FoV clean of any GN-z11 emission. We combine these  spectra to generate the 1D median and standard deviation of the local background for channels 1 and 2. The median, which is compatible with zero, is subtracted to the GN-z11 spectra with the goal of removing any systematic residual feature left in the MRS calibration process. The standard deviation is assumed to be the 1\,$\sigma$ uncertainty of the GN-z11 spectra. As the GN-z11 is considered spatially unresolved on the MRS observations, we have implemented an aperture correction to derive the total [O\,III]\,5008$\AA$ and H$\alpha$ spectra of GN-z11. For the selected apertures in channels 1 and 2, the aperture correction factor is 1.64 following the MRS PSF (\citealt{Argyriou+23}; Patapis in prep.).

We detect the optical rest-frame [O\,III]\,5008$\AA$ and H$\alpha$ lines in the MRS spectrum of GN-z11 and establish upper limits to the H$\beta$, [O\,III]\,4960$\AA$, [N\,II]\,6585$\AA$, and [S\,II]\,6718,6733$\AA$ lines (see Figure \ref{fig:Emission_Lines}). The flux of [O\,III]\,5008$\AA$ and H$\alpha$ is calculated by integrating each emission line in the velocity range of $-$250\,<\,$v$\,[km\,s$^{-1}$]\,<\,300, and and taking as a reference a redshift of 10.602. This velocity range delimits the spectral range at which the [O\,III]\,5008$\AA$ and H$\alpha$ lines have a positive flux (see red-filled area in Figure \ref{fig:Emission_Lines}). The integrated [O\,III]\,5008$\AA$ and H$\alpha$ lines fluxes are 136\,$\pm$\,14 and 68\,$\pm$\,9\,$\times$\,10$^{-19}$\,erg\,s$^{-1}$\,cm$^{-2}$ with a signal-to-noise ratios of 10 and 8, respectively. The 3$\sigma$ upper limits of H$\beta$, [O\,III]\,4960$\AA$, [N\,II]\,6585$\AA$, and [S\,II]\,6718,6733$\AA$ emission lines are calculated using the same velocity range as for [O\,III]\,5008$\AA$ and H$\alpha$. The non-detection of [O\,III]\,4960$\AA$ line is compatible with the theoretical [O\,III]5008-to-[O\,III]4960 flux ratio of 2.98. The H$\beta$ upper limit is also compatible with a non-detection considering the H$\alpha$ flux, case B recombination, and no dust attenuation (see $\S$\,\ref{Sect:SFR}). The emission line fluxes and upper limits of GN-z11 are summarized in Table \ref{tab:EL_Fluxes}.

\begin{figure*}
\centering
   \includegraphics[width=\linewidth]{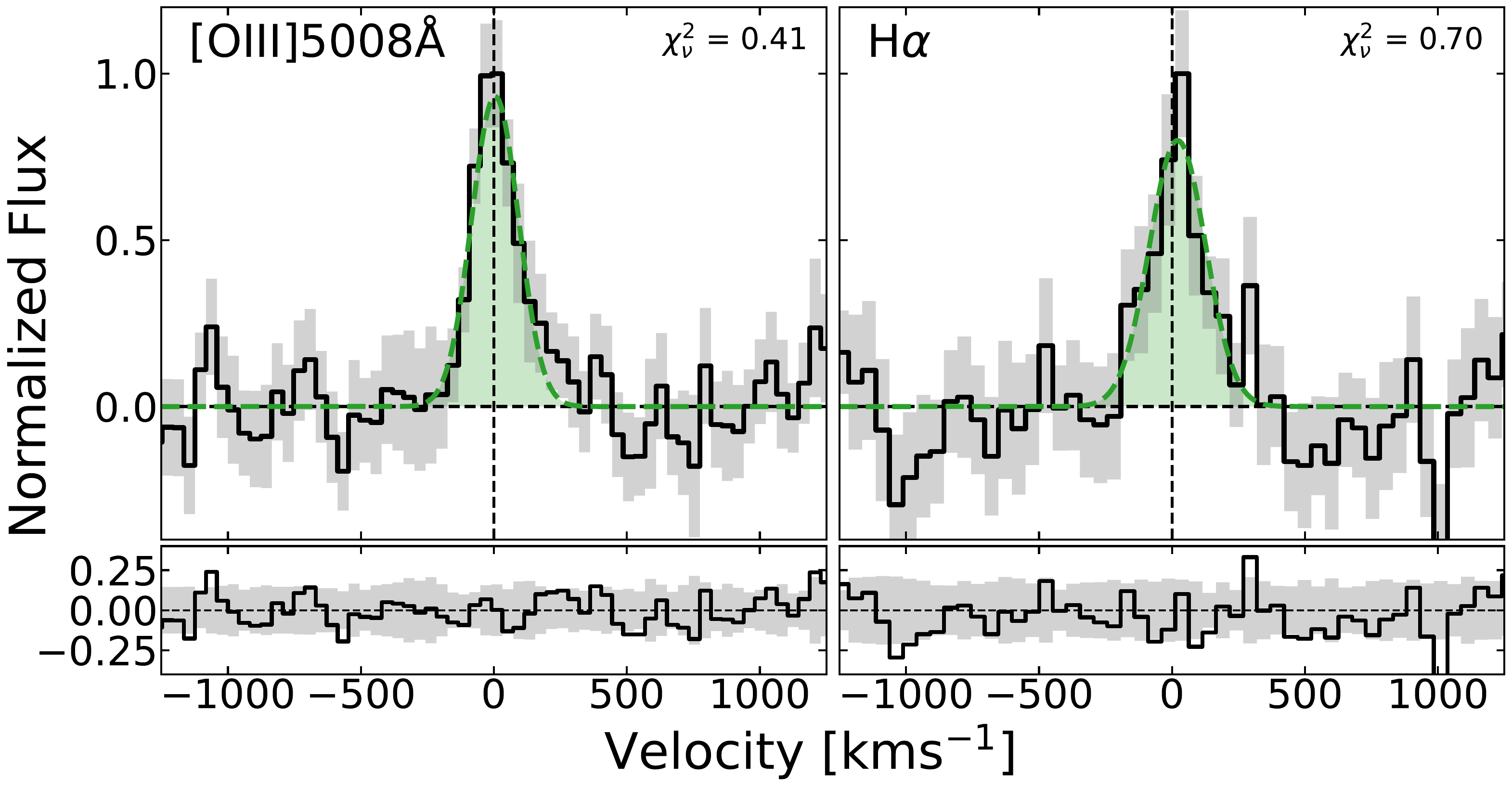}
      \caption{MRS [O\,III]\,5008$\AA$ and H$\alpha$ emission line fits. Left and right panels shows the one-component Gaussian fits, together with the fit residuals, for the [O\,III]\,5008$\AA$ and H$\alpha$ emission lines, respectively. Back continuous line: 1D extracted MRS spectrum. Gray area: $\pm1\sigma$ uncertainty calculated from the standard deviation of the local background. Green dashed line: one-components Gaussian function that best fits the spectra. Black vertical dashed line: wavelength at the peak of each emission line considering a redshift of 10.602. The $\chi_{\nu}^{2}$ for each emission line fit calculated in the velocity range, $-$500\,<\,$v$\,[km\,s$^{-1}$]\,<\,500, is included.}
         \label{fig:Emission_Lines_fit}
\end{figure*}

\subsection{[O\,III]\,5008$\AA$ and H$\alpha$ emission line profiles}\label{Sect:line_profiles}

The MRS provides the highest spectral resolution of all the JWST instruments, with a resolving power from 4000 to 3000 in channels 1 and 2 \citep{Labiano+21, Jones+23}. Specifically, the spectral resolution (FWHM) at the observed wavelengths of the redshifted [O\,III]\,5008$\AA$ and H$\alpha$ lines corresponds to 86\,$\pm$\,9\,km\,s$^{-1}$ and 91\,$\pm\,$9\,km\,s$^{-1}$, respectively. This spectral resolution, together with the significant detection (8\,$-$\,10$\sigma$) of the lines, allows to explore the emission line profiles, and therefore the kinematic properties of GN-z11.

The [O\,III]\,5008$\AA$ and H$\alpha$ emission lines are spectrally resolved (see Figure~\ref{fig:Emission_Lines}). We have performed one-component line fit to characterize the [O\,III]\,5008$\AA$ and H$\alpha$ emission lines profiles. For the fit, we consider the two emission lines independent, each modeled by a single Gaussian component without any initial constrains (see Figure \ref{fig:Emission_Lines_fit}). The final uncertainties in the fit parameters, such as FWHM and redshift, are obtained as the standard deviation of all the individual measurements of 1000 bootstrapped spectra after adding to the original spectrum a random Gaussian noise equal to the RMS. Table \ref{tab:EL_Fluxes} presents the [O\,III]\,5008$\AA$ and H$\alpha$ FWHMs and redshifts derived from the fits. The redshift of GN-z11 is 10.6022\,$\pm$\,0.0004, calculated from the [O\,III]\,5008$\AA$ emission line. It is compatible with the value derived from the H$\alpha$, and agrees within the uncertainties with the one previously reported based on the combination of the R100 and R1000 NIRSpec observations (10.6034\,$\pm$\,0.0013, \citealt{Bunker+23}).

The [O\,III]\,5008$\AA$ line profile shows a slight flux excess at velocities ranging from 100 to 350\,km\,s$^{-1}$ (see Figure \ref{fig:Emission_Lines_fit}), that could trace the presence of a redshifted faint secondary component as already identified in other high-$z$ galaxies (e.g. \citealt{Topping+24,Tang+23,Marconcini2024}). However, the significance of this secondary component is low (2$\sigma$ for a two-component Gaussian fit), and its confirmation requires deeper MRS observations.

\begin{table}[!ht]
\caption{Integrated emission line fluxes and ratios, results of the [O\,III]\,5008$\AA$ and H$\alpha$ emission line fits, and derived physical properties of GN-z11.}
\centering
\begin{tabular}{lcc}
\hline
F(H$\beta$)$^{a,b,c}$  & $<$\,44 \\
F([O\,III]\,4960\AA)$^{a,c}$  & $<$\,44 \\
F([O\,III]\,5008\AA)$^{a,c}$  & 136\,$\pm$\,14 \\ 
F(H$\alpha$)$^{a,c}$  & 68\,$\pm$\,9 \\
F([N\,II]\,6585\AA)$^{a,b,c}$  & $<$\,28 \\
F([S\,II]\,6718,6733\AA)$^{a,b,c}$  & $<$\,38 \\
Redshift$_{\mathrm{[O\,III]\,5008\AA}}^{d}$ & 10.6022\,$\pm$\,0.0004 \\
Redshift$_{\mathrm{H\alpha}}^{d}$ & 10.6029\,$\pm$\,0.0006 \\
FWHM$_{\mathrm{[O\,III]\,5008\AA}}$\,[km\,s$^{-1}$]$^{d}$ & 189\,$\pm$\,25 \\
FWHM$_{\mathrm{H\alpha}}$\,[km\,s$^{-1}$]$^{d}$ & 231\,$\pm$\,52 \\
R2$^{e}$  &  -0.45\,$\pm$\,0.06\\
R3$^{e}$  & 0.75\,$\pm$\,0.07\\
R23$^{e}$ &  0.89\,$\pm$\,0.07\\
O32$^{e}$ &  1.32\,$\pm$\,0.05 \\
O33$^{e}$ & 1.77\,$\pm$\,0.14 \\
O3H$\gamma$$^{e}$ & -0.71\,$\pm$\,0.14 \\ 
N2$^{e}$ & $<$$-$0.4 \\
S2$^{e}$ & $<$$-$0.3 \\
$T_{\mathrm{e}}$\,(O$^{++}$)\,[K] & 14000\,$\pm$\,2100 \\
12\,+\,$\log$(O/H) &  7.91\,$\pm$\,0.07\\
$\log$($U$)$^{f}$ &  $-$1.81\,$\pm$\,0.03 \\
SFR(H$\alpha$)\,[$M_{\odot}$\,yr$^{-1}$]$^{g}$ & 24\,$\pm$\,3\\
$\log$(sSFR)\,[yr$^{-1}$]$^{h}$ & $-$7.6\\
$\log(\zeta_\mathrm{ion})$ [$\mathrm{Hz\,erg^{-1}}$] & 25.66\,$\pm$\,0.06 \\
$f_\mathrm{esc}(\mathrm{Ly\alpha})$ & 0.04\,$\pm$\,0.01 \\
M$_{\mathrm{dyn}}$\,[$\times$\,10$^{9}$\,$M_{\odot}$] & 1.1\,$\pm$\,0.4 \\
\hline
\label{tab:EL_Fluxes}
\end{tabular}
\tablefoot{$^{a}$ All line fluxes are given in units of 10$^{-19}$\,erg\,s$^{-1}$\,cm$^{-2}$. $^{b}$ 3$\sigma$ upper limits. $^{c}$ Fluxes and upper limits are calculated by integrating the spectra in the velocity range of $-$250\,<\,$v$\,[km\,s$^{-1}$]\,<\,300, and taking a redshift of 10.602 as a reference. $^{d}$ FWHMs and redshifts are derived from the one-component Gaussian fit. $^{e}$ Definitions for the line ratios are given in $\S$\,\ref{Sect:line_ratios}. $^{f}$ The log($U$) is computed using the O32 ratio and the calibration presented in \citet{Berg+19b} for 0.1\,$Z_\mathrm{\odot}$. $^{g}$ SFR is derived considering the H$\alpha$ luminosity and 0.1\,$Z_{\odot}$ \citep{Theios+19}. $^{h}$ The sSFR is derived from the H$\alpha$ SFR and the mean of stellar masses from the SED-fitting of the NIRSpec spectrum \citep{Bunker+23} and NIRCam photometry \citep{Tacchella+23}.}
\end{table}

\subsection{NIRSpec ancillary emission line fluxes}
\label{Sect:NIRSpec_ancillary}

To derive line ratios and diagnostic diagrams involving the [O\,II]3727,3730$\AA$, [O\,III]\,4364$\AA$, and H$\gamma$ lines, we use the published lines fluxes from the NIRSpec R100 spectrum measured using a 5-pixel extraction (see table 1 from \citealt{Bunker+23}). The agreement between the NIRSpec R100 spectrum and the NIRCam photometry \citep{Tacchella+23} after aperture corrections has been demonstrated to be excellent, and within 5\% \citep{Bunker+23}. Therefore, we do not implement any normalization factor to the NIRSpec line fluxes when combining with the aperture corrected MRS [O\,III]\,5008$\AA$ and H$\alpha$ fluxes. This work uses only the relative uncertainties associated to the specific observations. The additional uncertainty due to the present knowledge of the NIRSpec and MIRI absolute spectrophotometry is not taken into account. In general, this is estimated at the 5\% and 10\% level for NIRSpec IFS and MRS JWST observing modes \citep{Rigby2023}\footnote{Additional information on the absolute flux calibrations in the JWST instruments: \url{https://jwst-docs.stsci.edu/jwst-calibration-status}}, respectively.

\section{Results}\label{Sect:results_dis}

\subsection{Dust attenuation, SFR, and burstiness}\label{Sect:SFR}

Previous analyses of NIRSpec spectroscopy and NIRCam photometry have derived a low (A$_{\mathrm{V}}$\,$=$\,0.17\,$\pm$\,0.03\,mag, \citealt{Bunker+23}) or even negligible (A$_{\mathrm{V}}$\,=\,0.08\,$^{+0.23}_{-0.06}$\,mag, \citealt{Tacchella+23}) dust attenuation in GN-z11. The MRS spectroscopy enables to revisit those measurements and extend them into the optical wavelengths using the H$\alpha$ emission line. The observed H$\alpha$/H$\beta$, H$\alpha$/H$\gamma$, and H$\alpha$/H$\delta$ line ratios are $>$\,1.5, 5.7\,$\pm$\,1.0 and 10.8\,$\pm$\,2.5, respectively. These Balmer line ratios agree, within the uncertainties, with case B recombination, 2.80\,$\pm$\,0.07, 5.93\,$\pm$\,0.09, and 10.68\,$\pm$\,0.17, respectively, for a $T_{\mathrm{e}}$\,=\,14000\,$\pm$\,2100\,K and n$_{\mathrm{e}}$\,=\,1000\,cm$^{-3}$ (see $\S$\,\ref{Sect:ISM_conditions}). These results are compatible with no, or large Super-Eddington ratios ($\lambda_{\mathrm{E}}$\,$>$\,29very little, dust attenuation in GN-z11, and therefore, an A$_{\mathrm{V}}$\,=\,0\,mag is assumed throughout this paper.

The instantaneous star formation rate (SFR) derived from the total H$\alpha$ luminosity, SFR(H$\alpha$), is 56\,$\pm$\,8, 34\,$\pm$\,5, 24\,$\pm$\,3, and 22\,$\pm$\,3\,$M_{\odot}$\,yr$^{-1}$ for metallicities equal to solar \citep{Kennicutt-Evans+12}, 0.28\,$Z_\odot$ \citep{Reddy+18}, 0.1\,$Z_\odot$ \citep{Theios+19}, and 0.05\,$Z_\odot$ \citep{Reddy+22}, respectively. These values are in agreement with those derived from the UV luminosity. Following \cite{Calzetti+13} and an UV luminosity density of 1.7\,$\times$\,10$^{29}$\,erg\,s$^{-1}$\,Hz$^{-1}$ at 1500\,$\AA$ \citep{Bunker+23}, we derive a SFR(UV) of 15, 22, and 51\,$M_{\odot}$\,yr$^{-1}$ for a solar metallicity and constant star formation over  100, 10, and 2\,Myr, respectively. The SFR(UV) decreases to 12\,$M_{\odot}$\,yr$^{-1}$ for a metallicity of 0.1\,$Z_\odot$ and a constant star formation over a 100\,Myr \citep{Theios+19}. The metallicity and star formation history (SFH) highly affect the derivation of the SFR, which is increased by a factor of two from 0.1\,Z$_{\odot}$ to solar metallicity, and by a factor of three from 100 to 2\,Myr in constant SFHs. For a metallicity of $\sim$\,0.1\,Z$_{\odot}$ and a constant SFH over 10\,Myr, the SFR in GN-z11 traced by the UV continuum and H$\alpha$ line is in the 18 and 24\,$M_{\odot}$\,yr$^{-1}$ range. These values are consistent with the SFR previously derived from NIRSpec spectroscopy \citep{Bunker+23} and NIRCam SED-fitting analysis \citep{Tacchella+23}, 18.8\,$^{+0.8}_{-0.7}$ and 21\,$^{+22}_{-10}$\,$M_{\odot}$\,yr$^{-1}$, respectively.

The SFR(H$\alpha$)-to-SFR(UV) ratio is often used to study the star-formation burstiness in galaxies (\citealt{2016ApJ...833..254S, Atek2022}, and references therein). The motivation is based on the different timescales for the star formation that H$\alpha$ (<\,10\,Myr) and UV continuum ($\sim$\,100\,Myr) trace. We obtain a SFR(H$\alpha$)-to-SFR(UV) ratio equal to 2.0\,$\pm$\,0.3 for a metallicity equal to 0.1\,$Z_\odot$, and considering the SFR(H$\alpha$) and SFR(UV) derived using a constant SFH over 10\,Myr and 100\,Myr, respectively. A SFR(H$\alpha$)-to-SFR(UV) ratio larger than one implies that GN-z11 is currently in a bursty phase, and suggest that its stellar population is dominated by young stars ($\leq$\,10\,Myr). However, the measured SFR(H$\alpha$)-to-SFR(UV) ratio of GN-z11 places it on somewhat lower star formation burstiness relative to $z$\,$>$\,8 galaxies with an average value of $\sim$\,4 \citep{langeroodi-bursty}.

\subsection{Photon production efficiency and Ly$\alpha$ escape fraction} \label{Sect:Photon_efficiency}

The ionizing photon production efficiency, $\log$($\zeta_\mathrm{ion}$), is given as the ratio of the ionizing to the non-ionizing UV flux. It can be derived from the H$\alpha$ and UV continuum luminosities, as long as the ionizing photon escape fraction ($f_\mathrm{esc,\,LyC}$) is known. The SED-fitting and Ly$\alpha$ profile analyses of GN-z11 gave $f_\mathrm{esc,\,LyC}$ equal to 0.03$^{+0.05}_{-0.02}$ \citep{Bunker+23} and 0.024 \citep{Hayes2023}, respectively. An indirect measurement of the $f_\mathrm{esc,\,LyC}$ using the UV slope ($\beta_{UV}$) provides a larger value of 0.11$\pm$0.09 \citep{Chisholm2022}, but compatible within the uncertainties. Given the low $f_\mathrm{esc,\,LyC}$ values measured by different methods, we assume a $f_\mathrm{esc,\,LyC}$\,=\,0 throughout this paper.

Following the prescription presented in \cite{Alvarez-Marquez+23-MACS}, an ionizing photon production efficiency of $\log$($\zeta_\mathrm{ion}$)\,=\,25.66\,$\pm$\,0.06\,Hz\,erg$^{-1}$ is obtained based on our MRS H$\alpha$ luminosity, and the UV luminosity density of 1.7\,$\times$\,10$^{29}$\,erg\,s$^{-1}$\,Hz$^{-1}$ at 1500\,$\AA$ \citep{Bunker+23}. This value agrees with the previous one derived from NIRSpec  H$\gamma$ luminosity (25.67\,$\pm$\,0.02\,Hz\,erg$^{-1}$; \citealt{Bunker+23}). The photon production efficiency of GN-z11 is significantly larger than the canonical value (25.2\,$\pm$\,0.1\,Hz\,erg$^{-1}$; \citealt{Robertson+23}), but follows the envelope of the value inferred from intermediate redshift galaxies of $\log$($\zeta_\mathrm{ion}$) and extrapolated as a function of redshift (25.46\,$\pm$\,0.28\,Hz\,erg$^{-1}$, \citealt{Matthee+17}). The $\log$($\zeta_\mathrm{ion}$) value in GN-z11 is typical of EoR sources (e.g. \citealt{Atek+24, Fujimoto+23, Tang+23, Morishita+23, Rinaldi+2024, Alvarez-Marquez+23-MACS, Morishita2024,Simmonds2024}), and of intermediate redshift, 2\,$<$\,$z$\,$<$\,5, galaxies with 
the highest specific SFR, $\log$(sSFR[yr$^{-1}$])\,$\sim$\,$-$7.5,  \citep{Castellano+23}. Comparing with galaxies at $z$\,$>$\,10, GN-z11 has similar values than GHZ2/GLASS-z12 galaxy at a redshift of 12.34 (25.7$_{-0.1}^{+0.2}$\,Hz\,erg$^{-1}$, \citealt{Calabro2024}), but it is significantly higher than MACS0647-JD galaxy at a redshift of 10.17 (25.3\,$\pm$\,0.1\,Hz\,erg$^{-1}$, \citealt{Hsiao+2024_MIRI}). The similarity of GN-z11 with the photon production efficiency of the resolved young stellar cluster detected in the strongly-lensed $z$\,$\sim$\,6 Sunrise arc \citep{Vanzella+23}, and the spatially resolved strong lensed galaxy RXCJ0600-z6-3, \citep{Gimenez-Arteaga2024}, suggests that GN-z11 could be formed by a combination of many young stellar clusters with ages at their maximum photon production efficiency (see $\S\,$\ref{Sec.Dis_SFR} for more details).

The Ly$\alpha$ escape fraction can be derived combining our H$\alpha$ with existing Ly$\alpha$ measurements.
The Ly$\alpha$ emission line was detected in medium-resolution (R\,$\approx$\,1000) NIRSpec observations with a total flux of 23\,$\pm$\,3\,$\times$\,10$^{-19}$\,erg\,s$^{-1}$\,cm$^{-2}$ \citep{Bunker+23}. Assuming a case B recombination Ly$\alpha$/H$\alpha$ line ratio of 9.28 (for $T_\mathrm{e}$\,=\,14000\,K and $n_\mathrm{e}$\,=\,1000\,cm$^{-3}$; see $\S$\,\ref{Sect:ISM_conditions}), we derive a Ly$\alpha$ escape fraction of 0.04\,$\pm$\,0.01 uncorrected by the effect of the intergalactic medium. This value is in agreement with the previously reported value using the H$\gamma$ (0.03$_{-0.02}^{+0.05}$, \citealt{Bunker+23}).

\subsection{Emission line ratios and diagnostic diagrams}\label{Sect:line_ratios}

We combine the MRS [O\,III]\,4960,5008$\AA$ and H$\alpha$ line fluxes and upper limits of [N\,II]\,6585$\AA$ and [S\,II]\,6718,6733$\AA$, with previous NIRSpec R100 [O\,III]\,4364\AA, [O\,II]3727,3730\AA, and H$\gamma$ line fluxes (see $\S$\,\ref{Sect:NIRSpec_ancillary} for more details) to derive commonly used line ratios. The following emission line ratios are used in the diagnostic diagrams:

\begin{eqnarray*}
    \mathrm{R2} &=& \log ([\mathrm{O\,II}]3727, 3730\AA/\mathrm{H}\beta) \\
    \mathrm{R3} &=& \log ([\mathrm{O\,III}]5008\AA/\mathrm{H}\beta) \\
    \mathrm{R23} &=& \log (([\mathrm{O\,III}]5008\AA + [\mathrm{O\,II}]3727, 3730\AA)/\mathrm{H}\beta) \\
    \mathrm{O32}  &=& \log ([\mathrm{O\,III}]4960,5008\AA/[\mathrm{O\,II}]3727, 3730\AA) \\
    \mathrm{O33}  &=& \log ([\mathrm{O\,III}]5008\AA/[\mathrm{O\,III}]4364\AA) \\
    \mathrm{O3H\gamma}  &=& \log ([\mathrm{O\,III}]4364\AA/\mathrm{H}\gamma) \\
    \mathrm{N2} &=& \log ([\mathrm{N\,II}]6585\AA/
    \mathrm{H}\alpha) \\
    \mathrm{S2} &=& \log ([\mathrm{S\,II}]6718,6733\AA/
    \mathrm{H}\alpha) \\
\end{eqnarray*}

\noindent
where H$\beta$ is derived from the H$\alpha$ assuming a case B recombination H$\alpha$/H$\beta$ line ratio of 2.80 (for $T_\mathrm{e}$\,=\,14000\,K and n$_\mathrm{e}$\,=\,1000\,cm$^{-3}$; see $\S$\,\ref{Sect:ISM_conditions}). The line ratios along with their associated uncertainties, are listed on Table~\ref{tab:EL_Fluxes}.

\begin{figure*}
\centering
   \includegraphics[width=0.95\linewidth]{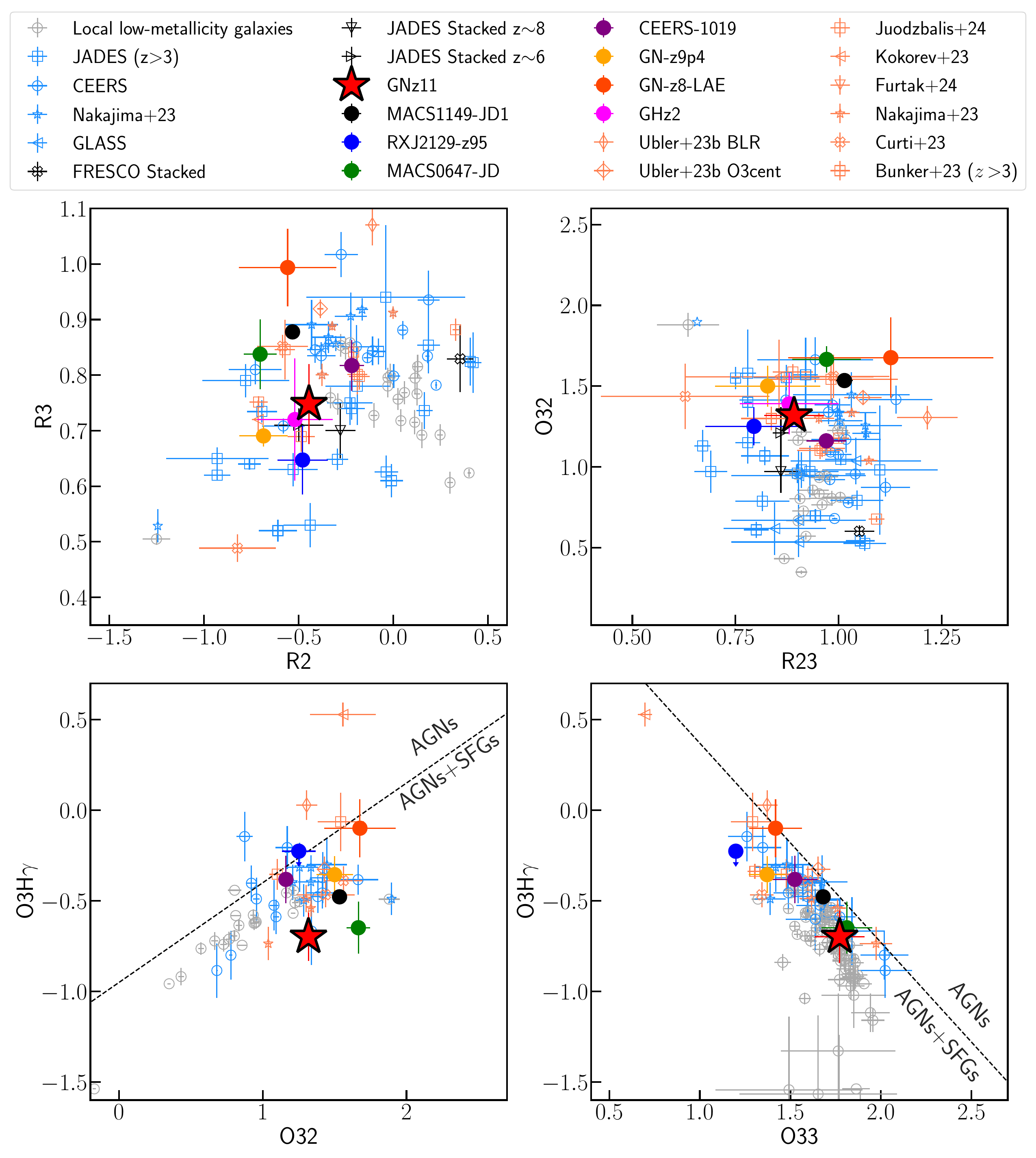}
      \caption{Line ratio diagrams. This figure shows the position of GN-z11 (red star) on the R2\,$-$\,R3, R23\,$-$\,O32, O32\,$-$\,O3H$\gamma$ and O32\,$-$\,O33 diagrams. JWST-detected $z$\,>\,8 galaxies are represented as filled circles. Blue and red markers represent the high-$z$ SFGs and type 1 AGNs from different samples, respectively. Further details of these samples are presented in $\S$\ref{Sect:line_ratios}. The local low-metallicity sub-sample (gray markers), drawn from \citet{Izotov+06,Izotov+11,Izotov+16,Izotov+19,Izotov+24}, is selected to be 12\,+\,log(O/H)\,<\,8. JADES sub-sample, drawn from \citet{Bunker+23} and \citet{Cameron2023}, is selected to be $z$\,$>$\,3. Dashed lines represent the separation between the AGN-dominant and the AGN+SFGs regions presented by \citet{Mazzolari+24}. Note that, when no H$\gamma$ data is available, we have considered H$\gamma$\,=\,H$\alpha$/5.93.}
         \label{fig:Line_ratios}
\end{figure*}

Figure~\ref{fig:Line_ratios} presents the position of GN-z11 on the R3$-$R2, O32$-$R23, O3H$\gamma-$O32 and O3H$\gamma-$O33 diagrams. For comparison, we include the line ratios of galaxies available at $z$\,$>$\,8: GN-z8-LAE ($z$\,$=$\,8.3; \citealt{Navarro-Carrera+24}), CEERS-1019 ($z$\,$=$\,8.7; \citealt{Marques-Chaves2024}), MACS1149-JD1 ($z$\,$=$\,9.1; \citealt{Stiavelli+23}), GN-z9p4 ($z$\,$=$\,9.4; \citealt{Schaerer+24}), RXJ2129-z95 ($z$\,$=$\,9.5; \citealt{Williams+23}), MACS0647-JD ($z$\,$=$\,10.2: \citealt{Abdurrouf+24}) and GHz2 ($z$\,$=$\,12.3; \citealt{Calabro2024}). Samples of SFGs and type 1 AGNs at different redshifts are also presented. On the one hand, we include $z$\,$>$\,3 SFGs identified in large JWST surveys (CEERS, \citealt{Sanders+24}; JADES, \citealt{Bunker+23,Cameron2023}; and GLASS, \citealt{Mascia+23}) as well as individual targets \citep{Nakajima+23} and the stacked FRESCO sub-sample of [O\,III] line emitters at $z$\,$>$\,6.8 \citep{Meyer+24}. For completeness, we also included the ratios for the JADES stacked sub-samples at $z$\,$\sim$\,6 and $z$\,$\sim$\,8 from \citet{Cameron2023}. On the other hand, we include a sample of high-$z$ ($z$\,$>$\,5) type 1 AGNs observed with JWST \citep{Bunker+23,Curti+23,Kokorev+23,Nakajima+23,Furtak+24,Juodvzbalis+24,Ubler+23b}. Finally, we also show a reference sample of known low-redshift metal-poor star-forming galaxies (12\,+\,log(O/H)\,$<$\,8.0; \citealt{Izotov+06,Izotov+11,Izotov+16,Izotov+24}). 

Although high-$z$ galaxies identified as type 1 AGNs and low-metallicity SFGs at high and low redshift have some overlap in the R3$-$R2 and O32$-$R23 diagrams, a large fraction of SFGs populate the region with R3$<$0.8 and O3$<$1.25 while most AGNs have values above these limits.  Type 1 AGNs and SFGs are also well separated in the recent O3H$\gamma-$O32 and O3H$\gamma-$O33 diagnostic diagrams \citep{Mazzolari+24}. Most confirmed type 1 AGNs show a O3H$\gamma$\,$>$\,-0.5 with a large fraction of the high-$z$ SFGs and the low-$z$ SFGs below that value. The position of GN-z11 in these diagrams, with ratios 0.75\,$\pm$\,0.07 (R3), 1.32\,$\pm$\,0.05 (O32) and $-$0.71\,$\pm$\,0.14 (O3H$\gamma$) within the range of values covered by local low-metallicity and high-$z$ SFGs, favours a starburst nature for the ionizing source. However, ionisation models are able to reproduce the range of observed line ratios covered by both AGNs and SFGs with the ionizing spectrum of an AGN when low-metallicity, high electron density, and large ionization parameters are invoked \citep{Nakajima+22, Calabro2024, Mazzolari+24}. Therefore the position of GN-z11 in these diagrams should not be taken in itself as the unambiguous identification of a starburst as the dominant ionizing source in this galaxy, and additional tracers should be considered.

Upper limits for the [N\,II]\,6585$\AA$ and [S\,II]\,6718,6733$\AA$ emission lines have also been obtained (see Table~\ref{tab:EL_Fluxes}). We can therefore explore the position of GN-z11 in the traditional BPT diagrams \citep{Baldwin+81, Kewley+01, Kauffmann+03}. In addition to the R3 value of 0.75, 3$\sigma$ upper limits for the N2 and S2 line ratios of $-$0.4 and $-$0.3 are derived, respectively. These values place GN-z11 in the starburst or composite region in the R3$-$N2 diagram, and in the starburst to Seyfert transition region in the R3$-$S2 diagram. GN-z11 has a gas-phase metallicity of 12\,+\,$\log$(O/H)\,$=$\,7.91. There is a strong decrease in the nitrogen and sulfur lines relative to hydrogen for lower metallicities (e.g. \citealt{Cameron2023}), and therefore the actual N2 and S2 upper limits can not place stringent constraints using these diagnostics. Deeper MRS observations are needed to measure the [N\,II]\,6585$\AA$ and [S\,II]\,6718,6733$\AA$ fluxes, or establish more strict upper limits, which would help to determine the exact position of GN-z11 in these traditional BPT diagrams.

\subsection{Gas-phase metallicity and physical conditions ($T_\mathrm{e}$, $U$)}\label{Sect:ISM_conditions}

The significant detection of the [O\,III]\,5008$\AA$ line with the MRS, together with the published [O\,III]\,4364$\AA$ flux \citep{Bunker+23}, allows us to derive the $T_\mathrm{e}$ of the O$^{++}$ region using the [O\,III]5008/[O\,III]4364 flux ratio and the metallicity applying the direct-$T_\mathrm{e}$ method.

We assume a n$_\mathrm{e}$\,=\,1000\,cm$^{-3}$ based on the redshift evolution of the electron density presented in \cite{Abdurrouf+24}. According to this relation, the electron density of the ISM in galaxies increases with redshift ($\propto$\,(1+$z$)$^{1.2\pm0.04}$). For the redshift of GN-z11 ($z$\,=\,10.602) the predicted electron density is $\sim$ 1000 cm$^{-3}$, in agreement with the value ($\log$(n$_\mathrm{e}$\,[cm$^{-3}$])\,=\,2.9\,$\pm$\,0.5) in MACS0647-JD at $z$\,=\,10.2 (\citealt{Abdurrouf+24}, see also \citealt{Isobe+23}). We do not consider any uncertainty in the n$_{\mathrm{e}}$ determination as the $T_{\mathrm{e}}$ is insensitive to density changes for n$_{\mathrm{e}}$\,$\lesssim$\,10$^{5}$\,cm$^{-3}$. The electron density we assume in this work for the optical emitting gas in GN-z11 is at least two orders of magnitude smaller than the one derived from the UV lines, which are larger than 10$^5$\,cm$^{-3}$, and as high as 10$^9$\,cm$^{-3}$ \citep{Maiolino2024_BH}. At these high densities, the measured [O\,III]\,5008/[O\,III]\,4363 ratio would be produced in a ionazed medium with very low electron temperatures, 8100\,K and 4600\,K for electron densities of 10$^{6}$ and 10$^{10}$\,cm$^{-3}$, respectively. However, these low temperatures are not compatible with those derived from the [O\,III]\,4363/[Ne\,III]\,3869 ratio, covering the range 10500 to 23600\,K for different metallicities \citep{Cameron2023_MNRAS523}. Differences in the electron densities derived from the UV and optical lines have been previously found in galaxies at low \citep{Mingozzi+22}, intermediate \citep{Acharyya+19}, and high (\citealt{Ji+Ubler2024}, Crespo-G\'omez in prep.) redshifts. These differences in the density are understood as the evidence for an stratified interstellar medium (ISM) where the lower ionization optical lines trace more extended, lower density ionized regions while the  higher ionization UV lines probe inner and denser ionized gas clouds. This stratification has been observed in galaxies with a well identified AGN (GS$\_$3073 at $z$\,$=$\,5.5, \citealt{Ji+Ubler2024}), and with an extremely compact starburst (RXCJ2248-ID at $z$\,$=$\,6.1, \citealt{Topping+24}; Crespo-G\'omez, in prep.). Thus, the stratification of the ionized ISM, covering a wide range of electron densities, appears to be independent of the ionizing source, either young massive starburst or AGN \citep{Pascale2023, Ji+2024}.

We obtain an electron temperature for the O$^{++}$ region of $T_\mathrm{e}$\,[O$^{++}$]\,=\,14000\,$\pm$\,2100\,K considering an O33 ratio of 1.77\,$\pm$\,0.14 and a n$_\mathrm{e}$ of 1000\,cm$^{-3}$ using the \texttt{getTemDen} module from \texttt{Pyneb} code \citep{Luridiana+15}. Following the $T_\mathrm{e}$\,[O$^{+}$]\,$=$\,0.7\,$\times$\, $T_\mathrm{e}$\,[O$^{++}$]\,+\,3000\,K relation presented by \citet{Campbell+86}, we obtain an electron temperature for the low-ionization zone of $T_\mathrm{e}$\,[O$^{+}$]\,$=$\,12800\,$\pm$\,1500\,K. We derive a compatible value (i.e. 13400\,K) when we applied the $T_\mathrm{e}$\,[O$^{++}$]\,$-$\,$T_\mathrm{e}$\,[O$^{+}$] relation for low-metallicity galaxies (i.e., $Z$\,$\sim$\,0.1\,$Z_\odot$) presented in \citet{Izotov+06}. We also used the \cite{genesis-metallicity} non-parametric calibration of the $T_\mathrm{e}$\,[O$^{++}$]--$T_\mathrm{e}$\,[O$^{+}$] relation to estimate the $T_\mathrm{e}$\,[O$^{+}$] based on the directly measured $T_\mathrm{e}$\,[O$^{++}$], R2, and R3. We find $T_\mathrm{e}$\,[O$^{+}$] = $15000 \pm 450\,$K.

The ionic O$^+$/H$^+$ and O$^{++}$/H$^+$ abundances are calculated using \texttt{Pyneb} for the $T_\mathrm{e}$ appropriate to each ion and the R2 and R3 line ratios (see Table~\ref{tab:EL_Fluxes}), respectively. We assume the total oxygen abundance to be the sum of these two phases, as the contribution of O$^{+++}$ has been found to be negligible even in high-ionization sources \citep{Berg+18,Berg+21}. This procedure yields a 12\,+\,$\log$(O/H) value of 7.91\,$\pm$\,0.07 which corresponds to 0.17\,$\pm$\,0.03\,$Z_\odot$, assuming a solar metallicity value of 12\,+\,log(O/H)\,=\,8.69 \citep{Asplund+09}.\footnote{If the electron density was as high as the lower limit derived from the UV lines (10$^5$\,cm$^{-3}$), the direct-T$_{\mathrm{e}}$ metallicity would move from 0.17 to 0.54\,$Z_\odot$.} We obtain similar results, 0.16\,$\pm$\,0.02\,$Z_\odot$, when using the expressions from \citet{Perez-Montero+17} based on the R2 and R3 ratios. A slightly lower metallicity, 0.08\,$\pm$\,0.02\,$Z_\odot$, is derived when using the metallicity indicator proposed by \citet{Izotov+19}, which combines R23 and O32 to improve the accuracy in the low-metallicity regime. Note that the \citet{Izotov+19} relation was derived for 12\,+\,log(O/H)\,$<$\,7.5 and therefore our value is based on an extrapolation. Additional empirical relations based on local analogs of high-$z$ galaxies presented in \citet{Bian+18} yield a slightly higher metallicity of 12\,+\,log(O/H)\,=\,8.05\,$\pm$\,0.15 (i.e. $\sim$\,0.23\,$Z_\odot$). Using the \texttt{genesis-metallicity} \citep{genesis-metallicity}, adopting the non-parametric electron temperature measurements mentioned above, we find 12\,+\,log(O/H)\,=\,7.86\,$\pm$\,0.11, fully compatible with the values from the direct-T$_{e}$ method.

For completeness, we consider the hypothesis that the ionization of the nebular emission traced by the oxygen lines could be due to an AGN and follow the so-called $T_\mathrm{e}$-AGN method (\citealt{Dors+20b} and \citealt{Dors+21}) to derive the metallicity. This method is a modification of the $T_\mathrm{e}$ direct method for HII regions considering a different $T_\mathrm{e}$\,[O$^{++}$]\,$-$\,$T_\mathrm{e}$\,[O$^{+}$] relation. Following this method and using the O32 ratio, we obtain $T_\mathrm{e}$\,[O$^{+}$]\,=\,9400\,K and $T_\mathrm{e}$\,[O$^{++}$]\,=\,14100\,K. 
If we consider that the O$^+$/H$^+$\,+\,O$^{++}$/H$^+$ abundance trace the $\sim$\,80$\%$ of the total O/H abundance in AGNs \citep{Dors+20a}, we obtain a metallicity of 12\,+\,log(O/H)\,=\,7.93. This value is in agreement with those obtained assuming HII ionisation and, therefore, both HII and AGN ionisation processes predict the same metallicity in GN-z11. 

Though we have considered different methods based on multiple line ratios in this section, we observe that all of them give similar metallicities with values ranging between 0.08 and 0.23\,$Z_\odot$. These results strengthen the value obtained with \texttt{Pyneb} (i.e. 12\,+\,log(O/H)\,=\,7.91\,$\pm$\,0.07), which we assume as the fiducial value. A combined ALMA and JWST analysis of JADES-GS-z14-0 galaxy at $z$\,$=$\,14.2 shows metallicity similar to GN-z11 (0.17\,Z$_{\odot}$; \citealt{Carniani+24_ALMA}). The bright galaxies (M$_{\mathrm{UV}}$\,=\,-21.8$-$-20.8) GN-z11 and JADES-GS-z14-0 already present an enriched ISM 300\,$-$\,440\,Myr after the Big Bang. This, together with their estimated high stellar and dynamical masses ($\sim$\,10$^{9}$\,$M_{\odot}$), suggest that massive and efficient stellar build up happened at even earlier times in the Universe.

The detection of [O\,III]\,5008$\AA$ for GN-z11 allow us to compute its O32 ratio (1.32\,$\pm$\,0.05), which is a commonly used tracer of the ionisation parameter. Based on O32\,$-$\,log($U$) expressions presented in \citet{Diaz+00} and \citet{Papovich+22}, we obtain an ionisation parameter log($U$) of -2.06\,$\pm$\,0.06 and -1.81\,$\pm$\,0.02, respectively. Using the metallicity-dependant calibration presented in \citet{Berg+19b}, based on \texttt{CLOUDY} models, we obtain a similar value ($-$1.97\,$\pm$\,0.06) adopting 0.1\,$Z_\odot$. These results agree with a previous value (log($U$)\,=\,$-$2.3\,$\pm$\,0.9) derived from SED fitting \citep{Bunker+23}. Besides, we observe that the ionisation parameter log($U$)\,$\sim$\,-2 of GN-z11 is in agreement with those values found in other $z$\,>\,8 galaxies. In particular, \citet{Zavala+2024} found a log($U$)\,=\,$-$1.8\,$\pm$\,0.3 for GHZ2 at $z$\,=\,12.3 while \citet{Hsiao+2024_MIRI} obtained a log($U$)\,=\,$-$2.0\,$\pm$\,0.1 for MACS0647-JD at $z$\,=\,10.2. 

\subsection{Ionized gas kinematics, gas and dynamical masses}\label{Sect:kinematic}

The [O\,III]\,5008$\AA$ and H$\alpha$ emission lines are well-modelled by a one-component Gaussian fit, with intrinsic FWHMs of 189\,$\pm$\,25 and 231\,$\pm$\,52\,km\,s$^{-1}$ (see Figure \ref{fig:Emission_Lines_fit}), respectively. These FWHMs are about a factor of two narrower than the FWHMs (430$-$470\,km\,s$^{-1}$) of the Mg\,II and N\,IV] UV emission lines obtained from a medium-resolution (R\,$\approx$\,1000) NIRSpec spectrum \citep{Maiolino2024_BH}. There is, therefore, no evidence of a dominant broad H$\alpha$ component associated with the Broad Line Regions (BLR) of a type 1 AGN with a BH mass of $\log(M_\mathrm{BH}[M_{\odot}])=6.2\pm0.3$, as previously inferred from the UV lines (see \citealt{Maiolino2024_BH} and $\S$\,\ref{Sec:disc_AGN} for a more detailed discussion). However, if a FWHM of 470$-$430\,km\,s$^{-1}$ and a BH mass of $\log(M_\mathrm{BH}[M_{\odot}])\,=\,6.2$ is assumed, the expected flux of a broad H$\alpha$ component associated with the BLR is about 2$-$3\,$\times$\,10$^{-18}$\,erg\,s$^{-1}$\,cm$^{-2}$. That H$\alpha$ component will be under the detection limit ($\approx$\,2$\sigma$) of the current MRS observation, and therefore, a non-dominant, weak broad H$\alpha$ component associated with the BLR cannot be ruled out in this work. This hypothetical component, if present, would represent a small fraction ($<$\,20$-$30\,\%) of the total flux of the H$\alpha$ line.    

A good correlation exists between the velocity dispersion of the stellar distribution and the width of the [O\,III]\,5008$\AA$ line, even in type 1 AGNs (e.g. \citealt{Greene+05a,Le2023}). The dynamical mass can therefore be derived from the width of the [O\,III]\,5008$\AA$ systemic component (FWHM of 189\,$\pm$\,25\,km\,s$^{-1}$) under the hypothesis that the mass enclosed within the compact size of GN-z11 (64\,$\pm$\,20\,pc, \citealt{Tacchella+23}) is virialized, and with a non significant rotational component. Following \cite{Alvarez-Marquez+23-MACS}, a dynamical mass of 1.1\,$\pm$\,0.4\,$\times$\,10$^9$\,$M_{\odot}$ is measured for GN-z11. This value agrees with the estimated stellar masses derived from the NIRSpec R100 spectrum (M$_*$\,=\,5.4$^{+0.8}_{-0.7}$\,$\times$\,10$^8$\,$M_{\odot}$; \citealt{Bunker+23}) and NIRCam multi-wavelength SED-fitting (M$_*$\,=\,1.3$^{+1.2}_{-0.8}$\,$\times$\,10$^9$\,$M_{\odot}$; \citealt{Tacchella+23}), and suggests that the dynamical mass in this compact source is dominated by the stellar component. An upper limit to the dynamical mass of 3.5\,$\pm$\,0.3\,$\times$\,10$^9$\,$M_{\odot}$ is obtained if the size of the [O\,III] emitting gas corresponds to that of the extended component (196\,$\pm$\,12\,pc) identified in the NIRCam images \citep{Tacchella+23}. 

\begin{figure*}
\centering
   \includegraphics[width=\linewidth]{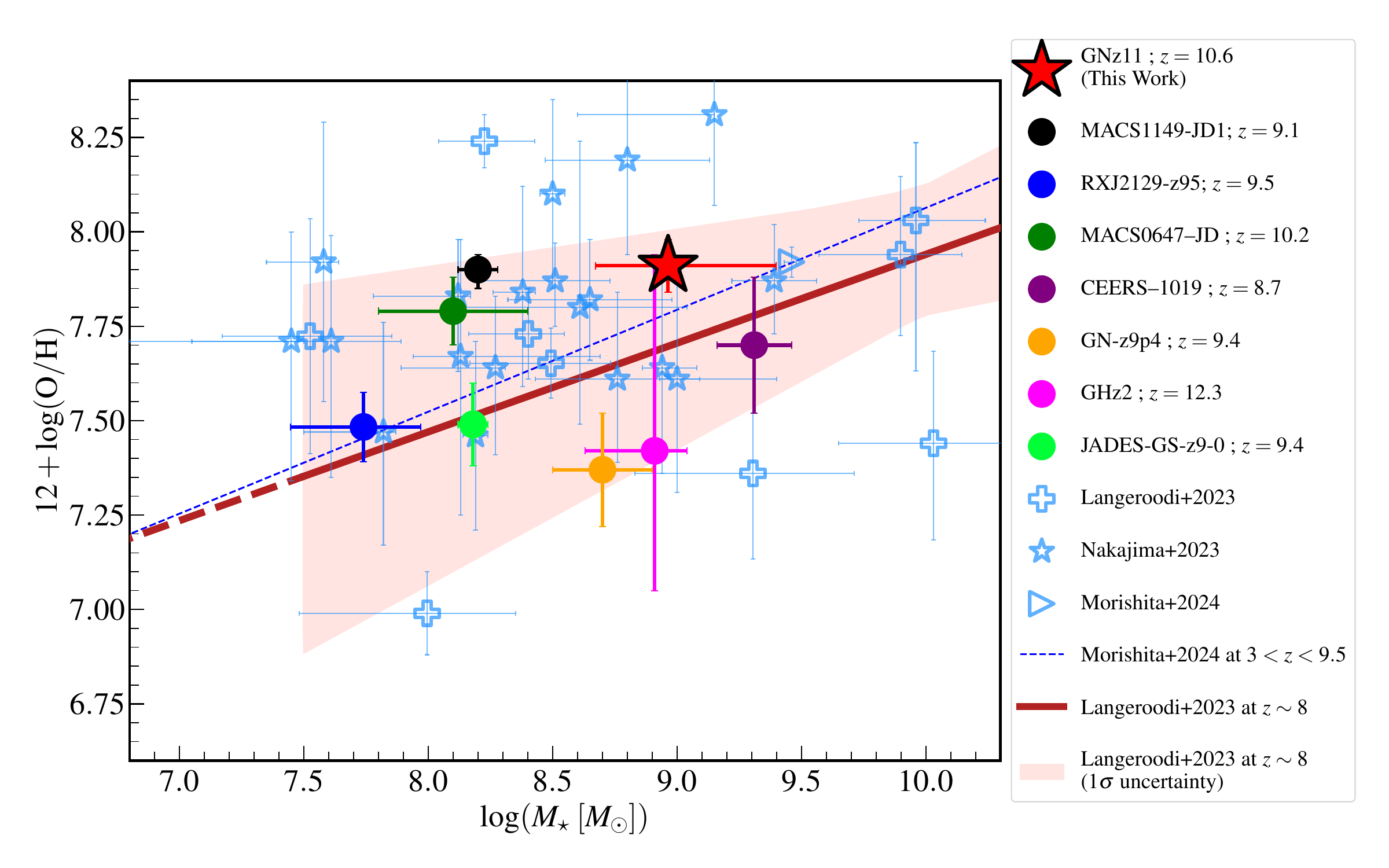}
      \caption{Mass-metallicity diagram for $z$\,$>$\,7 galaxies. The large red star shows the measured stellar mass and gas-phase metallicity for GN-z11. The colored circles show the location of the remaining $z$\,$>$\,9 galaxies, compiled from the literature: MACS1149-JD1 \citep{Stiavelli+23}, RXJ2129-z95 \citep{Williams+23}, MACS0647–JD \citep{Hsiao+2024_MIRI}, CEERS–1019 \citep{Nakajima+23, Marques-Chaves2024}, GN-z9p4 \citep{Schaerer-Rui2024}, GHz2 \citep{Zavala+2024}, and JADES-GS-z9-0 \citep{2024A&A...681A..70L, Curti+24}. The blue empty symbols indicate the mass and metallicity of the 7\,$<$\,$z$\,$<$\,9 galaxies, compiled from \cite{Langeroodi2023} (pluses), \cite{Nakajima+23} (stars), \cite{2024A&A...684A..75C} (squares), and \cite{Morishita2024} (triangles). The blue dashed line shows the best-fit mass-metallicity relation based on a sample of direct-method metallicity measurements at 3\,$<$\,$z$\,$<$\,9.5 \citep{Morishita2024}. The red solid line and the red shaded region show the best-fit mass-metallicity relation and its uncertainty at $z \sim 8$ as measured by \cite{Langeroodi2023}. GN-z11 is in agreement with the best-fit mass-metallicity relation at $z \sim 8$, consistent with little to no evolution in the normalization of the mass-metallicity relation at $z$\,$>$\,6. Note that the stellar mass of GN-z11 is a mean of the values presented in \citet{Bunker+23} and \citet{Tacchella+23}.}
         \label{fig:Mass-metallicity}
\end{figure*} 

The total amount of ionized gas can be derived from the H$\alpha$ luminosity and assuming all the H$\alpha$ emitting gas is emitted in the low density regions. Following \cite{Dopita03}, and considering an electron density of 10$^3$\,cm$^{-3}$, a total ionized gas mass of $\sim$\,2\,$\times$\,10$^7$\,$M_{\odot}$ is obtained. The mass in ionized gas represents a small fraction ($<$\,2\,\%) of the total mass in the galaxy and is larger than the 3$\sigma$ upper limit of the dust mass ($\log(\mathrm{M_{dust}}[M_{\odot}])<6.9$) obtained from deep NOEMA imaging \citep{Fudamoto2024}. The amount of molecular mass based on the 3$\sigma$ [C\,II]158$\mu$m upper limit corresponds to $<$\,2\,$\times$\,10$^9$\,$M_{\odot}$ \citep{Fudamoto2024}. In addition, the size (Str\"omgren radius) of the ionized nebula can also be estimated in the ideal case of a medium with a constant density, and assuming an ionizing escaping fraction equal to zero. For an average electron density of 10$^3$\,cm$^{-3}$, a radius of 64\,pc is obtained. Although this size is in good agreement with the effective radius of GN-z11 \citep{Tacchella+23}, it could represent a lower limit to the radius of an ionized nebula with a lower density and lower surface brightness ($\sim$\,300\,pc for a density of 10$^2$\,cm$^{-3}$). Evidence for an extended ionized nebula around GN-z11 at distances of 0.8\,kpc \citep{Bunker+23} and even low surface brightness up to 0\arcsec.4 (i.e., 1.6\,kpc) have been claimed from NIRSpec Ly$\alpha$ integral field spectroscopy \citep{Maiolino-nebula2024}. Deeper integral field spectroscopy with JWST is required to identify the presence and structure of this low surface brightness, extended ionized nebula with optical lines. 

\subsection{Mass-metallicity relation}\label{Sect:mass-metallicity}

We investigate the chemical evolutionary stage of GN-z11 in the context of its location on the mass-metallicity diagram. Figure \ref{fig:Mass-metallicity} shows the position of GN-z11 and other $z$\,$>$\,7 galaxies in the mass-metallicity diagram. This includes galaxies at cosmic dawn such as MACS1149-JD1 at $z$\,$=$\,9.11 \citep{Stiavelli+23}, RXJ2129-z95 at $z$\,$=$\,9.51 \citep{Williams+23}, MACS0647–JD at $z$\,$=$\,10.17 \citep{Hsiao+2024_MIRI}, CEERS–1019 at $z$\,$=$\,8.7 \citep{Nakajima+23, Marques-Chaves2024}, GN-z9p4 at $z$\,$=$\,9.38 \citep{Schaerer-Rui2024}, GHz2 at $z$\,$=$\,12.3 \citep{Zavala+2024}, JADES-GS-z9-0 at $z$\,$=$\,9.43 \citep{2024A&A...681A..70L, Curti+24}, and other 7\,$<$\,$z$\,$<$\,9 galaxies from \cite{Langeroodi2023}, \cite{Nakajima+23}, \cite{2024A&A...684A..75C} and \cite{Morishita2024}. The highest-$z$ inference of the mass-metallicity relation available in the literature, at $z$\,$\sim$\,8 \citep{Langeroodi2023}, is shown as the red line and the red shaded region. 

The stellar mass and gas-phase metallicity of GN-z11 are in agreement with the best-fit mass-metallicity relation at $z$\,$\sim$\,8. In general, this applies to the $z$\,$>$\,9 galaxies shown in Figure \ref{fig:Mass-metallicity}, all of which are scattered around the $z$\,$\sim$\,8 relation and are consistent with this relation within $1\sigma$. This suggests that the normalization of the mass-metallicity relation does not evolve noticeably at $z$\,$>$\,8. This interpretation is consistent with previous results based on numerical and semi-analytical simulations \citep[see e.g.,][]{2016MNRAS.456.2140M, 2023MNRAS.518.3557U, 2024ApJ...967L..41M} as well as recent observational JWST NIRSpec results \citep{Langeroodi2023, 2024A&A...684A..75C, Nakajima+23}, all of which suggest that the redshift evolution of the normalization of the mass-metallicity relation significantly slows down at redshifts above $z$\,=\,6.   

\section{Discussion. The nature of GN-z11}\label{Sect4:disc}

The new MRS measurements of the [O\,III]\,5008$\AA$ and H$\alpha$ lines establish the electron temperature and metallicity of the ionized gas. The derived direct-$T_{\mathrm{e}}$ gas-phase metallicity of GN-z11 follows the mass-metallicity relation for galaxies above redshift 8. The ratios of the optical emission lines in GN-z11 are in the range of values typically measured in low-$z$ low-metallicity and high-$z$ star-forming galaxies, favouring the starburst nature of GN-z11 (see also \citealt{Tacchella+23, Bunker+23}). However, previous diagnostics based on the ratios of some ultraviolet lines and their widths suggest that GN-z11 could host an AGN (\citealt{Maiolino2024_BH}, \citealt{Ji+2024}). In this section, we discuss the AGN and starburst scenarios of GN-z11 from the perspective of the new MRS observations.

\subsection{A type 1 AGN?}\label{Sec:disc_AGN}

The presence of a type 1 AGN in GN-z11 has been proposed based on the detection of some UV emission lines in the medium-resolution (R\,$\approx$\,1000) NIRSpec spectrum \citep{Maiolino2024_BH}. The lines trace the presence of relatively broad semi-forbidden (FWHM\,=\,470\,$\pm$\,50\,km\,s$^{-1}$, N\,IV]\,1487$\AA$) and permitted lines (FWHM\,=\,430\,$\pm$\,65\,km\,s$^{-1}$, Mg\,II\,2800$\AA$), and some lines emitted in regions with high electron densities (>\,10$^5$\,cm$^{-3}$, [N\,IV]\,1483,1486$\AA$, N\,III] multiplets). These properties are compatible with the scenario where the line emission comes from a Broad Line Region (BLR) around a black hole with a mass $\log$(M$_\mathrm{BH}$[$M_{\odot}]$)\,=\,6.2$\pm$0.3. The differences in the width (FWHM) of the UV lines, with the semi-forbidden (N\,IV]\,1486$\AA$) and permitted (Mg\,II\,2800$\AA$) lines broader than the optical forbidden lines ([O\,II]\,3727,3729$\AA$ and [Ne\,III]\,3870$\AA$), have been interpreted as evidence for a Narrow Line Seyfert 1 (NLS1) type AGN in GN-z11 \citep{Maiolino2024_BH}. Further, to explain the continuum at rest-frame wavelengths 3000-4000\,$\AA$, a scenario invoking the presence of two different ionized regions is considered: a hot and dense BLR dominating the broad Balmer emission and a colder, less-dense gas emitting the narrow forbidden lines and also contributing (but sub-dominant) to the Balmer emission \citep{Ji+2024}. The presence of high energy photons (above 64\,eV) has been identified by the detection of the  [Ne\,IV]\,2422,2424$\AA$ emission line. This is considered as an unambiguous tracer of AGN radiation \citep{Maiolino2024_BH}. However, recent NIRSpec integral field spectroscopy have identified extended [MgIV]4.487$\mu$m (ionization potential of 80 eV) emission in some low$-z$ luminous infrared starburst galaxies with high SFR surface densities \citep{Pereira-Santaella2024}. The line ratios are consistent with the presence of shocks with velocities of about 100$-$130 km s$^{-1}$. Such shocks could also produce strong [Ne\,IV]\,2422,2424$\AA$ emission, around 10\% of H$\beta$ emission (Pereira-Santaella, private comm.). Thus, while AGNs naturally produce high energetic radiation, shocks could also produce high energy photons capable of ionizing lines such as [Ne\,IV]\,2422,2424$\AA$. 

In the following we explore the presence of a dominant type 1 AGN and the scenario of two ionized regions in the light of our new MRS observations. We first discuss the implications derived from the hypothesis that the H$\alpha$ line is fully emitted in a BLR around a massive black hole. We further discuss whether the MRS data favors the hypothesis of two different ionized regions.

\subsubsection{Width of the H$\alpha$ line emission and BH mass}

The high spectral resolution of the MRS allows us to resolve the H$\alpha$ line, and establish its intrinsic line width and luminosity. The mass of the potential BH can be estimated following the well established empirical relationship between the BH mass and the H$\alpha$ FWHM and luminosity (\citealt{Shen2011}, see also \citealt{Bosman2024}). A good agreement has been demonstrated between the BH mass estimates based on the Mg\,II\,2800$\AA$ and H$\alpha$ lines in the 10$^7$\,$-$\,10$^9$\,$M_{\odot}$ mass range \citep{Matsuoka2013}. This has been also confirmed recently for the luminous quasar J1120+0641 at a redshift of 7.1. The H$\alpha$ MRS measurements show a very good agreement with previous Mg\,II estimates (1.52\,$\pm$\,0.17)\,$\times$\,10$^9$\,$M_{\odot}$ and (1.35\,$\pm$\,0.04)\,$\times$\,10$^9$\,$M_{\odot}$ \citep{Bosman2024}, respectively. 

Under the hypothesis that the luminosity and width of the H$\alpha$ line in GN-z11 were entirely produced in a high-density Broad Line Region (BLR) due to the radiation from an AGN, the BH mass in GN-z11 would be 6$_{-3}^{+4}$\,$\times$\,10$^5$\,$M_{\odot}$ for an H$\alpha$ FWHM of 231\,$\pm$\,52\,km\,s$^{-1}$.  The derived BH mass is about a factor of three below the value of $\log$(M$_\mathrm{BH}$[$M_{\odot}]$)\,=\,6.2$\pm$0.3 \citep{Maiolino2024_BH} estimated from the lower resolution NIRSpec spectrum and the Mg\,II\,2800$\AA$ line, but compatible within the uncertainties. The agreement apparently supports the presence of a low-mass type 1 AGN in the nucleus of GN-z11. Note that BH mass value would be correct only under the assumption that the known H$\alpha$ to BH mass relation is still valid for a narrow H$\alpha$ line such as measured in GN-z11. The same applies to the empirical relation between the mass of a BH and the Mg\,II\,2800$\AA$ line properties established for low-redshift type 1 AGNs \citep{Vestergaard2009}. However, there is no empirical evidence that the emission line to BH mass relation can be directly extrapolated to the narrow H$\alpha$ emission line detected in GN-z11. As presented in the following sections, the empirical evidence for a $\sim$\,0.6$-$1.4\,$\times$\,10$^6$\,$M_{\odot}$ BH with radiation properties similar to those of low-$z$ AGNs is not supported by the luminosities of the [O\,III]5008$\AA$ and H$\alpha$ lines.

\subsubsection{H$\alpha$ versus X-ray and optical luminosities}

Following the scenario where the H$\alpha$, optical continuum, and X-ray 2-10\,keV emission is dominated by an AGN, the presence  of a type 1 AGN appears to be in contradiction with two well known relations for low-$z$ AGNs: (i) the H$\alpha$ to the 2-10\,keV X-ray emission \citep{Ho2001,Jin2012}, and (ii) the H$\alpha$ to the optical continuum flux at 5100\,$\AA$ \citep{Greene+05b}.

GN-z11 has not been detected in X-rays but a 3$\sigma$ upper limit of 3\,$\times$\,10$^{43}$\,erg\,s$^{-1}$ has been derived for its 2$-$10 keV luminosity ($L_\mathrm{X}$(2-10\,keV)) assuming the typical photon index for Narrow Line Seyfert 1 galaxies (NLS1, \citealt{Maiolino-Xray2024}). Following the $L_\mathrm{X}$(2-10\,keV)-to-$L$(H$\alpha$) relation, the expected  $L_\mathrm{X}$(2-10\,keV) would be in the range of 1.2$-$1.8\,$\times$\,10$^{44}$\,erg\,s$^{-1}$, i.e. a factor 4 to 6 larger than the current 3$\sigma$ upper limit for GN-z11. However, the estimated X-ray emission in high-$z$ type 1 AGNs is highly uncertain \citep{Maiolino-Xray2024} and therefore the discrepancy identified in GN-z11 has to be taken with caution. Recent analysis of deep Chandra X-ray 2-10\,keV observations for the high-$z$ JWST-detected type 1 AGNs indicates that most of them are not detected, placing strong upper limits in their $L_\mathrm{X}$(2-10\,keV) luminosities \citep{Maiolino2024_BH}. These AGNs appear to be one to two orders of magnitude fainter than their low-$z$ counterparts.  A similar conclusion is derived for Little Red Dots (LRDs) in the lensing cluster A2744 containing AGNs \citep{Ananna2024}. The X-ray undetected sources imply that either the BH are less massive than previously thought or the accretion of material proceeds without a strong X-ray emitting corona. In summary, the results derived for GN-z11, as well as for the high-$z$ JWST-detected AGNs suggest that, if a type 1 AGN is present, the X-ray spectrum does not follow the standard SED of low-$z$ counterparts. On the other hand, type 2 AGNs are underluminous in X-ray relative to H$\alpha$, with a median $L_\mathrm{X}$(2-10\,keV)/$L$(H$\alpha$)\,=\,2 \citep{Ho2001}.  GN-z11 with a $L_\mathrm{X}$(2-10\,keV)/$L$(H$\alpha$)\,<\,2.8 could also be consistent with hosting a type 2 low luminosity AGN. Thus, based on the Halpha to X-ray emission, neither the presence nor the non-existence of an AGN can be concluded. 

In the optical, there is a well established correlation between the H$\alpha$ and the 5100\,$\AA$ luminosities for type 1 AGNs and covering a wide range of H$\alpha$ luminosities (10$^{41}$ to 10$^{44}$\,erg\,s$^{-1}$, \citealt{Greene+05b}). This correlation indicates that the optical continuum measured at 5100\,$\AA$ traces the far-UV ionizing continuum producing the optical emission lines. Following \citet{Greene+05b}, the predicted flux at the redshifted optical continuum is 0.24\,$\mu$Jy, well above the extrapolated continuum flux of 0.10\,$\mu$Jy from the NIRCam imaging \citep{Tacchella+23}, and NIRSpec spectrum \citep{Bunker+23}. This result does not support the idea that GN-z11 has a standard type 1 AGN in its nucleus dominating the nebular emission traced by the H$\alpha$ line. Further arguments supporting this conclusion come from the predicted AGN bolometric luminosity and accretion rate derived from the continuum and line flux measurements.

\subsubsection{AGN bolometric luminosities and BH Eddington ratios}

Assuming the observed UV-optical continuum and optical emission lines are produced by the radiation of an AGN, its bolometric luminosity can be derived applying different bolometric correction factors to the continuum luminosities \citep{Netzer2019}, and the H$\beta$ and [O\,III]\,5008$\AA$ lines \citep{Pennell2017,Netzer2019}. Taking NIRCam F150W and F356W \citep{Tacchella+23} observed fluxes and the 5.9\,$\mu$m flux extrapolated from the best SED model \citep{Tacchella+23,Bunker+23} as proxies of the rest-frame luminosity at 1400\,$\AA$, 3000\,$\AA$, and 5100\,$\AA$, respectively, a bolometric luminosity of $\sim$\,1.3\,$\times$\,10$^{45}$\,erg\,s$^{-1}$ is obtained, in agreement with a previous estimate \citep{Maiolino2024_BH}. However, the AGN bolometric luminosity derived from the H$\beta$ (given as H$\alpha$/2.80) and [O\,III]\,5008$\AA$ luminosities corresponds to 22\,$-$\,33\,$\times$\,10$^{45}$\,erg\,s$^{-1}$, i.e. factors 17\,$-$\,25 larger than expected from the continuum emission. This already indicates that, even if a 6\,$\times$\,10$^5$\,$M_{\odot}$ mass AGN is present in GN-z11, it can not dominate the nebular emission as measured by the optical emission lines. On the contrary, appears to play a minor role in the ionization of the surrounding interstellar medium.  Moreover, the predicted Eddington ratio for accretion derived from the emission line luminosities also supports this conclusion. Following \citet{Bosman2024}:
\begin{equation}
\lambda_\mathrm{Edd}= L_\mathrm{bol}/L_\mathrm{Edd}=  0.796 \times 10^{-38} \times L_\mathrm{bol}(\mathrm{erg\,s^{-1}}) \times M_\mathrm{BH}^{-1}(M_{\odot})
\end{equation}
\noindent the derived Eddington ratio has an extremely large (unrealistic) value of $\sim$\,290\,$-$\,440 for a BH mass of 6\,$\times$\,10$^5$\,$M_{\odot}$. These accretion ratios are factors 30 to 60 higher than those derived from the continuum windows, $\sim$\,7.5\,$-$\,9 above Eddington (see also \citealt{Maiolino2024_BH}) assuming the UV-optical continuum is fully produced by an AGN radiation. Recent cosmological simulations of the formation of galaxies and massive black holes do not support the existence of black holes radiating at super-Eddington ratios at high redshifts \citep{Bhatt2024}. These simulations predict an extremely low probability ($<$\,0.2\,\%) of having a black hole at $z$\,=\,10$-$11 with a mass of 10$^6$\,$M_{\odot}$ emitting at rates about five times Eddington. Note, however, that these simulations do not exclude the presence of a massive black hole that could be a minor contributor to the luminosity of GN-z11.

\subsubsection{Gas metallicity and AGN radiation}

AGN models with a low optical to X-ray power low index ($\alpha$\,$=$\,$-$\,1.5) reproduce the observed high-ionization [Ne\,IV]2422,2424/[Ne\,III]3870 and He\,II1640/H$\delta$ line ratios \citep{Isobe-metallicity2023}. 
An estimate of the metallicity of the ionized gas based on these AGN models, on the value for the ionization parameter ($U$) derived from the ultraviolet N\,IV]1483,1486/N\,III]1750 ratio and on the [O\,III]4363$\AA$/[O\,II] for the metallicity, predicts oxygen abundances close and above solar, i.e. 12+log(O/H)= 8.58-9.23 \citep{Isobe-metallicity2023}. This value is clearly inconsistent with our metallicity measurements (7.91\,$-$\,7.93; $\S$\ref{Sect:ISM_conditions}) from the [O\,III] and using the $T_\mathrm{e}$ and $T_\mathrm{e}$-AGN direct methods. So, the metallicity derivation based on some AGN radiation models and UV lines does not support the presence of an AGN. 

\subsubsection{Stratification and kinematics of the line emitting regions}

Under the hypothesis of an AGN, a scenario invoking the presence of two different ionized regions has been proposed  to explain the observed continuum at rest-frame wavelengths 3000-4000\,$\AA$ \citep{Ji+2024}: a hot and dense ionized gas in the BLR would dominate the Balmer emission and a colder, less-dense gas in the Narrow Line Region (NLR) or host galaxy would emit the forbidden lines and also contributing (but sub-dominant) to the Balmer emission. The differences in the width (FWHM) of the UV lines, with the semi-forbidden (N\,IV]\,1486$\AA$; 470\,$\pm$\,50\,km\,s$^{-1}$) and permitted (Mg\,II\,2800$\AA$; 430\,$\pm$\,65\,km\,s$^{-1}$) lines broader (430-470\,km\,s$^{-1}$) relative to the forbidden lines ([O\,II]\,3727,3729$\AA$; 368\,$\pm$\,50\,km\,s$^{-1}$, [Ne\,III]\,3870$\AA$; 340\,$\pm$\,30\,km\,s$^{-1}$, and [Ne\,IV]\,2422,2424$\AA$; 380\,$\pm$\,100\,km\,s$^{-1}$) could support this scenario and has been suggested as evidence for a Narrow Line Seyfert 1 (NLS1) type AGN in GN-z11 \citep{Maiolino2024_BH}. However, our [O\,III]\,5008$\AA$ and H$\alpha$ spectroscopy does not support neither the NLS1 nature of GN-z11 nor the presence of two ionized regions, with different kinematics. As already mentioned  (see $\S$\ref{Sect:kinematic}), the [O\,III]\,5008$\AA$ line width FWHM (189\,$\pm$\,25\,km\,s$^{-1}$) is a factor of about 2 narrower than the previous NIRSpec values derived for the forbidden lines.  This discrepancy is understood as due to the lower spectral resolution and detection level of the forbidden lines identified in the NIRSpec spectrum. In particular, the spectral resolution of the MRS corresponds to a FWHM of 86\,$\pm$\,9 km s$^{-1}$, a factor 2 to 3 better than the NIRSpec spectrum (FWHM estimated between 143 and 273 km s$^{-1}$ \citealt{Maiolino2024_BH}). In addition, the width of the H$\alpha$ line is consistent with that of the [O\,III]\,5008$\AA$ line, supporting the conclusion that the kinematics of the ionized gas traced by both lines is not associated with the velocities expected in broad line AGNs.
 
The density stratification of the ionized medium in compact high-$z$ galaxies could be common, and not necessarily associated with the presence of an AGN. Regions of high density (6.4$-$31\,$\times$\,10$^4$\,cm$^{-3}$) have been identified in the lensed starburst galaxy RXCJ2248-ID at a redshift of 6.1 from the UV lines \citep{Topping+24} while densities of about 4\,$\times$\,10$^2$\,cm$^{-3}$ are derived from the optical and infrared [O\,III] line ratios (Crespo-G\'omez et al., in prep.). Moreover, NIRCam rest-frame UV images of RXCJ2248-ID show the presence of two compact clumps of less than 22\,pc (effective radius) and separated by 220\,pc while the broad line components are detected in both [O\,III]\,5008$\AA$ and H$\alpha$ lines from NIRSpec slit spectroscopy \citep{Topping+24}.  Further NIRSpec integral field spectroscopy reveals that the optical line emission, including the broad component, is dominated by one of the clumps (Crespo-G\'omez et al., in prep). In summary, the scenario of a  (0.6$-$1.4\,$\times$\,10$^6$\,$M_{\odot}$) type 1 accreting BH located at the center of GN-z11, acting as the dominant energy source ionizing the surrounding interstellar medium, is not supported by the kinematics, luminosity and metallicity measurements derived from our new high spectral resolution H$\alpha$ and [O\,III]\,5008$\AA$ spectroscopy. 

\begin{figure*}
\centering
   \includegraphics[ width=\linewidth]{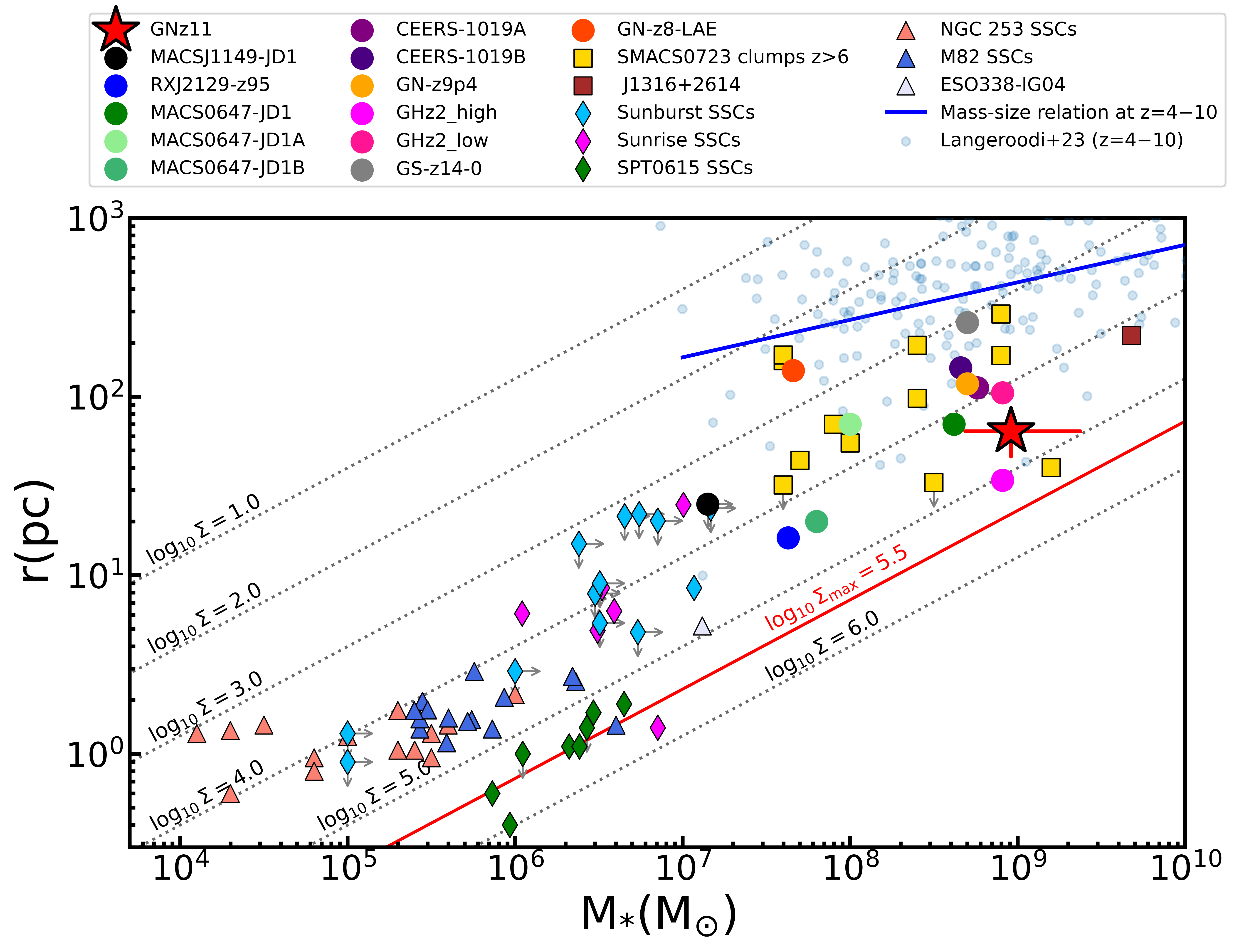}
      \caption{Mass-radius relation for GN-z11 including young star clusters in nearby starbursts (NGC253,  \citealt{Leroy2018}; M82, \citealt{McCrady2003, McCrady-Graham2007}), low-$z$ blue compact (ESO338-IG04, \citealt{Ostlin2007}) and low-metallicity (SBS0335-052E, \citealt{Adamo2010}) galaxies. 
      Also represented are the values for high-$z$ clusters (Sunburst, \citealt{Vanzella-Sunburst2022}; Sunrise, \citealt{Vanzella+23}; SPT0615-JD1, \citealt{Adamo+24}), clumps (SMACS0723, \citealt{Claeyssens+Adamo2023}), extremely UV-bright SFG (J1316+2614, \citealt{Marques-Chaves+24_SFE}), and luminous galaxies at redshifts above 8 (GN-z8-LAE, \citealt{Navarro-Carrera+24}; CEERS-1019, \citealt{Marques-Chaves2024}; MACS1149-JD1, \citealt{Bradac+24}; GN-z9p4, \citep{Curti+24, Schaerer-Rui2024}; RXJ2129-z95, \citealt{Williams+23}; MACS0647-JD, \citealt{Hsiao+23-NIRCam}; GHz2, \citealt{Calabro2024}; GS-z14-0, \citealt{Helton2024}). The mass$-$size relation derived for $z$\,=\,4$-$10 galaxies identified with JWST \citep{Langeroodi-mass-size2023} is shown (blue line) as reference. The dotted lines represent constant stellar mass surface density in units of $M_{\odot}$\,pc$^{-2}$. The line of log$_{10}\Sigma$\,=\,5.5 (in red) indicates the observed maximum value in clusters and nucleus of galaxies, and also predicted in dense systems under ineffective feedback regulated conditions \citep{Grudic2019}. Note that the stellar mass of GN-z11 is the average of the values presented in \citet{Bunker+23} and \citet{Tacchella+23}.} 
         \label{fig:Mass-size-clusters}
\end{figure*}   

\subsection{GN-z11 as an extreme, low-metallicity feedback-free starburst}\label{Sec.Dis_SFR}

Under the starburst scenario, the MRS [O\,III]\,5008$\AA$ and H$\alpha$ luminosities in GN-z11 are compatible with a low-metallicity (0.17\,$\pm$\,0.03\,$Z_{\odot}$) starburst forming stars at a rate of 24\,$\pm$\,3\,$M_{\odot}$\,yr$^{-1}$ over the last few million years. The dynamical mass (1.1\,$\pm$\,0.4\,$\times$\,10$^9$\,$M_{\odot}$) derived from the [O\,III]\,5008$\AA$ line supports the conclusion that the mass is dominated by the stellar component. Combined with the very compact size of the galaxy ($R_\mathrm{e}$\,=\,64\,$\pm$\,20\,pc; \citealt{Tacchella+23}), these results characterize GN-z11 as a system with extremely large SFR and stellar mass surface densities, $\Sigma_\mathrm{SFR}$\,=\,9.33\,$\times$\,10$^2$\,$M_{\odot}$\,yr$^{-1}$\,kpc$^{-2}$ and $\Sigma_\mathrm{*}$\,=\,3.6\,$\times$\,10$^4$\,$M_{\odot}$\,pc$^{-2}$, respectively (see Table~\ref{tab:Extreme_SB}). GN-z11 compactness, mass and surface densities are within the range of those measured in other extreme galaxies recently detected with JWST at redshifts above 8 (see Figure \ref{fig:Mass-size-clusters}): GN-z8-LAE ($z$\,$=$\,8.3; \citealt{Navarro-Carrera+24}), CEERS-1019 ($z$\,=\,8.7; \citealt{Marques-Chaves2024}), MACS1149-JD1 ($z$\,=\,9.1; \citealt{Bradac+24, Stiavelli+23, Alvarez-Marquez+23-MACS}), GN-z9p4 ($z$\,=\,9.4; \citealt{Schaerer-Rui2024}), RXJ2129 ($z$\,=\,9.5; \citealt{Williams+23}), MACS0647-JD ($z$\,=\,10.2; \citealt{Hsiao+23-NIRCam,Hsiao+23-NIRSpec}),  GHz2 ($z$\,=\,12.3; \citealt{Zavala+2024,Calabro2024,Castellano2024}), and GS-z14-0 ($z$\,=\,14.3; \citealt{Helton2024}). All these galaxies are far more compact (x3$-$10) than the average high-$z$ galaxy of the same stellar mass according to the mass$-$size relation derived for galaxies at redshifts 4 to 10 (see Figure \ref{fig:Mass-size-clusters}, \citealt{Langeroodi-mass-size2023}).

The stellar surface density of GN-z11 and of these galaxies when the Universe was less than 600 Myr old, (see Figure \ref{fig:Mass-size-clusters}), is in the range measured in compact clusters detected in strongly-lensed high-$z$ galaxies like the Sunburst ($z$\,=\,2.37, \citealt{Vanzella-Sunburst2022}), and Sunrise ($z$\,$\sim$\,6,\citealt{Vanzella2023}), clumps in high-$z$ ($z$\,$\sim$\,6.4\,$-$\,8.5) lensed galaxies behind SMACS0723 \citep{Claeyssens+Adamo2023}, and  the Cosmic Gems arc (SPT0615-JD1 at $z$\,$\sim$\,10.2, \citealt{Adamo+24}), and only a factor $\sim$\,10\,$-$\,100 below the maximum density observed in compact stellar clusters and nuclei of galaxies ($\Sigma_\mathrm{*}$\,=\,3\,$\times$\,10$^5$\,$M_{\odot}$\,pc$^{-2}$ \citealt{Hopkins2010, Grudic2019, Crocker2018}). Young massive clusters with $\Sigma_\mathrm{*}$\,=\,10$^4$\,$-$\,10$^5$\,$M_{\odot}$\,pc$^{-2}$ have also been identified in the nearby starburst galaxies M82 (\citealt{McCrady2003}, \citealt{McCrady-Graham2007}) and NGC 253 \citep{Leroy2018}, as well as in the blue compact dwarf galaxy ESO338-IG04 \citep{Ostlin2007}. 

\begin{table*}[!ht]
\caption{Physical parameters characterizing the starburst in GN-z11 and high-$z$ galaxies in the early Universe}
\centering
\resizebox{\textwidth}{!}{
\begin{tabular}{lcccccccc}
Galaxy & Redshift & Metallicity & SFR & $\log$(M$_\mathrm{*}$) & Re & $\Sigma_\mathrm{SFR}$ & $\Sigma_\mathrm{*}$ & $C_5$\\
& & 12+log(O/H) & ($M_{\odot}$\,yr$^{-1}$) & ($M_{\odot}$) & (pc) & (10$^3$\,$M_{\odot}$\,yr$^{-1}$\,kpc$^{-2}$) & (10$^4$\,$M_{\odot}$ pc$^{-2}$) & (10$^5$\,$M_{\odot}$\,pc$^{-1}$) \\
\hline
GN-z11 & 10.6 & 7.91 & 24 & 8.96 & 64 & 0.933 & 3.6 & 85\\
GN-z8-LAE & 8.3 & 7.85 & 11.3 & 7.66 & 143 & 0.088 & 0.036 & 2 \\ 
CEERS-1019A & 8.7 & 7.70 & 148 & 8.76 & 112 & 1.86 & 0.72 & 30 \\
CEERS-1019B & 8.7 & 7.70 & 83 & 8.66 & 145 & 0.645 & 0.355 & 19 \\
MACSJ1149-JD1 & 9.1 & 7.88 & 0.4 & 7.15 & $<$\,25 & $>$\,0.10 & $>$\,0.36 & $>$\,3\\
GN-z9p4 & 9.4 & 7.35 & 64 & 8.7 & 118 & 0.732 & 0.573 & 25\\
RXJ2129-z95 & 9.5 & 7.48 & 1.7 & 7.63 & 16.2 & 1.03 & 2.6 & 16 \\
MACS0647-JD1 & 10.2 & 7.79 & 5.0 & 8.62 & 70 & 0.162 & 1.35 & 35 \\
MACS0647-JD1A & 10.2 & 7.79 & 1 & 8 & 70 & 0.033 & 0.325 & 8\\
MACS0647-JD1B & 10.2 & 7.79 & 0.6 & 7.8 & 20 & 0.239 &  2.5 & 19 \\
GHz2 & 12.3 & 7.40 & 9 & 8.91 & 34$-$105 & 0.1$-$1.0 & 1$-$10 & 45$-$141 \\
GS-z14-0 & 14.3 &  6.9 & 25 & 8.7 & 260 & 0.059 & 0.12 & 11\\
\hline
\label{tab:Extreme_SB}
\end{tabular}}
\tablefoot{GN-z11: metallicity and SFR are from this work, stellar mass is a mean of the values presented in \citet{Bunker+23} and \citet{Tacchella+23}, effective radius (\citet{Tacchella+23}; 
GN-z8-LAE: \citet{Navarro-Carrera+24}; 
CEERS-1019: \citet{Marques-Chaves2024}; 
MACS1149-JD1:  metallicity \citep{Stiavelli+23}, size and stellar mass for clump 2 \citep{Bradac+24}; GN-z9p4: \citet{Schaerer-Rui2024}; RXJ2129: \citet{Williams+23}; MACS0647-JD: stellar mass \citep{Hsiao+23-NIRCam}, metallicity \citep{Hsiao+23-NIRSpec,Hsiao+2024_MIRI}, SFR \citep{Hsiao+2024_MIRI}; GHz2: metallicity and SFR \citep{Zavala+2024}, size, stellar mass, SFR and stellar mass surface brightness \citep{Calabro2024}; GS-z14-0: \citet{Helton2024}. The SFR and stellar mass surface densities are derived from the corresponding quantities divided by 2$\pi$$R_\mathrm{e}^{2}$. The compactness index ($C_5$) is defined as $C_5$\,=\,(M$_{*}$/10$^5$\,$M_{\odot}$)/$R_{\mathrm{hm}}$[pc] with $R_{\mathrm{hm}}$\,=\,1.7\,$\times$\,$R_\mathrm{e}$ as the half-mass radius, \citep{Krause2016}. The uncertainties in the different parameters are given in the references.}
\end{table*}

Thus, since the effective radius of GN-z11 and high-$z$ galaxies may represent an upper limit to their size, their already high surface densities could trace a young, massive cluster or, more likely, the assembly of many young stellar clusters as predicted in radiation-hydrodynamic simulations \citep{Garcia2023}. Indeed, the unusually super-solar N/O derived for GN-z11 and the similarities with the abundance patterns of globular clusters, have led several authors to suggest a link between GN-z11, and a few other high-$z$ strong N-emitters, with the formation of globular clusters in the early Universe (\citealt{Senchyna2024}, \citealt{Charbonnel2023}, \citealt{Isobe-metallicity2023}, \citealt{Marques-Chaves2024}, \citealt{Schaerer-Rui2024}). Its compactness (in terms of $\Sigma_\mathrm{*}$) further supports this scenario. Such assemblies of massive clusters have already been identified in low-$z$ star-forming galaxies as well as at high-$z$. Star-forming clumps with median effective radius of 58\,pc have been measured in the Ly$\alpha$ Reference Sample (LARS) of galaxies \citep{Messa2019}. These clumps are stellar cluster complexes containing multiple individual clusters, each with radius of 1 to 10\,pc. This appears to also be the case in the SPT0615-JD1 galaxy at $z$\,$=$\,10.2, where five clusters, each with effective radius of 1\,pc, ages of less than 35\,Myr, and intrinsic stellar mass of 0.6\,$-$\,3\,$\times$\,10$^6$\,$M_{\odot}$ have been identified within 70\,pc \citep{Adamo+24}.  

In addition, the star formation surface density in GN-z11, and other luminous galaxies at redshift above 8.5 (see Table~\ref{tab:Extreme_SB}), is unusually high, with some galaxies just below the maximum Eddington rate expected in radiation supported starbursts ($\Sigma_{\mathrm{SFR}}$\,$=$\,1\,$\times$\,10$^3$\,$M_{\odot}$\,kpc$^{-2}$,\citealt{Thompson+05}, \citealt{Andrews+Thompson2011}), and well above those of high-$z$ galaxies identified by JWST at redshifts above 5 (1\,$<$\,$\Sigma_\mathrm{SFR}$\,$<$\,300\,$M_{\odot}$\,kpc$^{-2}$, \citealt{Morishita-Stiavelli2024}). The $\Sigma_{\mathrm{SFR}}$ are only similar, and even higher, to those measured in the nuclear regions of low-$z$ luminous infrared galaxies \citep{Sanchez-Garcia2022} and in high-$z$ massive, dusty star-forming galaxies (e.g. \citealt{Hodge2015}, \citealt{Oteo2016}).  

How could systems like GN-z11 be generated at such an early epoch in the Universe? According to recent analytic models, massive low-metallicity galaxies at redshifts of 10 and above are expected to have densities of a few $\times$\,10$^{3}$\,cm$^{-3}$, and mass surface densities of $\sim$\,3\,$\times$\,10$^3$\,$M_{\odot}$\,pc$^{-2}$. For these high densities, the free-fall time is very short, less than 1\,Myr, and the gas accreted from the halos of these galaxies is converted to stars with a high efficiency, close to one, and with high rates of $\sim$\,65\,$M_{\odot}$\,yr$^{-1}$, i.e. galaxies would be in a free-fall starburst phase \citep{Dekel2023,Li+Dekel2023}. Moreover, due to the extreme mass concentration, the stellar winds and supernovae explosions would not be able to overcome the binding gravitational energy, outflows are expected weak and not be able to halt the star formation process, i.e. galaxies would be in a feedback-free burst (FFB) phase. In fact, the values of the compactness index ($C_5$\,=\,($M_{*}$/10$^5$ $M_{\odot}$)/$R_{\mathrm{hm}}$[pc]) with half-mass  radius $R_{\mathrm{hm}}$\,=\,1.7\,$\times$\,$R_\mathrm{e}$ \citep{Krause2016} measured in GN-z11 and other high-$z$ galaxies (see Table~\ref{tab:Extreme_SB}) are well above the limits ($C_5$\,$\sim$\,1$-$10) and the combined effect of stellar winds and supernovae would not be able to expel the gas efficiently. Only invoking the presence of energetic hypernovae, the expulsion of gas could take place for compactness index above 30, i.e. in the range of values measured in these high-$z$ galaxies. The presence of bright hypernovae and/or pair-instability supernovae have been invoked to explain the O/Fe abundance in GN-z11 \citep{Nakane2024}. The free-fall, feedback-free starburst scenario, galaxies at dawn are expected to be compact ($R_\mathrm{e}$\,$\sim$\,0.3\,kpc), bursty on timescales of 10\,Myr, and consisting of thousands of star clusters, each with masses of 10$^6$\,$M_{\odot}$, at the end of this phase spanning about 100\,Myr \citep{Li+Dekel2023}. In addition, these early starbursts would have a low gas mass fraction ($<$\,5\%) and a low metallicity (0.1\,$Z_{\odot}$).

The properties measured in GN-z11 are consistent with those expected in the free-fall, feedback-free starburst scenario. The metallicity measured in GN-z11 is 0.17\,$\pm$\,0.03\,$Z_{\odot}$, and the stellar mass is, within the uncertainties, the main (even dominant) component of the dynamical mass (see $\S$\ref{Sect:kinematic}). The recent star formation rate traced by the H$\alpha$ line corresponds to 24\,$\pm$\,3\,$M_{\odot}$ yr$^{-1}$ (see $\S$\ref{Sect:SFR}) in a small region with an effective radius of 64\,$\pm$\,20 pc \citep{Tacchella+23}. The predicted electron density of the ionized medium is about 10$^3$\,cm$^{-3}$ derived from  the variation of the electron density with redshift, and consistent with the density measured in the galaxy MACS0647-JD1 at redshift 10.2 (see section $\S$\ref{Sect:ISM_conditions}). Following \citet{Dekel2023}, a free-fall time of 1.6 Myr is obtained for GN-z11 assuming an average gas density of 10$^3$ cm$^{-3}$. A similar value, $\sim$\,1.9 Myr, is derived if a total gas mass equal to that of the ionized gas, i.e. 2 $\times$ 10$^7$ $M_{\odot}$ (see $\S$\ref{Sect:kinematic}) is enclosed in the effective radius of 64 pc. This free-fall time will be even smaller ($\sim$ 0.4 Myr) if a mass equal to the total stellar mass is considered. Finally, a compactness index of 85 is derived for GN-z11, this $C_5$ value requires a very high SF efficiency ($\sim$\,0.7, \citealt{Marques-Chaves+24_SFE}) for gas expulsion even if the presence of hypernovae is invoked.

The negligible nebular dust attenuation measured with the available hydrogen Balmer lines (see $\S$\,\ref{Sect:SFR}) is also compatible with the low dust attenuation predicted in the FFB scenario.  Since dust is mostly produced by supernovae, no significant dust is expected during the short FFB phase. During the post FFB phase, the supernovae outflows would remove the dust and residual gas, generating a compact, blue luminous galaxy \citep{Dekel2023, Li+Dekel2023}. Moreover, according to this scenario, dust attenuation would diminish with redshift, favoring the detection of UV luminous galaxies early in the Universe, at redshifts 10 and above \citep{Li+Dekel2023}. The predictions of the low dust attenuation are not unique of FFBs. Alternative scenarios invoke the existence of radiative-driven dusty outflows in super-Eddington luminous starburst that would clear the galaxy from dust \citep{Ferrara23, Fiore23, Ziparo23}. According to this scenario, the radiation pressure efficiency increases by large factors (100-1000) in dusty environments, and therefore the effective Eddington luminosity is reduced by similar factors \citep{Fiore23}. Thus, for a given starburst, the condition of a super-Eddington dusty outflow translates into a minimum specific star formation rate (sSFR\,$=$\,SFR/M$_{*}$) of 13\,Gyr$^{-1}$; \citealt{Fiore23}). GN-z11 with a sSFR of 25\,Gyr$^{-1}$ (see Table \ref{tab:EL_Fluxes}) would be experiencing a radiation-driven dusty outflow clearing the galaxy from dust. Very deep NOEMA observations have not been able to detect neither [C\,II]158$\mu$m nor continuum dust emission at rest-frame 160$\mu$m \citep{Fudamoto2024}, suggesting that GN-z11 is a dust-poor system. However, the derived upper limits for the continuum emission are still above the predicted fluxes under the dusty outflow scenario. Thus, the origin of the low dust attenuation in GN-z11 derived from the SED and hydrogen Balmer lines is still unclear. Deeper NOEMA continuum observations are needed to constrain the temperature and amount of dust. 

The MRS spectroscopy shows a flux excess in the red part of the [O\,III]\,5008$\AA$ emission line profile, suggesting the presence of a faint secondary red component at velocities 100\,$-$\,300\,km\,s$^{-1}$ detected at a low significance ($<$\,2$\sigma$). SNe produced in earlier generations of clusters in the compact starburst regions during the 100\,Myr long phase, i.e. clusters in a post feedback-free phase, would generate winds with extremely high velocities of 700 km\,s$^{-1}$ to 3300 km\,s$^{-1}$ for star formation efficiencies of 0.2 to 1 \citep{Li+Dekel2023}. These winds could be traced by broad emission lines with FWHM of 1400 km\,s$^{-1}$ to 6600 km\,s$^{-1}$. Detection of such predicted broad lines is not possible with the current set of data and requires deeper spectroscopy with JWST. Note that the prediction of this scenario applies to all (collisional and recombination) optical emission lines, i.e. [O\,III]5008 and H$\alpha$, and therefore would be able to distinguish high-velocity winds/outflows in the dense ($\sim$ 10$^3$ cm$^{-3}$) interstellar medium of the galaAt the sensitivity of our MRS spectrum,xy, from the emission of an extremely dense ($>$ 10$^9$ cm$^{-3}$) broad line region around a low mass black hole, present only in the H$\alpha$ recombination line.
 
In summary, GN-z11 shows characteristics that are not unique, and are shared with other galaxies identified by JWST in the early universe, at cosmic dawn, i.e. at redshifts $\sim$ 8.5 and above. The surface density of the stellar mass and star formation in these galaxies is very extreme and close to that observed in young, massive stellar clusters. Under the starburst hypothesis, we speculate about a potential formation scenario, the feedback-free phase, involving a high efficient star formation, even close to one, in a low metallicity, high dense environment. While the properties measured in GN-z11 are consistent with the predictions of the feedback-free scenario, a confirmation of the predicted high velocity winds in the [O\,III] and H$\alpha$ lines, if the starburst includes previous generations of clusters in a post feedback-free phase, requires deeper spectroscopy with JWST.

\section{Summary and conclusions}\label{Sect:conclusion_Summary}

This paper reports new MIRI medium resolution spectroscopic observations of the high-$z$ galaxy GN-z11 covering the rest-frame optical spectrum. We detect the [O\,III]\,5008$\AA$ and H$\alpha$ emission lines with signal-to-noise ratios of 10 and 8, respectively. The combination of the MRS [O\,III]\,5008$\AA$ and H$\alpha$ lines, and previously reported UV and optical lines from NIRSpec observations \citep{Bunker+23}, allows us to study the ISM physical properties and the ionization nature of GN-z11. The main results are summarized as follows:

\begin{itemize}

\item The [O\,III]\,5008$\AA$ and H$\alpha$ emission lines are resolved and well-modeled by a one Gaussian component with FWHMs of 189\,$\pm$\,25 and 231\,$\pm$\,52\,km\,s$^{-1}$, respectively. Their FWHMs are about a factor of two narrower than the FWHMs of the semi-forbidden (FWHM\,=\,470\,$\pm$\,50\,km\,s$^{-1}$, N\,IV]\,1487$\AA$) and permitted ultraviolet lines (FWHM\,=\,430\,$\pm$\,65\,km\,s$^{-1}$, Mg\,II\,2800$\AA$) previously reported using medium-resolution (R\,$\approx$\,1000) NIRSpec spectroscopy. There is, therefore, no evidence of a dominant broad H$\alpha$ component associated with the Broad Line Regions (BLR) of a type 1 AGN. However, a broad ($\sim$\,430$-$470\,km\,s$^{-1}$) and weak ($<$\,2$-$3\,$\times$\,10$^{-18}$\,erg\,s$^{-1}$\,cm$^{-2}$) H$\alpha$ line, if present, would be at the 2$\sigma$ detection level in our current MRS observations. Therefore, a minor broad H$\alpha$ component associated with the BLR of an AGN cannot be ruled out in this work, and deeper MRS spectroscopy is required for its final confirmation.  

\item The ISM is characterized by a negligible or zero dust attenuation, an electron temperature of 14000\,$\pm$\,2100\,K, an ionization parameter (U) in the $-$1.81 to $-$2.06 range, and a direct-$T_\mathrm{e}$ gas-phase metallicity of 12\,$+$\,$\log$(O/H)\,=\,7.91\,$\pm$\,0.07 (0.17\,$\pm$\,0.03\,Z$_{\odot}$). GN-z11 is located in the SFG/AGN region in the R3$-$R2, O32$-$R23, O3H$\gamma-$O32 and O3H$\gamma-$O33 diagrams, although more compatible with local low-metallicity and high-$z$ SFGs. A total ionized gas mass of $\sim$\,2\,$\times$\,10$^7$\,$M_{\odot}$ is inferred from the H$\alpha$ luminosity. Its Str\"omgren sphere is estimated to have a radius of 64\,pc for an average electron density of 10$^3$\,cm$^{-3}$, similar to the compact size of the UV continuum source \citep{Tacchella+23}.

\item A dynamical mass of (1.1\,$\pm$\,0.4)\,$\times$\,10$^9$\,$M_{\odot}$ is derived assuming GN-z11 is virialized. This value is consistent within the uncertainties with previous stellar mass estimates \citep{Bunker+23,Tacchella+23}, suggesting that the dynamical mass in GN-z11 is dominated by the stellar component. The stellar mass and direct-$T_\mathrm{e}$ gas-phase metallicity place GN-z11 on the best-fit mass-metallicity relation for galaxies at $z$\,$\sim$\,8. 
 
\item Assuming a young starburst is the ionizing source in GN-z11, a SFR of  24\,$\pm$\,3\,$M_{\odot}$\,yr$^{-1}$ is derived from the H$\alpha$ luminosity and considering a metallicity of 0.1\,$Z_{\odot}$. The H$\alpha$ SFR agrees within the uncertainties with the UV SFR, and with the previous values derived from multi-wavelength SED-fitting \citep{Tacchella+23} and other Balmer lines \citep{Bunker+23}. The photon production efficiency and the Ly$\alpha$ escape fraction of GN-z11 are $\log$($\zeta_\mathrm{ion}$)\,$=$\,25.66\,$\pm$\,0.06\,Hz\,erg$^{-1}$ and $f_\mathrm{esc,\,LyC}$\,$=$\,0.04\,$\pm$\,0.01, respectively. These values are in very good agreement with previous derivations \citep{Bunker+23}, and are within the range measured in most of the EoR galaxies observed to date. 

\item The presence of an accreting black hole dominating the optical continuum and emission lines of GN-z11, and following the relations well established for low-$z$ AGNs, is not supported by the MRS H$\alpha$ and [O\,III]\,5008$\AA$ luminosities and line profiles. If the H$\alpha$ line is entirely produced in a high-density BLR and the standard H$\alpha-$BH mass relation of low-$z$ type 1 AGNs applies, a black hole mass is estimated to be 6$_{-3}^{+4}$\,$\times$\,10$^5$\,$M_{\odot}$, a factor of two lower than the one inferred by UV emission lines $\log$(M$_\mathrm{BH}$[$M_{\odot}]$)\,=\,6.2$\pm$0.3 \citep{Maiolino2024_BH}. If the UV-optical continuum and optical emission lines were produced by an accreting BH, the AGN bolometric luminosities derived from the [O\,III]\,5008$\AA$ and H$\alpha$ lines would be factors of 17\,$-$\,25 higher than those derived from the UV and optical continuum. Moreover, the Eddington ratio for accretion ($\lambda_\mathrm{E}$) derived from the [O\,III]\,5008$\AA$ and H$\alpha$ luminosities would have extremely large (unrealistic) values, 290\,$-$\,440. In addition, the presence of an AGN radiation dominating the H$\alpha$ emission is not supported by the measured $L$(H$\alpha$)/$L_\mathrm{X}$(2-10\,keV) and $L$(H$\alpha$)/$L$(5100\,$\AA$) ratios, deviating from the well established relations in low-$z$ AGNs. 
 
\item The stellar mass and size places GN-z11 well below the mass-size relation derived for galaxies at redshifts 4 to 10 \citep{Langeroodi-mass-size2023}. The corresponding stellar mass surface density of 3.6\,$\times$\,10$^4$\,$M_{\odot}$\,pc$^{-2}$, is close to that of the densest stellar clusters and young clusters detected in intermediate and high-$z$ galaxies. This high stellar surface density, together with its extreme SFR surface density (933 $M_{\odot}$\,kpc$^{-2}$) and low gas-phase metallicity (0.17$\pm$0.03\,$Z_{\odot}$), suggest the presence of multiple young, low-metallicity stellar clusters as the main ionization source.

\item The high SFR and stellar mass surface densities ($\Sigma_\mathrm{SFR}$ and $\Sigma_\mathrm{*}$) in GN-z11, together with its small size ($R_\mathrm{e}$), compactness ($C_5$), and low metallicity are consistent with the scenario of a free-fall, feedback-free starburst at cosmic dawn \citep{Dekel2023}. 

\end{itemize}

The nature of GN-z11 and its dominant energy source have been investigated with the new [O\,III]\,5008$\AA$ and H$\alpha$ measurements. The properties of GN-z11, its compactness, low metallicity and high stellar mass surface brightness are shared by other luminous galaxies at cosmic dawn, i.e. at redshifts above 8.5. We propose that these galaxies could be transitioning through a highly efficient star formation phase, converting the available low-metallicity gas into stars during short periods of time due to a free-fall, feedback-free process, as recently suggested by \cite{Dekel2023} and \cite{Li+Dekel2023}. Since the feedback-free phase is bursty, this scenario predicts very broad ($>$ 1000 km s$^{-1}$) weak secondary components in both [O\,III]\,5008$\AA$ and H$\alpha$ lines due to supernovae feedback from previous generations of clusters within the compact starburst region. The MRS spectroscopy shows a flux excess in the red part of the [O\,III]\,5008$\AA$ emission line profile, suggesting the presence of a possible secondary line component detected at a low significance ($<$\,2$\sigma$). Deeper rest-frame optical spectroscopy and imaging of GN-z11 with JWST are needed to further elucidate the presence of the predicted broad components in both the forbidden and recombination lines, and older stellar populations from previous bursts. 

\begin{acknowledgements}
The  authors  thank to the  anonymous referee for useful  comments. J.A-M., L.Colina, A.C.G., C.P.-J., L.Costantin acknowledge support by grants PIB2021-127718NB-100, A.A-H. by grant PID2021-124665NB-I00, and P.G.P-G and L.Costantin by grant PID2022-139567NBI00 from the Spanish Ministry of Science and Innovation/State Agency of Research MCIN/AEI/10.13039/501100011033 and by “ERDF A way of making Europe”. A.C.G. acknowledges support by JWST contract B0215/JWST-GO-02926. J.M., A.B., and G.O. acknowledges support from the Swedish National Space Administration (SNSA). K.I.C., P.R., and E.I acknowledge funding from the Netherlands Research, the Netherlands Research School for Astronomy (NOVA), and the Dutch Research Council (NWO) through the award of the Vici Grant VI.C.212.036. LB and F.W. acknowledge support from the ERC Advanced Grant 740246 (Cosmic\_Gas). R.A.M. acknowledges support from the Swiss National Science Foundation (SNSF) through project grant 200020\_207349. This work was supported by research grants (VIL16599, VIL54489) from VILLUM FONDEN. The project that gave rise to these results received the support of a fellowship from the ``la Caixa” Foundation (ID 100010434). The fellowship code is LCF/BQ/PR24/12050015. This work is based on observations made with the NASA/ESA/CSA James Webb Space Telescope. The data were obtained from the Mikulski Archive for Space Telescopes at the Space Telescope Science Institute, which is operated by the Association of Universities for Research in Astronomy, Inc., under NASA contract NAS 5-03127 for \textit{JWST}; and from the \href{https://jwst.esac.esa.int/archive/}{European \textit{JWST} archive (e\textit{JWST})} operated by the ESDC. 

This research made use of Photutils, an Astropy package for detection and photometry of astronomical sources \citep{larry_bradley_2022_6825092}.

\end{acknowledgements}

\bibliographystyle{aa} 
\bibliography{bibliography.bib} 

\begin{thebibliography}{180}
\expandafter\ifx\csname natexlab\endcsname\relax\def\natexlab#1{#1}\fi

\bibitem[{{Abdurro'uf} {et~al.}(2024){Abdurro'uf}, {Larson}, {Coe}, {Hsiao}, {{\'A}lvarez-M{\'a}rquez}, {Crespo G{\'o}mez}, {Adamo}, {Bhatawdekar}, {Bik}, {Bradley}, {Conselice}, {Dayal}, {Diego}, {Fujimoto}, {Furtak}, {Hutchison}, {Jung}, {Killi}, {Kokorev}, {Mingozzi}, {Norman}, {Resseguier}, {Ricotti}, {Rigby}, {Vanzella}, {Welch}, {Windhorst}, {Xu}, \& {Zitrin}}]{Abdurrouf+24}
{Abdurro'uf}, {Larson}, R.~L., {Coe}, D., {et~al.} 2024, \apj, 973, 47

\bibitem[{{Acharyya} {et~al.}(2019){Acharyya}, {Kewley}, {Rigby}, {Bayliss}, {Bian}, {Nicholls}, {Federrath}, {Kaasinen}, {Florian}, \& {Blanc}}]{Acharyya+19}
{Acharyya}, A., {Kewley}, L.~J., {Rigby}, J.~R., {et~al.} 2019, \mnras, 488, 5862

\bibitem[{{Adamo} {et~al.}(2024){Adamo}, {Bradley}, {Vanzella}, {Claeyssens}, {Welch}, {Diego}, {Mahler}, {Oguri}, {Sharon}, {Abdurro'uf}, {Hsiao}, {Xu}, {Messa}, {Lassen}, {Zackrisson}, {Brammer}, {Coe}, {Kokorev}, {Ricotti}, {Zitrin}, {Fujimoto}, {Inoue}, {Resseguier}, {Rigby}, {Jim{\'e}nez-Teja}, {Windhorst}, {Hashimoto}, \& {Tamura}}]{Adamo+24}
{Adamo}, A., {Bradley}, L.~D., {Vanzella}, E., {et~al.} 2024, \nat, 632, 513

\bibitem[{{Adamo} {et~al.}(2010){Adamo}, {Zackrisson}, {{\"O}stlin}, \& {Hayes}}]{Adamo2010}
{Adamo}, A., {Zackrisson}, E., {{\"O}stlin}, G., \& {Hayes}, M. 2010, \apj, 725, 1620

\bibitem[{{{\'A}lvarez-M{\'a}rquez} {et~al.}(2024){{\'A}lvarez-M{\'a}rquez}, {Colina}, {Crespo G{\'o}mez}, {Rinaldi}, {Melinder}, {{\"O}stlin}, {Annunziatella}, {Labiano}, {Bik}, {Bosman}, {Greve}, {Wright}, {Alonso-Herrero}, {Boogaard}, {Azollini}, {Caputi}, {Costantin}, {Eckart}, {Garc{\'\i}a-Mar{\'\i}n}, {Gillman}, {Hjorth}, {Iani}, {Ilbert}, {Jermann}, {Langeroodi}, {Meyer}, {Pei{\ss}ker}, {P{\'e}rez-Gonz{\'a}lez}, {Pye}, {Tikkanen}, {Topinka}, {van der Werf}, {Walter}, {Henning}, \& {Ray}}]{Alvarez-Marquez+23-MACS}
{{\'A}lvarez-M{\'a}rquez}, J., {Colina}, L., {Crespo G{\'o}mez}, A., {et~al.} 2024, \aap, 686, A85

\bibitem[{{{\'A}lvarez-M{\'a}rquez} {et~al.}(2019){{\'A}lvarez-M{\'a}rquez}, {Colina}, {Marques-Chaves}, {Ceverino}, {Alonso-Herrero}, {Caputi}, {Garc{\'\i}a-Mar{\'\i}n}, {Labiano}, {Le F{\`e}vre}, {Norgaard-Nielsen}, {{\"O}stlin}, {P{\'e}rez-Gonz{\'a}lez}, {Pye}, {Tikkanen}, {van der Werf}, {Walter}, \& {Wright}}]{Alvarez-Marquez+19_mrs}
{{\'A}lvarez-M{\'a}rquez}, J., {Colina}, L., {Marques-Chaves}, R., {et~al.} 2019, \aap, 629, A9

\bibitem[{{{\'A}lvarez-M{\'a}rquez} {et~al.}(2023){{\'A}lvarez-M{\'a}rquez}, {Crespo G{\'o}mez}, {Colina}, {Neeleman}, {Walter}, {Labiano}, {P{\'e}rez-Gonz{\'a}lez}, {Bik}, {Noorgaard-Nielsen}, {Ostlin}, {Wright}, {Alonso-Herrero}, {Azollini}, {Caputi}, {Eckart}, {Le F{\`e}vre}, {Garc{\'\i}a-Mar{\'\i}n}, {Greve}, {Hjorth}, {Ilbert}, {Kendrew}, {Pye}, {Tikkanen}, {Topinka}, {van der Werf}, {Ward}, {van Dishoeck}, {G{\"u}del}, {Henning}, {Lagage}, {Ray}, \& {Waelkens}}]{Alvarez-Marquez+23-SPT}
{{\'A}lvarez-M{\'a}rquez}, J., {Crespo G{\'o}mez}, A., {Colina}, L., {et~al.} 2023, \aap, 671, A105

\bibitem[{{Ananna} {et~al.}(2024){Ananna}, {Bogd{\'a}n}, {Kov{\'a}cs}, {Natarajan}, \& {Hickox}}]{Ananna2024}
{Ananna}, T.~T., {Bogd{\'a}n}, {\'A}., {Kov{\'a}cs}, O.~E., {Natarajan}, P., \& {Hickox}, R.~C. 2024, \apjl, 969, L18

\bibitem[{{Andrews} \& {Thompson}(2011)}]{Andrews+Thompson2011}
{Andrews}, B.~H. \& {Thompson}, T.~A. 2011, \apj, 727, 97

\bibitem[{{Argyriou} {et~al.}(2023){Argyriou}, {Glasse}, {Law}, {Labiano}, {{\'A}lvarez-M{\'a}rquez}, {Patapis}, {Kavanagh}, {Gasman}, {Mueller}, {Larson}, {Vandenbussche}, {Glauser}, {Royer}, {Dicken}, {Harkett}, {Sargent}, {Engesser}, {Jones}, {Kendrew}, {Noriega-Crespo}, {Brandl}, {Rieke}, {Wright}, {Lee}, \& {Wells}}]{Argyriou+23}
{Argyriou}, I., {Glasse}, A., {Law}, D.~R., {et~al.} 2023, \aap, 675, A111

\bibitem[{{Arrabal Haro} {et~al.}(2023){Arrabal Haro}, {Dickinson}, {Finkelstein}, {Fujimoto}, {Fern{\'a}ndez}, {Kartaltepe}, {Jung}, {Cole}, {Burgarella}, {Chworowsky}, {Hutchison}, {Morales}, {Papovich}, {Simons}, {Amor{\'\i}n}, {Backhaus}, {Bagley}, {Bisigello}, {Calabr{\`o}}, {Castellano}, {Cleri}, {Dav{\'e}}, {Dekel}, {Ferguson}, {Fontana}, {Gawiser}, {Giavalisco}, {Harish}, {Hathi}, {Hirschmann}, {Holwerda}, {Huertas-Company}, {Koekemoer}, {Larson}, {Lucas}, {Mobasher}, {P{\'e}rez-Gonz{\'a}lez}, {Pirzkal}, {Rose}, {Santini}, {Trump}, {de la Vega}, {Wang}, {Weiner}, {Wilkins}, {Yang}, {Yung}, \& {Zavala}}]{Arrabal-Haro+23}
{Arrabal Haro}, P., {Dickinson}, M., {Finkelstein}, S.~L., {et~al.} 2023, \apjl, 951, L22

\bibitem[{{Asplund} {et~al.}(2009){Asplund}, {Grevesse}, {Sauval}, \& {Scott}}]{Asplund+09}
{Asplund}, M., {Grevesse}, N., {Sauval}, A.~J., \& {Scott}, P. 2009, \araa, 47, 481

\bibitem[{{Atek} {et~al.}(2022){Atek}, {Furtak}, {Oesch}, {van Dokkum}, {Reddy}, {Contini}, {Illingworth}, \& {Wilkins}}]{Atek2022}
{Atek}, H., {Furtak}, L.~J., {Oesch}, P., {et~al.} 2022, \mnras, 511, 4464

\bibitem[{{Atek} {et~al.}(2024){Atek}, {Labb{\'e}}, {Furtak}, {Chemerynska}, {Fujimoto}, {Setton}, {Miller}, {Oesch}, {Bezanson}, {Price}, {Dayal}, {Zitrin}, {Kokorev}, {Weaver}, {Brammer}, {Dokkum}, {Williams}, {Cutler}, {Feldmann}, {Fudamoto}, {Greene}, {Leja}, {Maseda}, {Muzzin}, {Pan}, {Papovich}, {Nelson}, {Nanayakkara}, {Stark}, {Stefanon}, {Suess}, {Wang}, \& {Whitaker}}]{Atek+24}
{Atek}, H., {Labb{\'e}}, I., {Furtak}, L.~J., {et~al.} 2024, \nat, 626, 975

\bibitem[{{Baldwin} {et~al.}(1981){Baldwin}, {Phillips}, \& {Terlevich}}]{Baldwin+81}
{Baldwin}, J.~A., {Phillips}, M.~M., \& {Terlevich}, R. 1981, \pasp, 93, 5

\bibitem[{{Berg} {et~al.}(2021){Berg}, {Chisholm}, {Erb}, {Skillman}, {Pogge}, \& {Olivier}}]{Berg+21}
{Berg}, D.~A., {Chisholm}, J., {Erb}, D.~K., {et~al.} 2021, \apj, 922, 170

\bibitem[{{Berg} {et~al.}(2018){Berg}, {Erb}, {Auger}, {Pettini}, \& {Brammer}}]{Berg+18}
{Berg}, D.~A., {Erb}, D.~K., {Auger}, M.~W., {Pettini}, M., \& {Brammer}, G.~B. 2018, \apj, 859, 164

\bibitem[{{Berg} {et~al.}(2019){Berg}, {Erb}, {Henry}, {Skillman}, \& {McQuinn}}]{Berg+19b}
{Berg}, D.~A., {Erb}, D.~K., {Henry}, R. B.~C., {Skillman}, E.~D., \& {McQuinn}, K. B.~W. 2019, \apj, 874, 93

\bibitem[{{Bhatt} {et~al.}(2024){Bhatt}, {Gallerani}, {Ferrara}, {Mazzucchelli}, {D'Odorico}, {Valentini}, {Zana}, {Farina}, \& {Chakraborty}}]{Bhatt2024}
{Bhatt}, M., {Gallerani}, S., {Ferrara}, A., {et~al.} 2024, \aap, 686, A141

\bibitem[{{Bian} {et~al.}(2018){Bian}, {Kewley}, \& {Dopita}}]{Bian+18}
{Bian}, F., {Kewley}, L.~J., \& {Dopita}, M.~A. 2018, \apj, 859, 175

\bibitem[{{Bosman} {et~al.}(2024){Bosman}, {{\'A}lvarez-M{\'a}rquez}, {Colina}, {Walter}, {Alonso-Herrero}, {Ward}, {{\"O}stlin}, {Greve}, {Wright}, {Bik}, {Boogaard}, {Caputi}, {Costantin}, {Eckart}, {Garc{\'\i}a-Mar{\'\i}n}, {Gillman}, {Hjorth}, {Iani}, {Ilbert}, {Jermann}, {Labiano}, {Langeroodi}, {Pei{\ss}ker}, {Rinaldi}, {Topinka}, {van der Werf}, {G{\"u}del}, {Henning}, {Lagage}, {Ray}, {van Dishoeck}, \& {Vandenbussche}}]{Bosman2024}
{Bosman}, S. E.~I., {{\'A}lvarez-M{\'a}rquez}, J., {Colina}, L., {et~al.} 2024, Nature Astronomy [\eprint[arXiv]{2307.14414}]

\bibitem[{{Boyett} {et~al.}(2024){Boyett}, {Bunker}, {Curtis-Lake}, {Chevallard}, {Cameron}, {Jones}, {Saxena}, {Charlot}, {Curti}, {Wallace}, {Arribas}, {Carniani}, {Willott}, {Alberts}, {Eisenstein}, {Hainline}, {Hausen}, {Johnson}, {Rieke}, {Robertson}, {Stark}, {Tacchella}, {Williams}, {Chen}, {Egami}, {Endsley}, {Laseter}, {Looser}, {Maseda}, {Smit}, \& {Witstok}}]{Boyett+24}
{Boyett}, K., {Bunker}, A.~J., {Curtis-Lake}, E., {et~al.} 2024, arXiv e-prints, arXiv:2401.16934

\bibitem[{{Brada{\v{c}}} {et~al.}(2024){Brada{\v{c}}}, {Strait}, {Mowla}, {Iyer}, {Noirot}, {Willott}, {Brammer}, {Abraham}, {Asada}, {Desprez}, {Estrada-Carpenter}, {Harshan}, {Martis}, {Matharu}, {Muzzin}, {Rihtar{\v{s}}i{\v{c}}}, {Sarrouh}, \& {Sawicki}}]{Bradac+24}
{Brada{\v{c}}}, M., {Strait}, V., {Mowla}, L., {et~al.} 2024, \apjl, 961, L21

\bibitem[{Bradley {et~al.}(2022)Bradley, Sip{\H o}cz, Robitaille, Tollerud, Vin{\'{\i}}cius, Deil, Barbary, Wilson, Busko, G{\"u}nther, Cara, Conseil, Bostroem, Droettboom, Bray, Bratholm, Lim, Barentsen, Craig, Pascual, Perren, Greco, Donath, de~Val-Borro, Kerzendorf, Bach, Weaver, D'Eugenio, Souchereau, \& Ferreira}]{larry_bradley_2022_6825092}
Bradley, L., Sip{\H o}cz, B., Robitaille, T., {et~al.} 2022, astropy/photutils: 1.5.0

\bibitem[{{Bunker} {et~al.}(2023){Bunker}, {Saxena}, {Cameron}, {Willott}, {Curtis-Lake}, {Jakobsen}, {Carniani}, {Smit}, {Maiolino}, {Witstok}, {Curti}, {D'Eugenio}, {Jones}, {Ferruit}, {Arribas}, {Charlot}, {Chevallard}, {Giardino}, {de Graaff}, {Looser}, {L{\"u}tzgendorf}, {Maseda}, {Rawle}, {Rix}, {Del Pino}, {Alberts}, {Egami}, {Eisenstein}, {Endsley}, {Hainline}, {Hausen}, {Johnson}, {Rieke}, {Rieke}, {Robertson}, {Shivaei}, {Stark}, {Sun}, {Tacchella}, {Tang}, {Williams}, {Willmer}, {Baker}, {Baum}, {Bhatawdekar}, {Bowler}, {Boyett}, {Chen}, {Circosta}, {Helton}, {Ji}, {Kumari}, {Lyu}, {Nelson}, {Parlanti}, {Perna}, {Sandles}, {Scholtz}, {Suess}, {Topping}, {{\"U}bler}, {Wallace}, \& {Whitler}}]{Bunker+23}
{Bunker}, A.~J., {Saxena}, A., {Cameron}, A.~J., {et~al.} 2023, \aap, 677, A88

\bibitem[{Bushouse {et~al.}(2024)Bushouse, Eisenhamer, Dencheva, Davies, Greenfield, Morrison, Hodge, Simon, Grumm, Droettboom, Slavich, Sosey, Pauly, Miller, Jedrzejewski, Hack, Davis, Crawford, Law, Gordon, Regan, Cara, MacDonald, Bradley, Shanahan, Jamieson, Teodoro, Williams, \& Pena-Guerrero}]{bushouse_1.14.0}
Bushouse, H., Eisenhamer, J., Dencheva, N., {et~al.} 2024, JWST Calibration Pipeline

\bibitem[{{Calabro} {et~al.}(2024){Calabro}, {Castellano}, {Zavala}, {Pentericci}, {Arrabal Haro}, {Bakx}, {Burgarella}, {Casey}, {Dickinson}, {Finkelstein}, {Fontana}, {Llerena}, {Mascia}, {Merlin}, {Mitsuhashi}, {Napolitano}, {Paris}, {Perez-Gonzalez}, {Roberts-Borsani}, {Santini}, {Treu}, \& {Vanzella}}]{Calabro2024}
{Calabro}, A., {Castellano}, M., {Zavala}, J.~A., {et~al.} 2024, arXiv e-prints, arXiv:2403.12683

\bibitem[{{Calzetti}(2013)}]{Calzetti+13}
{Calzetti}, D. 2013, in Secular Evolution of Galaxies, ed. J.~{Falc{\'o}n-Barroso} \& J.~H. {Knapen}, 419

\bibitem[{{Cameron} {et~al.}(2023{\natexlab{a}}){Cameron}, {Katz}, {Rey}, \& {Saxena}}]{Cameron2023_MNRAS523}
{Cameron}, A.~J., {Katz}, H., {Rey}, M.~P., \& {Saxena}, A. 2023{\natexlab{a}}, \mnras, 523, 3516

\bibitem[{{Cameron} {et~al.}(2023{\natexlab{b}}){Cameron}, {Saxena}, {Bunker}, {D'Eugenio}, {Carniani}, {Maiolino}, {Curtis-Lake}, {Ferruit}, {Jakobsen}, {Arribas}, {Bonaventura}, {Charlot}, {Chevallard}, {Curti}, {Looser}, {Maseda}, {Rawle}, {Rodr{\'\i}guez Del Pino}, {Smit}, {{\"U}bler}, {Willott}, {Witstok}, {Egami}, {Eisenstein}, {Johnson}, {Hainline}, {Rieke}, {Robertson}, {Stark}, {Tacchella}, {Williams}, {Willmer}, {Bhatawdekar}, {Bowler}, {Boyett}, {Circosta}, {Helton}, {Jones}, {Kumari}, {Ji}, {Nelson}, {Parlanti}, {Sandles}, {Scholtz}, \& {Sun}}]{Cameron2023}
{Cameron}, A.~J., {Saxena}, A., {Bunker}, A.~J., {et~al.} 2023{\natexlab{b}}, \aap, 677, A115

\bibitem[{{Campbell} {et~al.}(1986){Campbell}, {Terlevich}, \& {Melnick}}]{Campbell+86}
{Campbell}, A., {Terlevich}, R., \& {Melnick}, J. 1986, \mnras, 223, 811

\bibitem[{{Carniani} {et~al.}(2024{\natexlab{a}}){Carniani}, {D'Eugenio}, {Ji}, {Parlanti}, {Scholtz}, {Sun}, {Venturi}, {Bakx}, {Curti}, {Maiolino}, {Tacchella}, {Zavala}, {Hainline}, {Witstok}, {Johnson}, {Alberts}, {Bunker}, {Charlot}, {Eisenstein}, {Helton}, {Jakobsen}, {Kumari}, {Robertson}, {Saxena}, {{\"U}bler}, {Williams}, {Willmer}, \& {Willott}}]{Carniani+24_ALMA}
{Carniani}, S., {D'Eugenio}, F., {Ji}, X., {et~al.} 2024{\natexlab{a}}, arXiv e-prints, arXiv:2409.20533

\bibitem[{{Carniani} {et~al.}(2024{\natexlab{b}}){Carniani}, {Hainline}, {D'Eugenio}, {Eisenstein}, {Jakobsen}, {Witstok}, {Johnson}, {Chevallard}, {Maiolino}, {Helton}, {Willott}, {Robertson}, {Alberts}, {Arribas}, {Baker}, {Bhatawdekar}, {Boyett}, {Bunker}, {Cameron}, {Cargile}, {Charlot}, {Curti}, {Curtis-Lake}, {Egami}, {Giardino}, {Isaak}, {Ji}, {Jones}, {Maseda}, {Parlanti}, {Rawle}, {Rieke}, {Rieke}, {Rodr{\'\i}guez Del Pino}, {Saxena}, {Scholtz}, {Smit}, {Sun}, {Tacchella}, {{\"U}bler}, {Venturi}, {Williams}, \& {Willmer}}]{Carniani+24}
{Carniani}, S., {Hainline}, K., {D'Eugenio}, F., {et~al.} 2024{\natexlab{b}}, arXiv e-prints, arXiv:2405.18485

\bibitem[{{Castellano} {et~al.}(2023){Castellano}, {Belfiori}, {Pentericci}, {Calabr{\`o}}, {Mascia}, {Napolitano}, {Caro}, {Charlot}, {Chevallard}, {Curtis Lake}, {Talia}, {Bongiorno}, {Fontana}, {Fynbo}, {Garilli}, {Guaita}, {McLure}, {Merlin}, {Mignoli}, {Moresco}, {Pompei}, {Pozzetti}, {Saldana Lopez}, {Saxena}, {Santini}, {Schaerer}, {Schreiber}, {Shapley}, {Vanzella}, \& {Zamorani}}]{Castellano+23}
{Castellano}, M., {Belfiori}, D., {Pentericci}, L., {et~al.} 2023, \aap, 675, A121

\bibitem[{{Castellano} {et~al.}(2024){Castellano}, {Napolitano}, {Fontana}, {Roberts-Borsani}, {Treu}, {Vanzella}, {Zavala}, {Arrabal Haro}, {Calabr{\`o}}, {Llerena}, {Mascia}, {Merlin}, {Paris}, {Pentericci}, {Santini}, {Bakx}, {Bergamini}, {Cupani}, {Dickinson}, {Filippenko}, {Glazebrook}, {Grillo}, {Kelly}, {Malkan}, {Mason}, {Morishita}, {Nanayakkara}, {Rosati}, {Sani}, {Wang}, \& {Yoon}}]{Castellano2024}
{Castellano}, M., {Napolitano}, L., {Fontana}, A., {et~al.} 2024, \apj, 972, 143

\bibitem[{{Chabrier}(2003)}]{Chabrier+03}
{Chabrier}, G. 2003, \pasp, 115, 763

\bibitem[{{Charbonnel} {et~al.}(2023){Charbonnel}, {Schaerer}, {Prantzos}, {Ram{\'\i}rez-Galeano}, {Fragos}, {Kuruvanthodi}, {Marques-Chaves}, \& {Gieles}}]{Charbonnel2023}
{Charbonnel}, C., {Schaerer}, D., {Prantzos}, N., {et~al.} 2023, \aap, 673, L7

\bibitem[{{Chisholm} {et~al.}(2022){Chisholm}, {Saldana-Lopez}, {Flury}, {Schaerer}, {Jaskot}, {Amor{\'\i}n}, {Atek}, {Finkelstein}, {Fleming}, {Ferguson}, {Fern{\'a}ndez}, {Giavalisco}, {Hayes}, {Heckman}, {Henry}, {Ji}, {Marques-Chaves}, {Mauerhofer}, {McCandliss}, {Oey}, {{\"O}stlin}, {Rutkowski}, {Scarlata}, {Thuan}, {Trebitsch}, {Wang}, {Worseck}, \& {Xu}}]{Chisholm2022}
{Chisholm}, J., {Saldana-Lopez}, A., {Flury}, S., {et~al.} 2022, \mnras, 517, 5104

\bibitem[{{Claeyssens} {et~al.}(2023){Claeyssens}, {Adamo}, {Richard}, {Mahler}, {Messa}, \& {Dessauges-Zavadsky}}]{Claeyssens+Adamo2023}
{Claeyssens}, A., {Adamo}, A., {Richard}, J., {et~al.} 2023, \mnras, 520, 2180

\bibitem[{{Crocker} {et~al.}(2018){Crocker}, {Krumholz}, {Thompson}, {Baumgardt}, \& {Mackey}}]{Crocker2018}
{Crocker}, R.~M., {Krumholz}, M.~R., {Thompson}, T.~A., {Baumgardt}, H., \& {Mackey}, D. 2018, \mnras, 481, 4895

\bibitem[{{Curti} {et~al.}(2023){Curti}, {D'Eugenio}, {Carniani}, {Maiolino}, {Sandles}, {Witstok}, {Baker}, {Bennett}, {Piotrowska}, {Tacchella}, {Charlot}, {Nakajima}, {Maheson}, {Mannucci}, {Amiri}, {Arribas}, {Belfiore}, {Bonaventura}, {Bunker}, {Chevallard}, {Cresci}, {Curtis-Lake}, {Hayden-Pawson}, {Jones}, {Kumari}, {Laseter}, {Looser}, {Marconi}, {Maseda}, {Scholtz}, {Smit}, {{\"U}bler}, \& {Wallace}}]{Curti+23}
{Curti}, M., {D'Eugenio}, F., {Carniani}, S., {et~al.} 2023, \mnras, 518, 425

\bibitem[{{Curti} {et~al.}(2024{\natexlab{a}}){Curti}, {Maiolino}, {Curtis-Lake}, {Chevallard}, {Carniani}, {D'Eugenio}, {Looser}, {Scholtz}, {Charlot}, {Cameron}, {{\"U}bler}, {Witstok}, {Boyett}, {Laseter}, {Sandles}, {Arribas}, {Bunker}, {Giardino}, {Maseda}, {Rawle}, {Rodr{\'\i}guez Del Pino}, {Smit}, {Willott}, {Eisenstein}, {Hausen}, {Johnson}, {Rieke}, {Robertson}, {Tacchella}, {Williams}, {Willmer}, {Baker}, {Bhatawdekar}, {Egami}, {Helton}, {Ji}, {Kumari}, {Perna}, {Shivaei}, \& {Sun}}]{2024A&A...684A..75C}
{Curti}, M., {Maiolino}, R., {Curtis-Lake}, E., {et~al.} 2024{\natexlab{a}}, \aap, 684, A75

\bibitem[{{Curti} {et~al.}(2024{\natexlab{b}}){Curti}, {Witstok}, {Jakobsen}, {Kobayashi}, {Curtis-Lake}, {Hainline}, {Ji}, {D'Eugenio}, {Chevallard}, {Maiolino}, {Scholtz}, {Carniani}, {Arribas}, {Baker}, {Bhatawdekar}, {Boyett}, {Bunker}, {Cameron}, {Cargile}, {Charlot}, {Eisenstein}, {Ji}, {Johnson}, {Kumari}, {Maseda}, {Robertson}, {Silcock}, {Tacchella}, {Ubler}, {Venturi}, {Williams}, {Willmer}, \& {Willott}}]{Curti+24}
{Curti}, M., {Witstok}, J., {Jakobsen}, P., {et~al.} 2024{\natexlab{b}}, arXiv e-prints, arXiv:2407.02575

\bibitem[{{Curtis-Lake} {et~al.}(2023){Curtis-Lake}, {Carniani}, {Cameron}, {Charlot}, {Jakobsen}, {Maiolino}, {Bunker}, {Witstok}, {Smit}, {Chevallard}, {Willott}, {Ferruit}, {Arribas}, {Bonaventura}, {Curti}, {D'Eugenio}, {Franx}, {Giardino}, {Looser}, {L{\"u}tzgendorf}, {Maseda}, {Rawle}, {Rix}, {Rodr{\'\i}guez del Pino}, {{\"U}bler}, {Sirianni}, {Dressler}, {Egami}, {Eisenstein}, {Endsley}, {Hainline}, {Hausen}, {Johnson}, {Rieke}, {Robertson}, {Shivaei}, {Stark}, {Tacchella}, {Williams}, {Willmer}, {Bhatawdekar}, {Bowler}, {Boyett}, {Chen}, {de Graaff}, {Helton}, {Hviding}, {Jones}, {Kumari}, {Lyu}, {Nelson}, {Perna}, {Sandles}, {Saxena}, {Suess}, {Sun}, {Topping}, {Wallace}, \& {Whitler}}]{Curtis-Lake+23}
{Curtis-Lake}, E., {Carniani}, S., {Cameron}, A., {et~al.} 2023, Nature Astronomy, 7, 622

\bibitem[{{D'Antona} {et~al.}(2023){D'Antona}, {Vesperini}, {Calura}, {Ventura}, {D'Ercole}, {Caloi}, {Marino}, {Milone}, {Dell'Agli}, \& {Tailo}}]{DAntona2023}
{D'Antona}, F., {Vesperini}, E., {Calura}, F., {et~al.} 2023, \aap, 680, L19

\bibitem[{{Dekel} {et~al.}(2023){Dekel}, {Sarkar}, {Birnboim}, {Mandelker}, \& {Li}}]{Dekel2023}
{Dekel}, A., {Sarkar}, K.~C., {Birnboim}, Y., {Mandelker}, N., \& {Li}, Z. 2023, \mnras, 523, 3201

\bibitem[{{D{\'\i}az} {et~al.}(2000){D{\'\i}az}, {Castellanos}, {Terlevich}, \& {Luisa Garc{\'\i}a-Vargas}}]{Diaz+00}
{D{\'\i}az}, A.~I., {Castellanos}, M., {Terlevich}, E., \& {Luisa Garc{\'\i}a-Vargas}, M. 2000, \mnras, 318, 462

\bibitem[{{Dopita} \& {Sutherland}(2003)}]{Dopita03}
{Dopita}, M.~A. \& {Sutherland}, R.~S. 2003, {Astrophysics of the diffuse universe}

\bibitem[{{Dors}(2021)}]{Dors+21}
{Dors}, O.~L. 2021, \mnras, 507, 466

\bibitem[{{Dors} {et~al.}(2020{\natexlab{a}}){Dors}, {Freitas-Lemes}, {Am{\^o}res}, {P{\'e}rez-Montero}, {Cardaci}, {H{\"a}gele}, {Armah}, {Krabbe}, \& {Fa{\'u}ndez-Abans}}]{Dors+20a}
{Dors}, O.~L., {Freitas-Lemes}, P., {Am{\^o}res}, E.~B., {et~al.} 2020{\natexlab{a}}, \mnras, 492, 468

\bibitem[{{Dors} {et~al.}(2020{\natexlab{b}}){Dors}, {Maiolino}, {Cardaci}, {H{\"a}gele}, {Krabbe}, {P{\'e}rez-Montero}, \& {Armah}}]{Dors+20b}
{Dors}, O.~L., {Maiolino}, R., {Cardaci}, M.~V., {et~al.} 2020{\natexlab{b}}, \mnras, 496, 3209

\bibitem[{{Ferrara} {et~al.}(2023){Ferrara}, {Pallottini}, \& {Dayal}}]{Ferrara23}
{Ferrara}, A., {Pallottini}, A., \& {Dayal}, P. 2023, \mnras, 522, 3986

\bibitem[{{Finkelstein} {et~al.}(2023){Finkelstein}, {Bagley}, {Ferguson}, {Wilkins}, {Kartaltepe}, {Papovich}, {Yung}, {Haro}, {Behroozi}, {Dickinson}, {Kocevski}, {Koekemoer}, {Larson}, {Le Bail}, {Morales}, {P{\'e}rez-Gonz{\'a}lez}, {Burgarella}, {Dav{\'e}}, {Hirschmann}, {Somerville}, {Wuyts}, {Bromm}, {Casey}, {Fontana}, {Fujimoto}, {Gardner}, {Giavalisco}, {Grazian}, {Grogin}, {Hathi}, {Hutchison}, {Jha}, {Jogee}, {Kewley}, {Kirkpatrick}, {Long}, {Lotz}, {Pentericci}, {Pierel}, {Pirzkal}, {Ravindranath}, {Ryan}, {Trump}, {Yang}, {Bhatawdekar}, {Bisigello}, {Buat}, {Calabr{\`o}}, {Castellano}, {Cleri}, {Cooper}, {Croton}, {Daddi}, {Dekel}, {Elbaz}, {Franco}, {Gawiser}, {Holwerda}, {Huertas-Company}, {Jaskot}, {Leung}, {Lucas}, {Mobasher}, {Pandya}, {Tacchella}, {Weiner}, \& {Zavala}}]{Finkelstein+23}
{Finkelstein}, S.~L., {Bagley}, M.~B., {Ferguson}, H.~C., {et~al.} 2023, \apjl, 946, L13

\bibitem[{{Fiore} {et~al.}(2023){Fiore}, {Ferrara}, {Bischetti}, {Feruglio}, \& {Travascio}}]{Fiore23}
{Fiore}, F., {Ferrara}, A., {Bischetti}, M., {Feruglio}, C., \& {Travascio}, A. 2023, \apjl, 943, L27

\bibitem[{{Fudamoto} {et~al.}(2024){Fudamoto}, {Oesch}, {Walter}, {Decarli}, {Carilli}, {Ferrara}, {Barrufet}, {Bouwens}, {Dessauges-Zavadsky}, {Nelson}, {Dannerbauer}, {Illingworth}, {Inoue}, {Marques-Chaves}, {P{\'e}rez-Fournon}, {Riechers}, {Schaerer}, {Smit}, {Sugahara}, \& {van der Werf}}]{Fudamoto2024}
{Fudamoto}, Y., {Oesch}, P.~A., {Walter}, F., {et~al.} 2024, \mnras, 530, 340

\bibitem[{{Fujimoto} {et~al.}(2023){Fujimoto}, {Arrabal Haro}, {Dickinson}, {Finkelstein}, {Kartaltepe}, {Larson}, {Burgarella}, {Bagley}, {Behroozi}, {Chworowsky}, {Hirschmann}, {Trump}, {Wilkins}, {Yung}, {Koekemoer}, {Papovich}, {Pirzkal}, {Ferguson}, {Fontana}, {Grogin}, {Grazian}, {Kewley}, {Kocevski}, {Lotz}, {Pentericci}, {Ravindranath}, {Somerville}, {Wilkins}, {Amor{\'\i}n}, {Backhaus}, {Calabr{\`o}}, {Casey}, {Cooper}, {Fern{\'a}ndez}, {Franco}, {Giavalisco}, {Hathi}, {Harish}, {Hutchison}, {Iyer}, {Jung}, {Lucas}, \& {Zavala}}]{Fujimoto+23}
{Fujimoto}, S., {Arrabal Haro}, P., {Dickinson}, M., {et~al.} 2023, \apjl, 949, L25

\bibitem[{{Furtak} {et~al.}(2024){Furtak}, {Labb{\'e}}, {Zitrin}, {Greene}, {Dayal}, {Chemerynska}, {Kokorev}, {Miller}, {Goulding}, {de Graaff}, {Bezanson}, {Brammer}, {Cutler}, {Leja}, {Pan}, {Price}, {Wang}, {Weaver}, {Whitaker}, {Atek}, {Bogd{\'a}n}, {Charlot}, {Curtis-Lake}, {van Dokkum}, {Endsley}, {Feldmann}, {Fudamoto}, {Fujimoto}, {Glazebrook}, {Juneau}, {Marchesini}, {Maseda}, {Nelson}, {Oesch}, {Plat}, {Setton}, {Stark}, \& {Williams}}]{Furtak+24}
{Furtak}, L.~J., {Labb{\'e}}, I., {Zitrin}, A., {et~al.} 2024, \nat, 628, 57

\bibitem[{{Garcia} {et~al.}(2023){Garcia}, {Ricotti}, {Sugimura}, \& {Park}}]{Garcia2023}
{Garcia}, F. A.~B., {Ricotti}, M., {Sugimura}, K., \& {Park}, J. 2023, \mnras, 522, 2495

\bibitem[{{Gardner} {et~al.}(2023){Gardner}, {Mather}, {Abbott}, {Abell}, {Abernathy}, {Abney}, {Abraham}, {Abraham}, {Abul-Huda}, {Acton}, {Adams}, {Adams}, {Adler}, {Adriaensen}, {Aguilar}, {Ahmed}, {Ahmed}, {Ahmed}, {Albat}, {Albert}, {Alberts}, {Aldridge}, {Allen}, {Allen}, {Altenburg}, {Altunc}, {Alvarez}, {{\'A}lvarez-M{\'a}rquez}, {de Oliveira}, {Ambrose}, {Anandakrishnan}, {Andersen}, {Anderson}, {Anderson}, {Anderson}, {Anderson}, {Aprea}, {Archer}, {Arenberg}, {Argyriou}, {Arribas}, {Artigau}, {Arvai}, {Atcheson}, {Atkinson}, {Averbukh}, {Aymergen}, {Bacinski}, {Baggett}, {Bagnasco}, {Baker}, {Balzano}, {Banks}, {Baran}, {Barker}, {Barrett}, {Barringer}, {Barto}, {Bast}, {Baudoz}, {Baum}, {Beatty}, {Beaulieu}, {Bechtold}, {Beck}, {Beddard}, {Beichman}, {Bellagama}, {Bely}, {Berger}, {Bergeron}, {Bernier}, {Bertch}, {Beskow}, {Betz}, {Biagetti}, {Birkmann}, {Bjorklund}, {Blackwood}, {Blazek}, {Blossfeld}, {Bluth}, {Boccaletti}, {Boegner}, {Bohlin}, {Boia}, {B{\"o}ker}, {Bonaventura}, {Bond},
  {Bosley}, {Boucarut}, {Bouchet}, {Bouwman}, {Bower}, {Bowers}, {Bowers}, {Boyce}, {Boyer}, {Boyer}, {Boyer}, {Boyer}, {Bradley}, {Brady}, {Brandl}, {Brannen}, {Breda}, {Bremmer}, {Brennan}, {Bresnahan}, {Bright}, {Broiles}, {Bromenschenkel}, {Brooks}, {Brooks}, {Brown}, {Brown}, {Brown}, {Bruce}, {Bryson}, {Bujanda}, {Bullock}, {Bunker}, {Bureo}, {Burt}, {Bush}, {Bushouse}, {Bussman}, {Cabaud}, {Cale}, {Calhoon}, {Calvani}, {Canipe}, {Caputo}, {Cara}, {Carey}, {Case}, {Cesari}, {Cetorelli}, {Chance}, {Chandler}, {Chaney}, {Chapman}, {Charlot}, {Chayer}, {Cheezum}, {Chen}, {Chen}, {Cherinka}, {Chichester}, {Chilton}, {Chittiraibalan}, {Clampin}, {Clark}, {Clark}, {Clark}, {Claybrooks}, {Cleveland}, {Cohen}, {Cohen}, {Col{\'o}n}, {Coleman}, {Colina}, {Comber}, {Comeau}, {Comer}, {Reis}, {Connolly}, {Conroy}, {Contos}, {Contreras}, {Cook}, {Cooper}, {Cooper}, {Correia}, {Correnti}, {Cossou}, {Costanza}, {Coulais}, {Cox}, {Coyle}, {Cracraft}, {Crew}, {Curtis}, {Cusveller}, {Maciel}, {Dailey}, {Daugeron},
  {Davidson}, {Davies}, {Davis}, {Davis}, {Day}, {de Chambure}, {de Jong}, {De Marchi}, {Dean}, {Decker}, {Delisa}, {Dell}, {Dellagatta}, {Dembinska}, {Demosthenes}, {Dencheva}, {Deneu}, {DePriest}, {Deschenes}, {Dethienne}, {Detre}, {Diaz}, {Dicken}, {DiFelice}, {Dillman}, {Disharoon}, {Dixon}, {Doggett}, {Dominguez}, {Donaldson}, {Doria-Warner}, {Santos}, {Doty}, {Douglas}, {Doyon}, {Dressler}, {Driggers}, {Driggers}, {Dunn}, {DuPrie}, {Dupuis}, {Durning}, {Dutta}, {Earl}, {Eccleston}, {Ecobichon}, {Egami}, {Ehrenwinkler}, {Eisenhamer}, {Eisenhower}, {Eisenstein}, {El Hamel}, {Elie}, {Elliott}, {Elliott}, {Engesser}, {Espinoza}, {Etienne}, {Etxaluze}, {Evans}, {Fabreguettes}, {Falcolini}, {Falini}, {Fatig}, {Feeney}, {Feinberg}, {Fels}, {Ferdous}, {Ferguson}, {Ferrarese}, {Ferreira}, {Ferruit}, {Ferry}, {Filippazzo}, {Firre}, {Fix}, {Flagey}, {Flanagan}, {Fleming}, {Florian}, {Flynn}, {Foiadelli}, {Fontaine}, {Fontanella}, {Forshay}, {Fortner}, {Fox}, {Framarini}, {Francisco}, {Franck}, {Franx}, {Franz},
  {Friedman}, {Friend}, {Frost}, {Fu}, {Fullerton}, {Gaillard}, {Galkin}, {Gallagher}, {Galyer}, {Garc{\'\i}a Mar{\'\i}n}, {Gardner}, {Garland}, {Garrett}, {Gasman}, {G{\'a}sp{\'a}r}, {Gastaud}, {Gaudreau}, {Gauthier}, {Geers}, {Geithner}, {Gennaro}, {Gerber}, {Gereau}, {Giampaoli}, {Giardino}, {Gibbons}, {Gilbert}, {Gilman}, {Girard}, {Giuliano}, {Gkountis}, {Glasse}, {Glassmire}, {Glauser}, {Glazer}, {Goldberg}, {Golimowski}, {Gonzaga}, {Gordon}, {Gordon}, {Goudfrooij}, {Gough}, {Graham}, {Grau}, {Green}, {Greene}, {Greene}, {Greenfield}, {Greenhouse}, {Greve}, {Greville}, {Grimaldi}, {Groe}, {Groebner}, {Grumm}, {Grundy}, {G{\"u}del}, {Guillard}, {Guldalian}, {Gunn}, {Gurule}, {Gutman}, {Guy}, {Guyot}, {Hack}, {Haderlein}, {Hagan}, {Hagedorn}, {Hainline}, {Haley}, {Hami}, {Hamilton}, {Hammann}, {Hammel}, {Hanley}, {Hansen}, {Hardy}, {Harnisch}, {Harr}, {Harris}, {Hart}, {Hartig}, {Hasan}, {Hashim}, {Hashimoto}, {Haskins}, {Hawkins}, {Hayden}, {Hayden}, {Healy}, {Hecht}, {Heeg}, {Hejal}, {Helm},
  {Hengemihle}, {Henning}, {Henry}, {Henry}, {Henshaw}, {Hernandez}, {Herrington}, {Heske}, {Hesman}, {Hickey}, {Hilbert}, {Hines}, {Hinz}, {Hirsch}, {Hitcho}, {Hodapp}, {Hodge}, {Hoffman}, {Holfeltz}, {Holler}, {Hoppa}, {Horner}, {Howard}, {Howard}, {Huber}, {Hunkeler}, {Hunter}, {Hunter}, {Hurd}, {Hurst}, {Hutchings}, {Hylan}, {Ignat}, {Illingworth}, {Irish}, {Isaacs}, {Jackson}, {Jaffe}, {Jahic}, {Jahromi}, {Jakobsen}, {James}, {James}, {James}, {Jamieson}, {Jandra}, {Jayawardhana}, {Jedrzejewski}, {Jeffers}, {Jensen}, {Joanne}, {Johns}, {Johnson}, {Johnson}, {Johnson}, {Johnson}, {Johnson}, {Johnson}, {Johnstone}, {Jollet}, {Jones}, {Jones}, {Jones}, {Jones}, {Jones}, {Jordan}, {Jordan}, {Jue}, {Jurkowski}, {Justis}, {Justtanont}, {Kaleida}, {Kalirai}, {Kalmanson}, {Kaltenegger}, {Kammerer}, {Kan}, {Kanarek}, {Kao}, {Karakla}, {Karl}, {Kassin}, {Kauffman}, {Kavanagh}, {Kelley}, {Kelly}, {Kendrew}, {Kennedy}, {Kenny}, {Keski-Kuha}, {Keyes}, {Khan}, {Kidwell}, {Kimble}, {King}, {King}, {Kinzel}, {Kirk},
  {Kirkpatrick}, {Klaassen}, {Klingemann}, {Klintworth}, {Knapp}, {Knight}, {Knollenberg}, {Knutsen}, {Koehler}, {Koekemoer}, {Kofler}, {Kontson}, {Kovacs}, {Kozhurina-Platais}, {Krause}, {Kriss}, {Krist}, {Kristoffersen}, {Krogel}, {Krueger}, {Kulp}, {Kumari}, {Kwan}, {Kyprianou}, {Labador}, {Labiano}, {Lafreni{\`e}re}, {Lagage}, {Laidler}, {Laine}, {Laird}, {Lajoie}, {Lallo}, {Lam}, {LaMassa}, {Lambros}, {Lampenfield}, {Lander}, {Langston}, {Larson}, {Larson}, {LaVerghetta}, {Law}, {Lawrence}, {Lee}, {Lee}, {Lee}, {Leisenring}, {Leveille}, {Levenson}, {Levi}, {Levine}, {Lewis}, {Lewis}, {Lewis}, {Libralato}, {Lidon}, {Liebrecht}, {Lightsey}, {Lilly}, {Lim}, {Lim}, {Ling}, {Link}, {Link}, {Lipinski}, {Liu}, {Lo}, {Lobmeyer}, {Logue}, {Long}, {Long}, {Long}, {Long}, {L{\'o}pez-Caniego}, {Lotz}, {Love-Pruitt}, {Lubskiy}, {Luers}, {Luetgens}, {Luevano}, {Lui}, {Lund}, {Lundquist}, {Lunine}, {L{\"u}tzgendorf}, {Lynch}, {MacDonald}, {MacDonald}, {Macias}, {Macklis}, {Maghami}, {Maharaja}, {Maiolino},
  {Makrygiannis}, {Malla}, {Malumuth}, {Manjavacas}, {Marini}, {Marrione}, {Marston}, {Martel}, {Martin}, {Martin}, {Martinez}, {Maschmann}, {Masci}, {Masetti}, {Maszkiewicz}, {Matthews}, {Matuskey}, {McBrayer}, {McCarthy}, {McCaughrean}, {McClare}, {McClare}, {McCloskey}, {McClurg}, {McCoy}, {McElwain}, {McGregor}, {McGuffey}, {McKay}, {McKenzie}, {McLean}, {McMaster}, {McNeil}, {De Meester}, {Mehalick}, {Meixner}, {Mel{\'e}ndez}, {Menzel}, {Menzel}, {Merz}, {Mesterharm}, {Meyer}, {Meyett}, {Meza}, {Midwinter}, {Milam}, {Miller}, {Miller}, {Miskey}, {Misselt}, {Mitchell}, {Mohan}, {Montoya}, {Moran}, {Morishita}, {Moro-Mart{\'\i}n}, {Morrison}, {Morrison}, {Morse}, {Moschos}, {Moseley}, {Mosier}, {Mosner}, {Mountain}, {Muckenthaler}, {Mueller}, {Mueller}, {Muhiem}, {M{\"u}hlmann}, {Mullally}, {Mullen}, {Munger}, {Murphy}, {Murray}, {Muzerolle}, {Mycroft}, {Myers}, {Myers}, {Myers}, {Myers}, {Myrick}, {Nagle}, {Nayak}, {Naylor}, {Neff}, {Nelan}, {Nella}, {Nguyen}, {Nguyen}, {Nickson}, {Nidhiry}, {Niedner},
  {Nieto-Santisteban}, {Nikolov}, {Nishisaka}, {Noriega-Crespo}, {Nota}, {O'Mara}, {Oboryshko}, {O'Brien}, {Ochs}, {Offenberg}, {Ogle}, {Ohl}, {Olmsted}, {Osborne}, {O'Shaughnessy}, {{\"O}stlin}, {O'Sullivan}, {Otor}, {Ottens}, {Ouellette}, {Outlaw}, {Owens}, {Pacifici}, {Page}, {Paranilam}, {Park}, {Parrish}, {Paschal}, {Patapis}, {Patel}, {Patrick}, {Pattishall}, {Paul}, {Paul}, {Pauly}, {Pavlovsky}, {Pe{\~n}a-Guerrero}, {Pedder}, {Peek}, {Pelham}, {Penanen}, {Perriello}, {Perrin}, {Perrine}, {Perrygo}, {Peslier}, {Petach}, {Peterson}, {Pfarr}, {Pierson}, {Pietraszkiewicz}, {Pilchen}, {Pipher}, {Pirzkal}, {Pitman}, {Player}, {Plesha}, {Plitzke}, {Pohner}, {Poletis}, {Pollizzi}, {Polster}, {Pontius}, {Pontoppidan}, {Porges}, {Potter}, {Prescott}, {Proffitt}, {Pueyo}, {Quispe Neira}, {Radich}, {Rager}, {Rameau}, {Ramey}, {Alarcon}, {Rampini}, {Rapp}, {Rashford}, {Rauscher}, {Ravindranath}, {Rawle}, {Rawlings}, {Ray}, {Regan}, {Rehm}, {Rehm}, {Reid}, {Reis}, {Renk}, {Reoch}, {Ressler}, {Rest}, {Reynolds},
  {Richon}, {Richon}, {Ridgaway}, {Riedel}, {Rieke}, {Rieke}, {Rifelli}, {Rigby}, {Riggs}, {Ringel}, {Ritchie}, {Rix}, {Robberto}, {Robinson}, {Robinson}, {Robinson}, {Rock}, {Rodriguez}, {Rodr{\'\i}guez del Pino}, {Roellig}, {Rohrbach}, {Roman}, {Romelfanger}, {Romo}, {Rosales}, {Rose}, {Roteliuk}, {Roth}, {Rothwell}, {Rouzaud}, {Rowe}, {Rowlands}, {Roy}, {Royer}, {Rui}, {Rumler}, {Rumpl}, {Russ}, {Ryan}, {Ryan}, {Saad}, {Sabata}, {Sabatino}, {Sabbi}, {Sabelhaus}, {Sabia}, {Sahu}, {Saif}, {Salvignol}, {Samara-Ratna}, {Samuelson}, {Sanders}, {Sappington}, {Sargent}, {Sauer}, {Savadkin}, {Sawicki}, {Schappell}, {Scheffer}, {Scheithauer}, {Scherer}, {Schiff}, {Schlawin}, {Schmeitzky}, {Schmitz}, {Schmude}, {Schneider}, {Schreiber}, {Schroeven-Deceuninck}, {Schultz}, {Schwab}, {Schwartz}, {Scoccimarro}, {Scott}, {Scott}, {Seaton}, {Seely}, {Seery}, {Seidleck}, {Sembach}, {Shanahan}, {Shaughnessy}, {Shaw}, {Shay}, {Sheehan}, {Sheth}, {Shih}, {Shivaei}, {Siegel}, {Sienkiewicz}, {Simmons}, {Simon}, {Sirianni},
  {Sivaramakrishnan}, {Slade}, {Sloan}, {Slocum}, {Slowinski}, {Smith}, {Smith}, {Smith}, {Smith}, {Smith}, {Smith}, {Smolik}, {Soderblom}, {Sohn}, {Sokol}, {Sonneborn}, {Sontag}, {Sooy}, {Soummer}, {Southwood}, {Spain}, {Sparmo}, {Speer}, {Spencer}, {Sprofera}, {Stallcup}, {Stanley}, {Stansberry}, {Stark}, {Starr}, {Stassi}, {Steck}, {Steeley}, {Stephens}, {Stephenson}, {Stewart}, {Stiavelli}, {}, {Strada}, {Straughn}, {Streetman}, {Strickland}, {Strobele}, {Stuhlinger}, {Stys}, {Such}, {Sukhatme}, {Sullivan}, {Sullivan}, {Sumner}, {Sun}, {Sunnquist}, {Swade}, {Swam}, {Swenton}, {Swoish}, {Tam Litten}, {Tamas}, {Tao}, {Taylor}, {Taylor}, {Plate}, {Van Tea}, {Teague}, {Telfer}, {Temim}, {Texter}, {Thatte}, {Thompson}, {Thompson}, {Thomson}, {Thronson}, {Tierney}, {Tikkanen}, {Tinnin}, {Tippet}, {Todd}, {Tran}, {Trauger}, {Trejo}, {Vinh Truong}, {Tsukamoto}, {Tufail}, {Tumlinson}, {Tustain}, {Tyra}, {Ubeda}, {Underwood}, {Uzzo}, {Vaclavik}, {Valenduc}, {Valenti}, {Van Campen}, {van de Wetering}, {Van Der
  Marel}, {van Haarlem}, {Vandenbussche}, {van Dishoeck}, {Vanterpool}, {Vernoy}, {Vila Costas}, {Volk}, {Voorzaat}, {Voyton}, {Vydra}, {Waddy}, {Waelkens}, {Wahlgren}, {Walker}, {Wander}, {Warfield}, {Warner}, {Wasiak}, {Wasiak}, {Wehner}, {Weiler}, {Weilert}, {Weiss}, {Wells}, {Welty}, {Wheate}, {Wheeler}, {White}, {Whitehouse}, {Whiteleather}, {Whitman}, {Williams}, {Willmer}, {Willott}, {Willoughby}, {Wilson}, {Wilson}, {Wilson}, {Windhorst}, {Wislowski}, {Wolfe}, {Wolfe}, {Wolff}, {Wondel}, {Woo}, {Woods}, {Worden}, {Workman}, {Wright}, {Wu}, {Wu}, {Wun}, {Wymer}, {Yadetie}, {Yan}, {Yang}, {Yates}, {Yeager}, {Yerger}, {Young}, {Young}, {Yu}, {Yu}, {Zak}, {Zeidler}, {Zepp}, {Zhou}, {Zincke}, {Zonak}, \& {Zondag}}]{Gardner+23}
{Gardner}, J.~P., {Mather}, J.~C., {Abbott}, R., {et~al.} 2023, \pasp, 135, 068001

\bibitem[{{Gim{\'e}nez-Arteaga} {et~al.}(2024){Gim{\'e}nez-Arteaga}, {Fujimoto}, {Valentino}, {Brammer}, {Mason}, {Rizzo}, {Rusakov}, {Colina}, {Prieto-Lyon}, {Oesch}, {Espada}, {Heintz}, {Knudsen}, {Dessauges-Zavadsky}, {Laporte}, {Lee}, {Magdis}, {Ono}, {Ao}, {Ouchi}, {Kohno}, \& {Koekemoer}}]{Gimenez-Arteaga2024}
{Gim{\'e}nez-Arteaga}, C., {Fujimoto}, S., {Valentino}, F., {et~al.} 2024, \aap, 686, A63

\bibitem[{{Greene} \& {Ho}(2005{\natexlab{a}})}]{Greene+05a}
{Greene}, J.~E. \& {Ho}, L.~C. 2005{\natexlab{a}}, \apj, 627, 721

\bibitem[{{Greene} \& {Ho}(2005{\natexlab{b}})}]{Greene+05b}
{Greene}, J.~E. \& {Ho}, L.~C. 2005{\natexlab{b}}, \apj, 630, 122

\bibitem[{{Greene} {et~al.}(2024){Greene}, {Labbe}, {Goulding}, {Furtak}, {Chemerynska}, {Kokorev}, {Dayal}, {Volonteri}, {Williams}, {Wang}, {Setton}, {Burgasser}, {Bezanson}, {Atek}, {Brammer}, {Cutler}, {Feldmann}, {Fujimoto}, {Glazebrook}, {de Graaff}, {Khullar}, {Leja}, {Marchesini}, {Maseda}, {Matthee}, {Miller}, {Naidu}, {Nanayakkara}, {Oesch}, {Pan}, {Papovich}, {Price}, {van Dokkum}, {Weaver}, {Whitaker}, \& {Zitrin}}]{Greene+24}
{Greene}, J.~E., {Labbe}, I., {Goulding}, A.~D., {et~al.} 2024, \apj, 964, 39

\bibitem[{{Grudi{\'c}} {et~al.}(2019){Grudi{\'c}}, {Hopkins}, {Quataert}, \& {Murray}}]{Grudic2019}
{Grudi{\'c}}, M.~Y., {Hopkins}, P.~F., {Quataert}, E., \& {Murray}, N. 2019, \mnras, 483, 5548

\bibitem[{{Harikane} {et~al.}(2023{\natexlab{a}}){Harikane}, {Ouchi}, {Oguri}, {Ono}, {Nakajima}, {Isobe}, {Umeda}, {Mawatari}, \& {Zhang}}]{Harikane+23a}
{Harikane}, Y., {Ouchi}, M., {Oguri}, M., {et~al.} 2023{\natexlab{a}}, \apjs, 265, 5

\bibitem[{{Harikane} {et~al.}(2023{\natexlab{b}}){Harikane}, {Zhang}, {Nakajima}, {Ouchi}, {Isobe}, {Ono}, {Hatano}, {Xu}, \& {Umeda}}]{Harikane+23b}
{Harikane}, Y., {Zhang}, Y., {Nakajima}, K., {et~al.} 2023{\natexlab{b}}, \apj, 959, 39

\bibitem[{{Hayes} \& {Scarlata}(2023)}]{Hayes2023}
{Hayes}, M.~J. \& {Scarlata}, C. 2023, \apjl, 954, L14

\bibitem[{{Heintz} {et~al.}(2023){Heintz}, {Brammer}, {Gim{\'e}nez-Arteaga}, {Strait}, {del P. Lagos}, {Vijayan}, {Matthee}, {Watson}, {Mason}, {Hutter}, {Toft}, {Fynbo}, \& {Oesch}}]{Heintz2023}
{Heintz}, K.~E., {Brammer}, G.~B., {Gim{\'e}nez-Arteaga}, C., {et~al.} 2023, Nature Astronomy, 7, 1517

\bibitem[{{Helton} {et~al.}(2024){Helton}, {Rieke}, {Alberts}, {Wu}, {Eisenstein}, {Hainline}, {Carniani}, {Ji}, {Baker}, {Bhatawdekar}, {Bunker}, {Cargile}, {Charlot}, {Chevallard}, {D'Eugenio}, {Egami}, {Johnson}, {Jones}, {Lyu}, {Maiolino}, {P{\'e}rez-Gonz{\'a}lez}, {Rieke}, {Robertson}, {Saxena}, {Scholtz}, {Shivaei}, {Sun}, {Tacchella}, {Whitler}, {Williams}, {Willmer}, {Willott}, {Witstok}, \& {Zhu}}]{Helton2024}
{Helton}, J.~M., {Rieke}, G.~H., {Alberts}, S., {et~al.} 2024, arXiv e-prints, arXiv:2405.18462

\bibitem[{{Ho} {et~al.}(2001){Ho}, {Feigelson}, {Townsley}, {Sambruna}, {Garmire}, {Brandt}, {Filippenko}, {Griffiths}, {Ptak}, \& {Sargent}}]{Ho2001}
{Ho}, L.~C., {Feigelson}, E.~D., {Townsley}, L.~K., {et~al.} 2001, \apjl, 549, L51

\bibitem[{{Hodge} {et~al.}(2015){Hodge}, {Riechers}, {Decarli}, {Walter}, {Carilli}, {Daddi}, \& {Dannerbauer}}]{Hodge2015}
{Hodge}, J.~A., {Riechers}, D., {Decarli}, R., {et~al.} 2015, \apjl, 798, L18

\bibitem[{{Hopkins} {et~al.}(2010){Hopkins}, {Murray}, {Quataert}, \& {Thompson}}]{Hopkins2010}
{Hopkins}, P.~F., {Murray}, N., {Quataert}, E., \& {Thompson}, T.~A. 2010, \mnras, 401, L19

\bibitem[{{Hsiao} {et~al.}(2024{\natexlab{a}}){Hsiao}, {Abdurro'uf}, {Coe}, {Larson}, {Jung}, {Mingozzi}, {Dayal}, {Kumari}, {Kokorev}, {Vikaeus}, {Brammer}, {Furtak}, {Adamo}, {Andrade-Santos}, {Antwi-Danso}, {Brada{\v{c}}}, {Bradley}, {Broadhurst}, {Carnall}, {Conselice}, {Diego}, {Donahue}, {Eldridge}, {Fujimoto}, {Henry}, {Hernandez}, {Hutchison}, {James}, {Norman}, {Park}, {Pirzkal}, {Postman}, {Ricotti}, {Rigby}, {Vanzella}, {Welch}, {Wilkins}, {Windhorst}, {Xu}, {Zackrisson}, \& {Zitrin}}]{Hsiao+23-NIRSpec}
{Hsiao}, T. Y.-Y., {Abdurro'uf}, {Coe}, D., {et~al.} 2024{\natexlab{a}}, \apj, 973, 8

\bibitem[{{Hsiao} {et~al.}(2024{\natexlab{b}}){Hsiao}, {{\'A}lvarez-M{\'a}rquez}, {Coe}, {Crespo G{\'o}mez}, {Abdurro'uf}, {Dayal}, {Larson}, {Bik}, {Blanco-Prieto}, {Colina}, {P{\'e}rez-Gonz{\'a}lez}, {Costantin}, {Prieto-Jim{\'e}nez}, {Adamo}, {Bradley}, {Conselice}, {Fujimoto}, {Furtak}, {Hutchison}, {James}, {Jim{\'e}nez-Teja}, {Jung}, {Kokorev}, {Mingozzi}, {Norman}, {Ricotti}, {Rigby}, {Sharon}, {Vanzella}, {Welch}, {Xu}, {Zackrisson}, \& {Zitrin}}]{Hsiao+2024_MIRI}
{Hsiao}, T. Y.-Y., {{\'A}lvarez-M{\'a}rquez}, J., {Coe}, D., {et~al.} 2024{\natexlab{b}}, \apj, 973, 81

\bibitem[{{Hsiao} {et~al.}(2023){Hsiao}, {Coe}, {Abdurro'uf}, {Whitler}, {Jung}, {Khullar}, {Meena}, {Dayal}, {Barrow}, {Santos-Olmsted}, {Casselman}, {Vanzella}, {Nonino}, {Jim{\'e}nez-Teja}, {Oguri}, {Stark}, {Furtak}, {Zitrin}, {Adamo}, {Brammer}, {Bradley}, {Diego}, {Zackrisson}, {Finkelstein}, {Windhorst}, {Bhatawdekar}, {Hutchison}, {Broadhurst}, {Dimauro}, {Andrade-Santos}, {Eldridge}, {Acebron}, {Avila}, {Bayliss}, {Ben{\'\i}tez}, {Binggeli}, {Bolan}, {Brada{\v{c}}}, {Carnall}, {Conselice}, {Donahue}, {Frye}, {Fujimoto}, {Henry}, {James}, {Kassin}, {Kewley}, {Larson}, {Lauer}, {Law}, {Mahler}, {Mainali}, {McCandliss}, {Nicholls}, {Pirzkal}, {Postman}, {Rigby}, {Ryan}, {Senchyna}, {Sharon}, {Shimizu}, {Strait}, {Tang}, {Trenti}, {Vikaeus}, \& {Welch}}]{Hsiao+23-NIRCam}
{Hsiao}, T. Y.-Y., {Coe}, D., {Abdurro'uf}, {et~al.} 2023, \apjl, 949, L34

\bibitem[{{Iani} {et~al.}(2024){Iani}, {Rinaldi}, {Caputi}, {Annunziatella}, {Langeroodi}, {Melinder}, {P{\'e}rez-Gonz{\'a}lez}, {{\'A}lvarez-M{\'a}rquez}, {Boogaard}, {Bosman}, {Costantin}, {Moutard}, {Colina}, {{\"O}stlin}, {Greve}, {Wright}, {Alonso-Herrero}, {Bik}, {Gillman}, {Crespo G{\'o}mez}, {Hjorth}, {Labiano}, {Pye}, {Tikkanen}, \& {van der Werf}}]{Iani-LRD2024}
{Iani}, E., {Rinaldi}, P., {Caputi}, K.~I., {et~al.} 2024, arXiv e-prints, arXiv:2406.18207

\bibitem[{{Isobe} {et~al.}(2023{\natexlab{a}}){Isobe}, {Ouchi}, {Nakajima}, {Harikane}, {Ono}, {Xu}, {Zhang}, \& {Umeda}}]{Isobe+23}
{Isobe}, Y., {Ouchi}, M., {Nakajima}, K., {et~al.} 2023{\natexlab{a}}, \apj, 956, 139

\bibitem[{{Isobe} {et~al.}(2023{\natexlab{b}}){Isobe}, {Ouchi}, {Tominaga}, {Watanabe}, {Nakajima}, {Umeda}, {Yajima}, {Harikane}, {Fukushima}, {Xu}, {Ono}, \& {Zhang}}]{Isobe-metallicity2023}
{Isobe}, Y., {Ouchi}, M., {Tominaga}, N., {et~al.} 2023{\natexlab{b}}, \apj, 959, 100

\bibitem[{{Izotov} {et~al.}(2019){Izotov}, {Guseva}, {Fricke}, \& {Henkel}}]{Izotov+19}
{Izotov}, Y.~I., {Guseva}, N.~G., {Fricke}, K.~J., \& {Henkel}, C. 2019, \aap, 623, A40

\bibitem[{{Izotov} {et~al.}(2016){Izotov}, {Schaerer}, {Thuan}, {Worseck}, {Guseva}, {Orlitov{\'a}}, \& {Verhamme}}]{Izotov+16}
{Izotov}, Y.~I., {Schaerer}, D., {Thuan}, T.~X., {et~al.} 2016, \mnras, 461, 3683

\bibitem[{{Izotov} {et~al.}(2006){Izotov}, {Stasi{\'n}ska}, {Meynet}, {Guseva}, \& {Thuan}}]{Izotov+06}
{Izotov}, Y.~I., {Stasi{\'n}ska}, G., {Meynet}, G., {Guseva}, N.~G., \& {Thuan}, T.~X. 2006, \aap, 448, 955

\bibitem[{{Izotov} \& {Thuan}(2011)}]{Izotov+11}
{Izotov}, Y.~I. \& {Thuan}, T.~X. 2011, \apj, 734, 82

\bibitem[{{Izotov} {et~al.}(2024){Izotov}, {Thuan}, \& {Guseva}}]{Izotov+24}
{Izotov}, Y.~I., {Thuan}, T.~X., \& {Guseva}, N.~G. 2024, \mnras, 527, 3486

\bibitem[{{Ji} {et~al.}(2024{\natexlab{a}}){Ji}, {Maiolino}, {Ferland}, {D'Eugenio}, {Bhatawdekar}, {Charlot}, {Chevallard}, {Curti}, {Curtis-Lake}, {Hainline}, {Ji}, {Robertson}, {Rodr{\'\i}guez Del Pino}, {Scholtz}, {Tacchella}, {Williams}, \& {Witstok}}]{Ji+2024}
{Ji}, X., {Maiolino}, R., {Ferland}, G., {et~al.} 2024{\natexlab{a}}, arXiv e-prints, arXiv:2405.05772

\bibitem[{{Ji} {et~al.}(2024{\natexlab{b}}){Ji}, {{\"U}bler}, {Maiolino}, {D'Eugenio}, {Arribas}, {Bunker}, {Charlot}, {Perna}, {Rodr{\'\i}guez Del Pino}, {B{\"o}ker}, {Cresci}, {Curti}, {Kumari}, \& {Lamperti}}]{Ji+Ubler2024}
{Ji}, X., {{\"U}bler}, H., {Maiolino}, R., {et~al.} 2024{\natexlab{b}}, arXiv e-prints, arXiv:2404.04148

\bibitem[{{Jin} {et~al.}(2012){Jin}, {Ward}, \& {Done}}]{Jin2012}
{Jin}, C., {Ward}, M., \& {Done}, C. 2012, \mnras, 422, 3268

\bibitem[{{Jones} {et~al.}(2023){Jones}, {{\'A}lvarez-M{\'a}rquez}, {Sloan}, {Kavanagh}, {Argyriou}, {Law}, {Labiano}, {Patapis}, {Mueller}, {Larson}, {Bright}, {Klaassen}, {Fox}, {Gasman}, {Geers}, {Glauser}, {Guillard}, {Nayak}, {Noriega-Crespo}, {Ressler}, {Sargent}, {Temim}, {Vandenbussche}, \& {Garc{\'\i}a Mar{\'\i}n}}]{Jones+23}
{Jones}, O.~C., {{\'A}lvarez-M{\'a}rquez}, J., {Sloan}, G.~C., {et~al.} 2023, \mnras, 523, 2519

\bibitem[{{Juod{\v{z}}balis} {et~al.}(2024){Juod{\v{z}}balis}, {Maiolino}, {Baker}, {Tacchella}, {Scholtz}, {D'Eugenio}, {Schneider}, {Trinca}, {Valiante}, {DeCoursey}, {Curti}, {Carniani}, {Chevallard}, {de Graaff}, {Arribas}, {Bennett}, {Bourne}, {Bunker}, {Charlot}, {Jiang}, {Koudmani}, {Perna}, {Robertson}, {Sijacki}, {{\"U}bler}, {Williams}, {Willott}, \& {Witstok}}]{Juodvzbalis+24}
{Juod{\v{z}}balis}, I., {Maiolino}, R., {Baker}, W.~M., {et~al.} 2024, arXiv e-prints, arXiv:2403.03872

\bibitem[{{Kauffmann} {et~al.}(2003){Kauffmann}, {Heckman}, {Tremonti}, {Brinchmann}, {Charlot}, {White}, {Ridgway}, {Brinkmann}, {Fukugita}, {Hall}, {Ivezi{\'c}}, {Richards}, \& {Schneider}}]{Kauffmann+03}
{Kauffmann}, G., {Heckman}, T.~M., {Tremonti}, C., {et~al.} 2003, \mnras, 346, 1055

\bibitem[{{Kennicutt} \& {Evans}(2012)}]{Kennicutt-Evans+12}
{Kennicutt}, R.~C. \& {Evans}, N.~J. 2012, \araa, 50, 531

\bibitem[{{Kewley} {et~al.}(2001){Kewley}, {Dopita}, {Sutherland}, {Heisler}, \& {Trevena}}]{Kewley+01}
{Kewley}, L.~J., {Dopita}, M.~A., {Sutherland}, R.~S., {Heisler}, C.~A., \& {Trevena}, J. 2001, \apj, 556, 121

\bibitem[{{Kobayashi} \& {Ferrara}(2024)}]{Kobayashi2024}
{Kobayashi}, C. \& {Ferrara}, A. 2024, \apjl, 962, L6

\bibitem[{{Kocevski} {et~al.}(2023){Kocevski}, {Barro}, {McGrath}, {Finkelstein}, {Bagley}, {Ferguson}, {Jogee}, {Yang}, {Dickinson}, {Hathi}, {Backhaus}, {Bell}, {Bisigello}, {Buat}, {Burgarella}, {Casey}, {Cleri}, {Cooper}, {Costantin}, {Croton}, {Daddi}, {Fontana}, {Fujimoto}, {Gardner}, {Gawiser}, {Giavalisco}, {Grazian}, {Grogin}, {Guo}, {Haro}, {Hirschmann}, {Holwerda}, {Huertas-Company}, {Hutchison}, {Iyer}, {Jones}, {Juneau}, {Kartaltepe}, {Kewley}, {Kirkpatrick}, {Koekemoer}, {Kurczynski}, {Le Bail}, {Long}, {Lotz}, {Lucas}, {Papovich}, {Pentericci}, {P{\'e}rez-Gonz{\'a}lez}, {Pirzkal}, {Rafelski}, {Ravindranath}, {Somerville}, {Straughn}, {Tacchella}, {Trump}, {Wilkins}, {Wuyts}, {Yung}, \& {Zavala}}]{Kocevski+23}
{Kocevski}, D.~D., {Barro}, G., {McGrath}, E.~J., {et~al.} 2023, \apjl, 946, L14

\bibitem[{{Kokorev} {et~al.}(2023){Kokorev}, {Fujimoto}, {Labbe}, {Greene}, {Bezanson}, {Dayal}, {Nelson}, {Atek}, {Brammer}, {Caputi}, {Chemerynska}, {Cutler}, {Feldmann}, {Fudamoto}, {Furtak}, {Goulding}, {de Graaff}, {Leja}, {Marchesini}, {Miller}, {Nanayakkara}, {Oesch}, {Pan}, {Price}, {Setton}, {Smit}, {Stefanon}, {Wang}, {Weaver}, {Whitaker}, {Williams}, \& {Zitrin}}]{Kokorev+23}
{Kokorev}, V., {Fujimoto}, S., {Labbe}, I., {et~al.} 2023, \apjl, 957, L7

\bibitem[{{Kokubo} \& {Harikane}(2024)}]{Kokubo-Harikane2024}
{Kokubo}, M. \& {Harikane}, Y. 2024, arXiv e-prints, arXiv:2407.04777

\bibitem[{{Krause} {et~al.}(2016){Krause}, {Charbonnel}, {Bastian}, \& {Diehl}}]{Krause2016}
{Krause}, M. G.~H., {Charbonnel}, C., {Bastian}, N., \& {Diehl}, R. 2016, \aap, 587, A53

\bibitem[{{Labiano} {et~al.}(2021){Labiano}, {Argyriou}, {{\'A}lvarez-M{\'a}rquez}, {Glasse}, {Glauser}, {Patapis}, {Law}, {Brandl}, {Justtanont}, {Lahuis}, {Mart{\'\i}nez-Galarza}, {Mueller}, {Noriega-Crespo}, {Royer}, {Shaughnessy}, \& {Vandenbussche}}]{Labiano+21}
{Labiano}, A., {Argyriou}, I., {{\'A}lvarez-M{\'a}rquez}, J., {et~al.} 2021, \aap, 656, A57

\bibitem[{{Langeroodi} \& {Hjorth}(2023)}]{Langeroodi-mass-size2023}
{Langeroodi}, D. \& {Hjorth}, J. 2023, arXiv e-prints, arXiv:2307.06336

\bibitem[{{Langeroodi} \& {Hjorth}(2024{\natexlab{a}})}]{genesis-metallicity}
{Langeroodi}, D. \& {Hjorth}, J. 2024{\natexlab{a}}, arXiv e-prints, arXiv:2409.07455

\bibitem[{{Langeroodi} \& {Hjorth}(2024{\natexlab{b}})}]{langeroodi-bursty}
{Langeroodi}, D. \& {Hjorth}, J. 2024{\natexlab{b}}, arXiv e-prints, arXiv:2404.13045

\bibitem[{{Langeroodi} {et~al.}(2023){Langeroodi}, {Hjorth}, {Chen}, {Kelly}, {Williams}, {Lin}, {Scarlata}, {Zitrin}, {Broadhurst}, {Diego}, {Huang}, {Filippenko}, {Foley}, {Jha}, {Koekemoer}, {Oguri}, {Perez-Fournon}, {Pierel}, {Poidevin}, \& {Strolger}}]{Langeroodi2023}
{Langeroodi}, D., {Hjorth}, J., {Chen}, W., {et~al.} 2023, \apj, 957, 39

\bibitem[{{Laseter} {et~al.}(2024){Laseter}, {Maseda}, {Curti}, {Maiolino}, {D'Eugenio}, {Cameron}, {Looser}, {Arribas}, {Baker}, {Bhatawdekar}, {Boyett}, {Bunker}, {Carniani}, {Charlot}, {Chevallard}, {Curtis-lake}, {Egami}, {Eisenstein}, {Hainline}, {Hausen}, {Ji}, {Kumari}, {Perna}, {Rawle}, {Rix}, {Robertson}, {Rodr{\'\i}guez Del Pino}, {Sandles}, {Scholtz}, {Smit}, {Tacchella}, {{\"U}bler}, {Williams}, {Willott}, \& {Witstok}}]{2024A&A...681A..70L}
{Laseter}, I.~H., {Maseda}, M.~V., {Curti}, M., {et~al.} 2024, \aap, 681, A70

\bibitem[{{Law} {et~al.}(2023){Law}, {Morrison}, {Argyriou}, {Patapis}, {{\'A}lvarez-M{\'a}rquez}, {Labiano}, \& {Vandenbussche}}]{Law+23}
{Law}, D.~D., {Morrison}, J.~E., {Argyriou}, I., {et~al.} 2023, \aj, 166, 45

\bibitem[{{Le} {et~al.}(2023){Le}, {Xue}, {Lin}, \& {Wang}}]{Le2023}
{Le}, H. A.~N., {Xue}, Y., {Lin}, X., \& {Wang}, Y. 2023, \apj, 945, 59

\bibitem[{{Leroy} {et~al.}(2018){Leroy}, {Bolatto}, {Ostriker}, {Walter}, {Gorski}, {Ginsburg}, {Krieger}, {Levy}, {Meier}, {Mills}, {Ott}, {Rosolowsky}, {Thompson}, {Veilleux}, \& {Zschaechner}}]{Leroy2018}
{Leroy}, A.~K., {Bolatto}, A.~D., {Ostriker}, E.~C., {et~al.} 2018, \apj, 869, 126

\bibitem[{{Li} {et~al.}(2023){Li}, {Dekel}, {Sarkar}, {Aung}, {Giavalisco}, {Mandelker}, \& {Tacchella}}]{Li+Dekel2023}
{Li}, Z., {Dekel}, A., {Sarkar}, K.~C., {et~al.} 2023, arXiv e-prints, arXiv:2311.14662

\bibitem[{{Luridiana} {et~al.}(2015){Luridiana}, {Morisset}, \& {Shaw}}]{Luridiana+15}
{Luridiana}, V., {Morisset}, C., \& {Shaw}, R.~A. 2015, \aap, 573, A42

\bibitem[{{Ma} {et~al.}(2016){Ma}, {Hopkins}, {Faucher-Gigu{\`e}re}, {Zolman}, {Muratov}, {Kere{\v{s}}}, \& {Quataert}}]{2016MNRAS.456.2140M}
{Ma}, X., {Hopkins}, P.~F., {Faucher-Gigu{\`e}re}, C.-A., {et~al.} 2016, \mnras, 456, 2140

\bibitem[{{Maiolino} {et~al.}(2024{\natexlab{a}}){Maiolino}, {Risaliti}, {Signorini}, {Trefoloni}, {Juodzbalis}, {Scholtz}, {Uebler}, {D'Eugenio}, {Carniani}, {Fabian}, {Ji}, {Mazzolari}, {Bertola}, {Brusa}, {Bunker}, {Charlot}, {Comastri}, {Cresci}, {DeCoursey}, {Egami}, {Fiore}, {Gilli}, {Perna}, {Tacchella}, \& {Venturi}}]{Maiolino-Xray2024}
{Maiolino}, R., {Risaliti}, G., {Signorini}, M., {et~al.} 2024{\natexlab{a}}, arXiv e-prints, arXiv:2405.00504

\bibitem[{{Maiolino} {et~al.}(2023){Maiolino}, {Scholtz}, {Curtis-Lake}, {Carniani}, {Baker}, {de Graaff}, {Tacchella}, {{\"U}bler}, {D'Eugenio}, {Witstok}, {Curti}, {Arribas}, {Bunker}, {Charlot}, {Chevallard}, {Eisenstein}, {Egami}, {Ji}, {Jones}, {Lyu}, {Rawle}, {Robertson}, {Rujopakarn}, {Perna}, {Sun}, {Venturi}, {Williams}, \& {Willott}}]{Maiolino+23b}
{Maiolino}, R., {Scholtz}, J., {Curtis-Lake}, E., {et~al.} 2023, arXiv e-prints, arXiv:2308.01230

\bibitem[{{Maiolino} {et~al.}(2024{\natexlab{b}}){Maiolino}, {Scholtz}, {Witstok}, {Carniani}, {D'Eugenio}, {de Graaff}, {{\"U}bler}, {Tacchella}, {Curtis-Lake}, {Arribas}, {Bunker}, {Charlot}, {Chevallard}, {Curti}, {Looser}, {Maseda}, {Rawle}, {Rodr{\'\i}guez del Pino}, {Willott}, {Egami}, {Eisenstein}, {Hainline}, {Robertson}, {Williams}, {Willmer}, {Baker}, {Boyett}, {DeCoursey}, {Fabian}, {Helton}, {Ji}, {Jones}, {Kumari}, {Laporte}, {Nelson}, {Perna}, {Sandles}, {Shivaei}, \& {Sun}}]{Maiolino2024_BH}
{Maiolino}, R., {Scholtz}, J., {Witstok}, J., {et~al.} 2024{\natexlab{b}}, \nat, 627, 59

\bibitem[{{Maiolino} {et~al.}(2024{\natexlab{c}}){Maiolino}, {{\"U}bler}, {Perna}, {Scholtz}, {D'Eugenio}, {Witten}, {Laporte}, {Witstok}, {Carniani}, {Tacchella}, {Baker}, {Arribas}, {Nakajima}, {Eisenstein}, {Bunker}, {Charlot}, {Cresci}, {Curti}, {Curtis-Lake}, {de Graaff}, {Egami}, {Ji}, {Johnson}, {Kumari}, {Looser}, {Maseda}, {Nelson}, {Robertson}, {Rodr{\'\i}guez Del Pino}, {Sandles}, {Simmonds}, {Smit}, {Sun}, {Venturi}, {Williams}, \& {Willmer}}]{Maiolino-nebula2024}
{Maiolino}, R., {{\"U}bler}, H., {Perna}, M., {et~al.} 2024{\natexlab{c}}, \aap, 687, A67

\bibitem[{{Marconcini} {et~al.}(2024){Marconcini}, {D'Eugenio}, {Maiolino}, {Arribas}, {Bunker}, {Carniani}, {Charlot}, {Perna}, {Rodr{\'\i}guez Del Pino}, {{\"U}bler}, {Willott}, {B{\"o}ker}, {Cresci}, {Curti}, {Jones}, {Lamperti}, {Parlanti}, \& {Venturi}}]{Marconcini2024}
{Marconcini}, C., {D'Eugenio}, F., {Maiolino}, R., {et~al.} 2024, \mnras, 533, 2488

\bibitem[{{Marques-Chaves} {et~al.}(2024{\natexlab{a}}){Marques-Chaves}, {Schaerer}, {Kuruvanthodi}, {Korber}, {Prantzos}, {Charbonnel}, {Weibel}, {Izotov}, {Messa}, {Brammer}, {Dessauges-Zavadsky}, \& {Oesch}}]{Marques-Chaves2024}
{Marques-Chaves}, R., {Schaerer}, D., {Kuruvanthodi}, A., {et~al.} 2024{\natexlab{a}}, \aap, 681, A30

\bibitem[{{Marques-Chaves} {et~al.}(2024{\natexlab{b}}){Marques-Chaves}, {Schaerer}, {Vanzella}, {Verhamme}, {Dessauges-Zavadsky}, {Chisholm}, {Leclercq}, {Upadhyaya}, {{\'A}lvarez-M{\'a}rquez}, {Colina}, {Garel}, \& {Messa}}]{Marques-Chaves+24_SFE}
{Marques-Chaves}, R., {Schaerer}, D., {Vanzella}, E., {et~al.} 2024{\natexlab{b}}, \aap, 691, A87

\bibitem[{{Marszewski} {et~al.}(2024){Marszewski}, {Sun}, {Faucher-Gigu{\`e}re}, {Hayward}, \& {Feldmann}}]{2024ApJ...967L..41M}
{Marszewski}, A., {Sun}, G., {Faucher-Gigu{\`e}re}, C.-A., {Hayward}, C.~C., \& {Feldmann}, R. 2024, \apjl, 967, L41

\bibitem[{{Mascia} {et~al.}(2023){Mascia}, {Pentericci}, {Calabr{\`o}}, {Treu}, {Santini}, {Yang}, {Napolitano}, {Roberts-Borsani}, {Bergamini}, {Grillo}, {Rosati}, {Vulcani}, {Castellano}, {Boyett}, {Fontana}, {Glazebrook}, {Henry}, {Mason}, {Merlin}, {Morishita}, {Nanayakkara}, {Paris}, {Roy}, {Williams}, {Wang}, {Brammer}, {Brada{\v{c}}}, {Chen}, {Kelly}, {Koekemoer}, {Trenti}, \& {Windhorst}}]{Mascia+23}
{Mascia}, S., {Pentericci}, L., {Calabr{\`o}}, A., {et~al.} 2023, \aap, 672, A155

\bibitem[{{Matsuoka} {et~al.}(2013){Matsuoka}, {Silverman}, {Schramm}, {Steinhardt}, {Nagao}, {Kartaltepe}, {Sanders}, {Treister}, {Hasinger}, {Akiyama}, {Ohta}, {Ueda}, {Bongiorno}, {Brandt}, {Brusa}, {Capak}, {Civano}, {Comastri}, {Elvis}, {Lilly}, {Mainieri}, {Masters}, {Mignoli}, {Salvato}, {Trump}, {Taniguchi}, {Zamorani}, {Alexander}, \& {Schawinski}}]{Matsuoka2013}
{Matsuoka}, K., {Silverman}, J.~D., {Schramm}, M., {et~al.} 2013, \apj, 771, 64

\bibitem[{{Matthee} {et~al.}(2023){Matthee}, {Mackenzie}, {Simcoe}, {Kashino}, {Lilly}, {Bordoloi}, \& {Eilers}}]{Matthee+23}
{Matthee}, J., {Mackenzie}, R., {Simcoe}, R.~A., {et~al.} 2023, \apj, 950, 67

\bibitem[{{Matthee} {et~al.}(2017){Matthee}, {Sobral}, {Best}, {Khostovan}, {Oteo}, {Bouwens}, \& {R{\"o}ttgering}}]{Matthee+17}
{Matthee}, J., {Sobral}, D., {Best}, P., {et~al.} 2017, \mnras, 465, 3637

\bibitem[{{Mazzolari} {et~al.}(2024){Mazzolari}, {{\"U}bler}, {Maiolino}, {Ji}, {Nakajima}, {Feltre}, {Scholtz}, {D'Eugenio}, {Curti}, {Mignoli}, \& {Marconi}}]{Mazzolari+24}
{Mazzolari}, G., {{\"U}bler}, H., {Maiolino}, R., {et~al.} 2024, arXiv e-prints, arXiv:2404.10811

\bibitem[{{McCrady} {et~al.}(2003){McCrady}, {Gilbert}, \& {Graham}}]{McCrady2003}
{McCrady}, N., {Gilbert}, A.~M., \& {Graham}, J.~R. 2003, \apj, 596, 240

\bibitem[{{McCrady} \& {Graham}(2007)}]{McCrady-Graham2007}
{McCrady}, N. \& {Graham}, J.~R. 2007, \apj, 663, 844

\bibitem[{{Messa} {et~al.}(2019){Messa}, {Adamo}, {{\"O}stlin}, {Melinder}, {Hayes}, {Bridge}, \& {Cannon}}]{Messa2019}
{Messa}, M., {Adamo}, A., {{\"O}stlin}, G., {et~al.} 2019, \mnras, 487, 4238

\bibitem[{{Meyer} {et~al.}(2024){Meyer}, {Oesch}, {Giovinazzo}, {Weibel}, {Brammer}, {Matthee}, {Naidu}, {Bouwens}, {Chisholm}, {Covelo-Paz}, {Fudamoto}, {Maseda}, {Nelson}, {Shivaei}, {Xiao}, {Herard-Demanche}, {Illingworth}, {Kerutt}, {Kramarenko}, {Labbe}, {Leonova}, {Magee}, {Matharu}, {Prieto Lyon}, {Reddy}, {Schaerer}, {Shapley}, {Stefanon}, {Wozniak}, \& {Wuyts}}]{Meyer+24}
{Meyer}, R.~A., {Oesch}, P.~A., {Giovinazzo}, E., {et~al.} 2024, arXiv e-prints, arXiv:2405.05111

\bibitem[{{Mingozzi} {et~al.}(2022){Mingozzi}, {James}, {Arellano-C{\'o}rdova}, {Berg}, {Senchyna}, {Chisholm}, {Brinchmann}, {Aloisi}, {Amor{\'\i}n}, {Charlot}, {Feltre}, {Hayes}, {Heckman}, {Henry}, {Hernandez}, {Kumari}, {Leitherer}, {Llerena}, {Martin}, {Nanayakkara}, {Ravindranath}, {Skillman}, {Sugahara}, {Wofford}, \& {Xu}}]{Mingozzi+22}
{Mingozzi}, M., {James}, B.~L., {Arellano-C{\'o}rdova}, K.~Z., {et~al.} 2022, \apj, 939, 110

\bibitem[{{Morishita} {et~al.}(2023){Morishita}, {Roberts-Borsani}, {Treu}, {Brammer}, {Mason}, {Trenti}, {Vulcani}, {Wang}, {Acebron}, {Bah{\'e}}, {Bergamini}, {Boyett}, {Bradac}, {Calabr{\`o}}, {Castellano}, {Chen}, {De Lucia}, {Filippenko}, {Fontana}, {Glazebrook}, {Grillo}, {Henry}, {Jones}, {Kelly}, {Koekemoer}, {Leethochawalit}, {Lu}, {Marchesini}, {Mascia}, {Mercurio}, {Merlin}, {Metha}, {Nanayakkara}, {Nonino}, {Paris}, {Pentericci}, {Rosati}, {Santini}, {Strait}, {Vanzella}, {Windhorst}, \& {Xie}}]{Morishita+23}
{Morishita}, T., {Roberts-Borsani}, G., {Treu}, T., {et~al.} 2023, \apjl, 947, L24

\bibitem[{{Morishita} {et~al.}(2024{\natexlab{a}}){Morishita}, {Stiavelli}, {Grillo}, {Rosati}, {Schuldt}, {Trenti}, {Bergamini}, {Boyett}, {Chary}, {Leethochawalit}, {Roberts-Borsani}, {Treu}, \& {Vanzella}}]{Morishita-Stiavelli2024}
{Morishita}, T., {Stiavelli}, M., {Grillo}, C., {et~al.} 2024{\natexlab{a}}, \apj, 971, 43

\bibitem[{{Morishita} {et~al.}(2024{\natexlab{b}}){Morishita}, {Stiavelli}, {Grillo}, {Rosati}, {Schuldt}, {Trenti}, {Bergamini}, {Boyett}, {Chary}, {Leethochawalit}, {Roberts-Borsani}, {Treu}, \& {Vanzella}}]{Morishita2024}
{Morishita}, T., {Stiavelli}, M., {Grillo}, C., {et~al.} 2024{\natexlab{b}}, arXiv e-prints, arXiv:2402.14084

\bibitem[{{Nakajima} {et~al.}(2023){Nakajima}, {Ouchi}, {Isobe}, {Harikane}, {Zhang}, {Ono}, {Umeda}, \& {Oguri}}]{Nakajima+23}
{Nakajima}, K., {Ouchi}, M., {Isobe}, Y., {et~al.} 2023, \apjs, 269, 33

\bibitem[{{Nakajima} {et~al.}(2022){Nakajima}, {Ouchi}, {Xu}, {Rauch}, {Harikane}, {Nishigaki}, {Isobe}, {Kusakabe}, {Nagao}, {Ono}, {Onodera}, {Sugahara}, {Kim}, {Komiyama}, {Lee}, \& {Zahedy}}]{Nakajima+22}
{Nakajima}, K., {Ouchi}, M., {Xu}, Y., {et~al.} 2022, \apjs, 262, 3

\bibitem[{{Nakane} {et~al.}(2024){Nakane}, {Ouchi}, {Nakajima}, {Harikane}, {Tominaga}, {Takahashi}, {Yanagisawa}, {Watanabe}, {Nomoto}, {Isobe}, {Kashino}, {Nishigaki}, {Ishigaki}, {Ono}, \& {Takeda}}]{Nakane2024}
{Nakane}, M., {Ouchi}, M., {Nakajima}, K., {et~al.} 2024, arXiv e-prints, arXiv:2407.14470

\bibitem[{{Navarro-Carrera} {et~al.}(2024){Navarro-Carrera}, {Caputi}, {Iani}, {Rinaldi}, {Kokorev}, \& {Kerutt}}]{Navarro-Carrera+24}
{Navarro-Carrera}, R., {Caputi}, K.~I., {Iani}, E., {et~al.} 2024, arXiv e-prints, arXiv:2407.14201

\bibitem[{{Netzer}(2019)}]{Netzer2019}
{Netzer}, H. 2019, \mnras, 488, 5185

\bibitem[{{Oesch} {et~al.}(2014){Oesch}, {Bouwens}, {Illingworth}, {Labb{\'e}}, {Smit}, {Franx}, {van Dokkum}, {Momcheva}, {Ashby}, {Fazio}, {Huang}, {Willner}, {Gonzalez}, {Magee}, {Trenti}, {Brammer}, {Skelton}, \& {Spitler}}]{Oesch+14}
{Oesch}, P.~A., {Bouwens}, R.~J., {Illingworth}, G.~D., {et~al.} 2014, \apj, 786, 108

\bibitem[{{Oesch} {et~al.}(2016){Oesch}, {Brammer}, {van Dokkum}, {Illingworth}, {Bouwens}, {Labb{\'e}}, {Franx}, {Momcheva}, {Ashby}, {Fazio}, {Gonzalez}, {Holden}, {Magee}, {Skelton}, {Smit}, {Spitler}, {Trenti}, \& {Willner}}]{Oesch+16}
{Oesch}, P.~A., {Brammer}, G., {van Dokkum}, P.~G., {et~al.} 2016, \apj, 819, 129

\bibitem[{{{\"O}stlin} {et~al.}(2007){{\"O}stlin}, {Cumming}, \& {Bergvall}}]{Ostlin2007}
{{\"O}stlin}, G., {Cumming}, R.~J., \& {Bergvall}, N. 2007, \aap, 461, 471

\bibitem[{{Oteo} {et~al.}(2016){Oteo}, {Ivison}, {Dunne}, {Smail}, {Swinbank}, {Zhang}, {Lewis}, {Maddox}, {Riechers}, {Serjeant}, {Van der Werf}, {Biggs}, {Bremer}, {Cigan}, {Clements}, {Cooray}, {Dannerbauer}, {Eales}, {Ibar}, {Messias}, {Micha{\l}owski}, {P{\'e}rez-Fournon}, \& {van Kampen}}]{Oteo2016}
{Oteo}, I., {Ivison}, R.~J., {Dunne}, L., {et~al.} 2016, \apj, 827, 34

\bibitem[{{Papovich} {et~al.}(2022){Papovich}, {Simons}, {Estrada-Carpenter}, {Matharu}, {Momcheva}, {Trump}, {Backhaus}, {Brammer}, {Cleri}, {Finkelstein}, {Giavalisco}, {Ji}, {Jung}, {Kewley}, {Nicholls}, {Pirzkal}, {Rafelski}, \& {Weiner}}]{Papovich+22}
{Papovich}, C., {Simons}, R.~C., {Estrada-Carpenter}, V., {et~al.} 2022, \apj, 937, 22

\bibitem[{{Pascale} {et~al.}(2023){Pascale}, {Dai}, {McKee}, \& {Tsang}}]{Pascale2023}
{Pascale}, M., {Dai}, L., {McKee}, C.~F., \& {Tsang}, B. T.~H. 2023, \apj, 957, 77

\bibitem[{{Pennell} {et~al.}(2017){Pennell}, {Runnoe}, \& {Brotherton}}]{Pennell2017}
{Pennell}, A., {Runnoe}, J.~C., \& {Brotherton}, M.~S. 2017, \mnras, 468, 1433

\bibitem[{{Pereira-Santaella} {et~al.}(2024){Pereira-Santaella}, {Garc{\'\i}a-Bernete}, {Gonz{\'a}lez-Alfonso}, {Alonso-Herrero}, {Colina}, {Garc{\'\i}a-Burillo}, {Rigopoulou}, {Arribas}, \& {Perna}}]{Pereira-Santaella2024}
{Pereira-Santaella}, M., {Garc{\'\i}a-Bernete}, I., {Gonz{\'a}lez-Alfonso}, E., {et~al.} 2024, \aap, 685, L13

\bibitem[{{P{\'e}rez-Gonz{\'a}lez} {et~al.}(2024){P{\'e}rez-Gonz{\'a}lez}, {Barro}, {Rieke}, {Lyu}, {Rieke}, {Alberts}, {Williams}, {Hainline}, {Sun}, {Pusk{\'a}s}, {Annunziatella}, {Baker}, {Bunker}, {Egami}, {Ji}, {Johnson}, {Robertson}, {Rodr{\'\i}guez Del Pino}, {Rujopakarn}, {Shivaei}, {Tacchella}, {Willmer}, \& {Willott}}]{PerezGonzalez-LRD2024}
{P{\'e}rez-Gonz{\'a}lez}, P.~G., {Barro}, G., {Rieke}, G.~H., {et~al.} 2024, \apj, 968, 4

\bibitem[{{P{\'e}rez-Gonz{\'a}lez} {et~al.}(2023){P{\'e}rez-Gonz{\'a}lez}, {Costantin}, {Langeroodi}, {Rinaldi}, {Annunziatella}, {Ilbert}, {Colina}, {N{\o}rgaard-Nielsen}, {Greve}, {{\"O}stlin}, {Wright}, {Alonso-Herrero}, {{\'A}lvarez-M{\'a}rquez}, {Caputi}, {Eckart}, {Le F{\`e}vre}, {Labiano}, {Garc{\'\i}a-Mar{\'\i}n}, {Hjorth}, {Kendrew}, {Pye}, {Tikkanen}, {van der Werf}, {Walter}, {Ward}, {Bik}, {Boogaard}, {Bosman}, {G{\'o}mez}, {Gillman}, {Iani}, {Jermann}, {Melinder}, {Meyer}, {Moutard}, {van Dishoek}, {Henning}, {Lagage}, {Guedel}, {Peissker}, {Ray}, {Vandenbussche}, {Garc{\'\i}a-Argum{\'a}nez}, \& {Mar{\'\i}a M{\'e}rida}}]{Perez-Gonzalez+23b}
{P{\'e}rez-Gonz{\'a}lez}, P.~G., {Costantin}, L., {Langeroodi}, D., {et~al.} 2023, \apjl, 951, L1

\bibitem[{{P{\'e}rez-Montero}(2017)}]{Perez-Montero+17}
{P{\'e}rez-Montero}, E. 2017, \pasp, 129, 043001

\bibitem[{{Planck Collaboration} {et~al.}(2020){Planck Collaboration}, {Aghanim}, {Akrami}, {Ashdown}, {Aumont}, {Baccigalupi}, {Ballardini}, {Banday}, {Barreiro}, {Bartolo}, {Basak}, {Battye}, {Benabed}, {Bernard}, {Bersanelli}, {Bielewicz}, {Bock}, {Bond}, {Borrill}, {Bouchet}, {Boulanger}, {Bucher}, {Burigana}, {Butler}, {Calabrese}, {Cardoso}, {Carron}, {Challinor}, {Chiang}, {Chluba}, {Colombo}, {Combet}, {Contreras}, {Crill}, {Cuttaia}, {de Bernardis}, {de Zotti}, {Delabrouille}, {Delouis}, {Di Valentino}, {Diego}, {Dor{\'e}}, {Douspis}, {Ducout}, {Dupac}, {Dusini}, {Efstathiou}, {Elsner}, {En{\ss}lin}, {Eriksen}, {Fantaye}, {Farhang}, {Fergusson}, {Fernandez-Cobos}, {Finelli}, {Forastieri}, {Frailis}, {Fraisse}, {Franceschi}, {Frolov}, {Galeotta}, {Galli}, {Ganga}, {G{\'e}nova-Santos}, {Gerbino}, {Ghosh}, {Gonz{\'a}lez-Nuevo}, {G{\'o}rski}, {Gratton}, {Gruppuso}, {Gudmundsson}, {Hamann}, {Handley}, {Hansen}, {Herranz}, {Hildebrandt}, {Hivon}, {Huang}, {Jaffe}, {Jones}, {Karakci}, {Keih{\"a}nen},
  {Keskitalo}, {Kiiveri}, {Kim}, {Kisner}, {Knox}, {Krachmalnicoff}, {Kunz}, {Kurki-Suonio}, {Lagache}, {Lamarre}, {Lasenby}, {Lattanzi}, {Lawrence}, {Le Jeune}, {Lemos}, {Lesgourgues}, {Levrier}, {Lewis}, {Liguori}, {Lilje}, {Lilley}, {Lindholm}, {L{\'o}pez-Caniego}, {Lubin}, {Ma}, {Mac{\'\i}as-P{\'e}rez}, {Maggio}, {Maino}, {Mandolesi}, {Mangilli}, {Marcos-Caballero}, {Maris}, {Martin}, {Martinelli}, {Mart{\'\i}nez-Gonz{\'a}lez}, {Matarrese}, {Mauri}, {McEwen}, {Meinhold}, {Melchiorri}, {Mennella}, {Migliaccio}, {Millea}, {Mitra}, {Miville-Desch{\^e}nes}, {Molinari}, {Montier}, {Morgante}, {Moss}, {Natoli}, {N{\o}rgaard-Nielsen}, {Pagano}, {Paoletti}, {Partridge}, {Patanchon}, {Peiris}, {Perrotta}, {Pettorino}, {Piacentini}, {Polastri}, {Polenta}, {Puget}, {Rachen}, {Reinecke}, {Remazeilles}, {Renzi}, {Rocha}, {Rosset}, {Roudier}, {Rubi{\~n}o-Mart{\'\i}n}, {Ruiz-Granados}, {Salvati}, {Sandri}, {Savelainen}, {Scott}, {Shellard}, {Sirignano}, {Sirri}, {Spencer}, {Sunyaev}, {Suur-Uski}, {Tauber}, {Tavagnacco},
  {Tenti}, {Toffolatti}, {Tomasi}, {Trombetti}, {Valenziano}, {Valiviita}, {Van Tent}, {Vibert}, {Vielva}, {Villa}, {Vittorio}, {Wandelt}, {Wehus}, {White}, {White}, {Zacchei}, \& {Zonca}}]{PlanckCollaboration18VI}
{Planck Collaboration}, {Aghanim}, N., {Akrami}, Y., {et~al.} 2020, \aap, 641, A6

\bibitem[{{Reddy} {et~al.}(2018){Reddy}, {Shapley}, {Sanders}, {Kriek}, {Coil}, {Shivaei}, {Freeman}, {Mobasher}, {Siana}, {Azadi}, {Fetherolf}, {Fornasini}, {Leung}, {Price}, {Zick}, \& {Barro}}]{Reddy+18}
{Reddy}, N.~A., {Shapley}, A.~E., {Sanders}, R.~L., {et~al.} 2018, \apj, 869, 92

\bibitem[{{Reddy} {et~al.}(2022){Reddy}, {Topping}, {Shapley}, {Steidel}, {Sanders}, {Du}, {Coil}, {Mobasher}, {Price}, \& {Shivaei}}]{Reddy+22}
{Reddy}, N.~A., {Topping}, M.~W., {Shapley}, A.~E., {et~al.} 2022, \apj, 926, 31

\bibitem[{{Rieke} {et~al.}(2015){Rieke}, {Wright}, {B{\"o}ker}, {Bouwman}, {Colina}, {Glasse}, {Gordon}, {Greene}, {G{\"u}del}, {Henning}, {Justtanont}, {Lagage}, {Meixner}, {N{\o}rgaard-Nielsen}, {Ray}, {Ressler}, {van Dishoeck}, \& {Waelkens}}]{Rieke+15}
{Rieke}, G.~H., {Wright}, G.~S., {B{\"o}ker}, T., {et~al.} 2015, \pasp, 127, 584

\bibitem[{{Rigby} {et~al.}(2023){Rigby}, {Perrin}, {McElwain}, {Kimble}, {Friedman}, {Lallo}, {Doyon}, {Feinberg}, {Ferruit}, {Glasse}, {Rieke}, {Rieke}, {Wright}, {Willott}, {Colon}, {Milam}, {Neff}, {Stark}, {Valenti}, {Abell}, {Abney}, {Abul-Huda}, {Acton}, {Adams}, {Adler}, {Aguilar}, {Ahmed}, {Albert}, {Alberts}, {Aldridge}, {Allen}, {Altenburg}, {{\'A}lvarez-M{\'a}rquez}, {Alves de Oliveira}, {Andersen}, {Anderson}, {Anderson}, {Argyriou}, {Armstrong}, {Arribas}, {Artigau}, {Arvai}, {Atkinson}, {Bacon}, {Bair}, {Banks}, {Barrientes}, {Barringer}, {Bartosik}, {Bast}, {Baudoz}, {Beatty}, {Bechtold}, {Beck}, {Bergeron}, {Bergkoetter}, {Bhatawdekar}, {Birkmann}, {Blazek}, {Blome}, {Boccaletti}, {B{\"o}ker}, {Boia}, {Bonaventura}, {Bond}, {Bosley}, {Boucarut}, {Bourque}, {Bouwman}, {Bower}, {Bowers}, {Boyer}, {Bradley}, {Brady}, {Braun}, {Breda}, {Bresnahan}, {Bright}, {Britt}, {Bromenschenkel}, {Brooks}, {Brooks}, {Brown}, {Brown}, {Brown}, {Bunker}, {Burger}, {Bushouse}, {Cale}, {Cameron}, {Cameron},
  {Canipe}, {Caplinger}, {Caputo}, {Cara}, {Carey}, {Carniani}, {Carrasquilla}, {Carruthers}, {Case}, {Catherine}, {Chance}, {Chapman}, {Charlot}, {Charlow}, {Chayer}, {Chen}, {Cherinka}, {Chichester}, {Chilton}, {Chonis}, {Clampin}, {Clark}, {Clark}, {Coe}, {Coleman}, {Comber}, {Comeau}, {Connolly}, {Cooper}, {Cooper}, {Coppock}, {Correnti}, {Cossou}, {Coulais}, {Coyle}, {Cracraft}, {Curti}, {Cuturic}, {Davis}, {Davis}, {Dean}, {DeLisa}, {deMeester}, {Dencheva}, {Dencheva}, {DePasquale}, {Deschenes}, {Hunor Detre}, {Diaz}, {Dicken}, {DiFelice}, {Dillman}, {Dixon}, {Doggett}, {Donaldson}, {Douglas}, {DuPrie}, {Dupuis}, {Durning}, {Easmin}, {Eck}, {Edeani}, {Egami}, {Ehrenwinkler}, {Eisenhamer}, {Eisenhower}, {Elie}, {Elliott}, {Elliott}, {Ellis}, {Engesser}, {Espinoza}, {Etienne}, {Etxaluze}, {Falini}, {Feeney}, {Ferry}, {Filippazzo}, {Fincham}, {Fix}, {Flagey}, {Florian}, {Flynn}, {Fontanella}, {Ford}, {Forshay}, {Fox}, {Franz}, {Fu}, {Fullerton}, {Galkin}, {Galyer}, {Garc{\'\i}a Mar{\'\i}n}, {Gardner},
  {Gardner}, {Garland}, {Garrett}, {Gasman}, {Gaspar}, {Gaudreau}, {Gauthier}, {Geers}, {Geithner}, {Gennaro}, {Giardino}, {Girard}, {Giuliano}, {Glassmire}, {Glauser}, {Glazer}, {Godfrey}, {Golimowski}, {Gollnitz}, {Gong}, {Gonzaga}, {Gordon}, {Gordon}, {Goudfrooij}, {Greene}, {Greenhouse}, {Grimaldi}, {Groebner}, {Grundy}, {Guillard}, {Gutman}, {Ha}, {Haderlein}, {Hagedorn}, {Hainline}, {Haley}, {Hami}, {Hamilton}, {Hammel}, {Hansen}, {Harkins}, {Harr}, {Hart}, {Hart}, {Hartig}, {Hashimoto}, {Haskins}, {Hathaway}, {Havey}, {Hayden}, {Hecht}, {Heller-Boyer}, {Henriques}, {Henry}, {Hermann}, {Hernandez}, {Hesman}, {Hicks}, {Hilbert}, {Hines}, {Hoffman}, {Holfeltz}, {Holler}, {Hoppa}, {Hott}, {Howard}, {Howard}, {Hunter}, {Hunter}, {Hurst}, {Husemann}, {Hustak}, {Ilinca Ignat}, {Illingworth}, {Irish}, {Jackson}, {Jahromi}, {Jakobsen}, {James}, {James}, {Januszewski}, {Jenkins}, {Jirdeh}, {Johnson}, {Johnson}, {Jones}, {Jones}, {Jones}, {Jones}, {Jordan}, {Jordan}, {Jurczyk}, {Jurling}, {Kaleida}, {Kalmanson},
  {Kammerer}, {Kang}, {Kao}, {Karakla}, {Kavanagh}, {Kelly}, {Kendrew}, {Kennedy}, {Kenny}, {Keski-kuha}, {Keyes}, {Kidwell}, {Kinzel}, {Kirk}, {Kirkpatrick}, {Kirshenblat}, {Klaassen}, {Knapp}, {Knight}, {Knollenberg}, {Koehler}, {Koekemoer}, {Kovacs}, {Kulp}, {Kumari}, {Kyprianou}, {La Massa}, {Labador}, {Labiano}, {Lagage}, {Lajoie}, {Lallo}, {Lam}, {Lamb}, {Lambros}, {Lampenfield}, {Langston}, {Larson}, {Law}, {Lawrence}, {Lee}, {Leisenring}, {Lepo}, {Leveille}, {Levenson}, {Levine}, {Levy}, {Lewis}, {Lewis}, {Libralato}, {Lightsey}, {Link}, {Liu}, {Lo}, {Lockwood}, {Logue}, {Long}, {Long}, {Loomis}, {Lopez-Caniego}, {Lorenzo Alvarez}, {Love-Pruitt}, {Lucy}, {Luetzgendorf}, {Maghami}, {Maiolino}, {Major}, {Malla}, {Malumuth}, {Manjavacas}, {Mannfolk}, {Marrione}, {Marston}, {Martel}, {Maschmann}, {Masci}, {Masciarelli}, {Maszkiewicz}, {Mather}, {McKenzie}, {McLean}, {McMaster}, {Melbourne}, {Mel{\'e}ndez}, {Menzel}, {Merz}, {Meyett}, {Meza}, {Miskey}, {Misselt}, {Moller}, {Morrison}, {Morse}, {Moseley},
  {Mosier}, {Mountain}, {Mueckay}, {Mueller}, {Mullally}, {Murphy}, {Murray}, {Murray}, {Mustelier}, {Muzerolle}, {Mycroft}, {Myers}, {Myrick}, {Nanavati}, {Nance}, {Nayak}, {Naylor}, {Nelan}, {Nickson}, {Nielson}, {Nieto-Santisteban}, {Nikolov}, {Noriega-Crespo}, {O'Shaughnessy}, {O'Sullivan}, {Ochs}, {Ogle}, {Oleszczuk}, {Olmsted}, {Osborne}, {Ottens}, {Owens}, {Pacifici}, {Pagan}, {Page}, {Park}, {Parrish}, {Patapis}, {Paul}, {Pauly}, {Pavlovsky}, {Pedder}, {Peek}, {Pena-Guerrero}, {Penanen}, {Perez}, {Perna}, {Perriello}, {Phillips}, {Pietraszkiewicz}, {Pinaud}, {Pirzkal}, {Pitman}, {Piwowar}, {Platais}, {Player}, {Plesha}, {Pollizi}, {Polster}, {Pontoppidan}, {Porterfield}, {Proffitt}, {Pueyo}, {Pulliam}, {Quirt}, {Quispe Neira}, {Ramos Alarcon}, {Ramsay}, {Rapp}, {Rapp}, {Rauscher}, {Ravindranath}, {Rawle}, {Regan}, {Reichard}, {Reis}, {Ressler}, {Rest}, {Reynolds}, {Rhue}, {Richon}, {Rickman}, {Ridgaway}, {Ritchie}, {Rix}, {Robberto}, {Robinson}, {Robinson}, {Robinson}, {Rock}, {Rodriguez}, {Rodriguez
  Del Pino}, {Roellig}, {Rohrbach}, {Roman}, {Romelfanger}, {Rose}, {Roteliuk}, {Roth}, {Rothwell}, {Rowlands}, {Roy}, {Royer}, {Royle}, {Rui}, {Rumler}, {Runnels}, {Russ}, {Rustamkulov}, {Ryden}, {Ryer}, {Sabata}, {Sabatke}, {Sabbi}, {Samuelson}, {Sapp}, {Sappington}, {Sargent}, {Sauer}, {Scheithauer}, {Schlawin}, {Schlitz}, {Schmitz}, {Schneider}, {Schreiber}, {Schulze}, {Schwab}, {Scott}, {Sembach}, {Shanahan}, {Shaughnessy}, {Shaw}, {Shawger}, {Shay}, {Sheehan}, {Shen}, {Sherman}, {Shiao}, {Shih}, {Shivaei}, {Sienkiewicz}, {Sing}, {Sirianni}, {Sivaramakrishnan}, {Skipper}, {Sloan}, {Slocum}, {Slowinski}, {Smith}, {Smith}, {Smith}, {Smith}, {Snyder}, {Soh}, {Sohn}, {Soto}, {Spencer}, {Stallcup}, {Stansberry}, {Starr}, {Starr}, {Stewart}, {Stiavelli}, {Straughn}, {Strickland}, {Stys}, {Summers}, {Sun}, {Sunnquist}, {Swade}, {Swam}, {Swaters}, {Swoish}, {Taylor}, {Taylor}, {Te Plate}, {Tea}, {Teague}, {Telfer}, {Temim}, {Thatte}, {Thompson}, {Thompson}, {Thomson}, {Tikkanen}, {Tippet}, {Todd}, {Toolan},
  {Tran}, {Trejo}, {Truong}, {Tsukamoto}, {Tustain}, {Tyra}, {Ubeda}, {Underwood}, {Uzzo}, {Van Campen}, {Vandal}, {Vandenbussche}, {Vila}, {Volk}, {Wahlgren}, {Waldman}, {Walker}, {Wander}, {Warfield}, {Warner}, {Wasiak}, {Watkins}, {Weaver}, {Weilert}, {Weiser}, {Weiss}, {Weissman}, {Welty}, {West}, {Wheate}, {Wheatley}, {Wheeler}, {White}, {Whiteaker}, {Whitehouse}, {Whiteleather}, {Whitman}, {Williams}, {Willmer}, {Willoughby}, {Wilson}, {Wirth}, {Wislowski}, {Wolf}, {Wolfe}, {Wolff}, {Workman}, {Wright}, {Wu}, {Wu}, {Wymer}, {Yates}, {Yeager}, {Yeates}, {Yerger}, {Yoon}, {Young}, {Yu}, {Zak}, {Zeidler}, {Zhou}, {Zielinski}, {Zincke}, \& {Zonak}}]{Rigby2023}
{Rigby}, J., {Perrin}, M., {McElwain}, M., {et~al.} 2023, \pasp, 135, 048001

\bibitem[{{Rinaldi} {et~al.}(2023){Rinaldi}, {Caputi}, {Costantin}, {Gillman}, {Iani}, {P{\'e}rez-Gonz{\'a}lez}, {{\"O}stlin}, {Colina}, {Greve}, {Noorgard-Nielsen}, {Wright}, {Alonso-Herrero}, {{\'A}lvarez-M{\'a}rquez}, {Eckart}, {Garc{\'\i}a-Mar{\'\i}n}, {Hjorth}, {Ilbert}, {Kendrew}, {Labiano}, {Le F{\`e}vre}, {Pye}, {Tikkanen}, {Walter}, {van der Werf}, {Ward}, {Annunziatella}, {Azzollini}, {Bik}, {Boogaard}, {Bosman}, {Crespo G{\'o}mez}, {Jermann}, {Langeroodi}, {Melinder}, {Meyer}, {Moutard}, {Peissker}, {Topinka}, {van Dishoeck}, {G{\"u}del}, {Henning}, {Lagage}, {Ray}, {Vandenbussche}, {Waelkens}, {Navarro-Carrera}, \& {Kokorev}}]{Rinaldi+23}
{Rinaldi}, P., {Caputi}, K.~I., {Costantin}, L., {et~al.} 2023, \apj, 952, 143

\bibitem[{{Rinaldi} {et~al.}(2024){Rinaldi}, {Caputi}, {Iani}, {Costantin}, {Gillman}, {Perez Gonzalez}, {{\"O}stlin}, {Colina}, {Greve}, {N{\o}rgard-Nielsen}, {Wright}, {{\'A}lvarez-M{\'a}rquez}, {Eckart}, {Garc{\'\i}a-Mar{\'\i}n}, {Hjorth}, {Ilbert}, {Kendrew}, {Labiano}, {Le F{\`e}vre}, {Pye}, {Tikkanen}, {Walter}, {van der Werf}, {Ward}, {Annunziatella}, {Azzollini}, {Bik}, {Boogaard}, {Bosman}, {Crespo G{\'o}mez}, {Jermann}, {Langeroodi}, {Melinder}, {Meyer}, {Moutard}, {Peissker}, {van Dishoeck}, {G{\"u}del}, {Henning}, {Lagage}, {Ray}, {Vandenbussche}, {Waelkens}, \& {Dayal}}]{Rinaldi+2024}
{Rinaldi}, P., {Caputi}, K.~I., {Iani}, E., {et~al.} 2024, \apj, 969, 12

\bibitem[{{Robertson} {et~al.}(2023){Robertson}, {Tacchella}, {Johnson}, {Hainline}, {Whitler}, {Eisenstein}, {Endsley}, {Rieke}, {Stark}, {Alberts}, {Dressler}, {Egami}, {Hausen}, {Rieke}, {Shivaei}, {Williams}, {Willmer}, {Arribas}, {Bonaventura}, {Bunker}, {Cameron}, {Carniani}, {Charlot}, {Chevallard}, {Curti}, {Curtis-Lake}, {D'Eugenio}, {Jakobsen}, {Looser}, {L{\"u}tzgendorf}, {Maiolino}, {Maseda}, {Rawle}, {Rix}, {Smit}, {{\"U}bler}, {Willott}, {Witstok}, {Baum}, {Bhatawdekar}, {Boyett}, {Chen}, {de Graaff}, {Florian}, {Helton}, {Hviding}, {Ji}, {Kumari}, {Lyu}, {Nelson}, {Sandles}, {Saxena}, {Suess}, {Sun}, {Topping}, \& {Wallace}}]{Robertson+23}
{Robertson}, B.~E., {Tacchella}, S., {Johnson}, B.~D., {et~al.} 2023, Nature Astronomy, 7, 611

\bibitem[{{S{\'a}nchez-Garc{\'\i}a} {et~al.}(2022){S{\'a}nchez-Garc{\'\i}a}, {Pereira-Santaella}, {Garc{\'\i}a-Burillo}, {Colina}, {Alonso-Herrero}, {Villar-Mart{\'\i}n}, {Saito}, {D{\'\i}az-Santos}, {Piqueras L{\'o}pez}, {Arribas}, {Bellocchi}, {Cazzoli}, \& {Labiano}}]{Sanchez-Garcia2022}
{S{\'a}nchez-Garc{\'\i}a}, M., {Pereira-Santaella}, M., {Garc{\'\i}a-Burillo}, S., {et~al.} 2022, \aap, 659, A102

\bibitem[{{Sanders} {et~al.}(2024){Sanders}, {Shapley}, {Topping}, {Reddy}, \& {Brammer}}]{Sanders+24}
{Sanders}, R.~L., {Shapley}, A.~E., {Topping}, M.~W., {Reddy}, N.~A., \& {Brammer}, G.~B. 2024, \apj, 962, 24

\bibitem[{{Schaerer} {et~al.}(2024){Schaerer}, {Marques-Chaves}, {Xiao}, \& {Korber}}]{Schaerer+24}
{Schaerer}, D., {Marques-Chaves}, R., {Xiao}, M., \& {Korber}, D. 2024, \aap, 687, L11

\bibitem[{Schaerer {et~al.}(2024)Schaerer, Marques-Chaves, Xiao, \& Korber}]{Schaerer-Rui2024}
Schaerer, D., Marques-Chaves, R., Xiao, M., \& Korber, D. 2024, Discovery of a new N-emitter in the epoch of reionization

\bibitem[{{Senchyna} {et~al.}(2024){Senchyna}, {Plat}, {Stark}, {Rudie}, {Berg}, {Charlot}, {James}, \& {Mingozzi}}]{Senchyna2024}
{Senchyna}, P., {Plat}, A., {Stark}, D.~P., {et~al.} 2024, \apj, 966, 92

\bibitem[{{Shen} {et~al.}(2011){Shen}, {Richards}, {Strauss}, {Hall}, {Schneider}, {Snedden}, {Bizyaev}, {Brewington}, {Malanushenko}, {Malanushenko}, {Oravetz}, {Pan}, \& {Simmons}}]{Shen2011}
{Shen}, Y., {Richards}, G.~T., {Strauss}, M.~A., {et~al.} 2011, \apjs, 194, 45

\bibitem[{{Simmonds} {et~al.}(2024){Simmonds}, {Tacchella}, {Hainline}, {Johnson}, {McClymont}, {Robertson}, {Saxena}, {Sun}, {Witten}, {Baker}, {Bhatawdekar}, {Boyett}, {Bunker}, {Charlot}, {Curtis-Lake}, {Egami}, {Eisenstein}, {Hausen}, {Maiolino}, {Maseda}, {Scholtz}, {Williams}, {Willott}, \& {Witstok}}]{Simmonds2024}
{Simmonds}, C., {Tacchella}, S., {Hainline}, K., {et~al.} 2024, \mnras, 527, 6139

\bibitem[{{Smit} {et~al.}(2016){Smit}, {Bouwens}, {Labb{\'e}}, {Franx}, {Wilkins}, \& {Oesch}}]{2016ApJ...833..254S}
{Smit}, R., {Bouwens}, R.~J., {Labb{\'e}}, I., {et~al.} 2016, \apj, 833, 254

\bibitem[{{Stiavelli} {et~al.}(2023){Stiavelli}, {Morishita}, {Chiaberge}, {Grillo}, {Leethochawalit}, {Rosati}, {Schuldt}, {Trenti}, \& {Treu}}]{Stiavelli+23}
{Stiavelli}, M., {Morishita}, T., {Chiaberge}, M., {et~al.} 2023, \apjl, 957, L18

\bibitem[{{Tacchella} {et~al.}(2023){Tacchella}, {Eisenstein}, {Hainline}, {Johnson}, {Baker}, {Helton}, {Robertson}, {Suess}, {Chen}, {Nelson}, {Pusk{\'a}s}, {Sun}, {Alberts}, {Egami}, {Hausen}, {Rieke}, {Rieke}, {Shivaei}, {Williams}, {Willmer}, {Bunker}, {Cameron}, {Carniani}, {Charlot}, {Curti}, {Curtis-Lake}, {Looser}, {Maiolino}, {Maseda}, {Rawle}, {Rix}, {Smit}, {{\"U}bler}, {Willott}, {Witstok}, {Baum}, {Bhatawdekar}, {Boyett}, {Danhaive}, {de Graaff}, {Endsley}, {Ji}, {Lyu}, {Sandles}, {Saxena}, {Scholtz}, {Topping}, \& {Whitler}}]{Tacchella+23}
{Tacchella}, S., {Eisenstein}, D.~J., {Hainline}, K., {et~al.} 2023, \apj, 952, 74

\bibitem[{{Tang} {et~al.}(2023){Tang}, {Stark}, {Chen}, {Mason}, {Topping}, {Endsley}, {Senchyna}, {Plat}, {Lu}, {Whitler}, {Robertson}, \& {Charlot}}]{Tang+23}
{Tang}, M., {Stark}, D.~P., {Chen}, Z., {et~al.} 2023, \mnras, 526, 1657

\bibitem[{{Theios} {et~al.}(2019){Theios}, {Steidel}, {Strom}, {Rudie}, {Trainor}, \& {Reddy}}]{Theios+19}
{Theios}, R.~L., {Steidel}, C.~C., {Strom}, A.~L., {et~al.} 2019, \apj, 871, 128

\bibitem[{{Thompson} {et~al.}(2005){Thompson}, {Quataert}, \& {Murray}}]{Thompson+05}
{Thompson}, T.~A., {Quataert}, E., \& {Murray}, N. 2005, \apj, 630, 167

\bibitem[{{Topping} {et~al.}(2024){Topping}, {Stark}, {Senchyna}, {Plat}, {Zitrin}, {Endsley}, {Charlot}, {Furtak}, {Maseda}, {Smit}, {Mainali}, {Chevallard}, {Molyneux}, \& {Rigby}}]{Topping+24}
{Topping}, M.~W., {Stark}, D.~P., {Senchyna}, P., {et~al.} 2024, \mnras, 529, 3301

\bibitem[{{{\"U}bler} {et~al.}(2024){{\"U}bler}, {Maiolino}, {P{\'e}rez-Gonz{\'a}lez}, {D'Eugenio}, {Perna}, {Curti}, {Arribas}, {Bunker}, {Carniani}, {Charlot}, {Rodr{\'\i}guez Del Pino}, {Baker}, {B{\"o}ker}, {Cresci}, {Dunlop}, {Grogin}, {Jones}, {Kumari}, {Lamperti}, {Laporte}, {Marshall}, {Mazzolari}, {Parlanti}, {Rawle}, {Scholtz}, {Venturi}, \& {Witstok}}]{Ubler+23b}
{{\"U}bler}, H., {Maiolino}, R., {P{\'e}rez-Gonz{\'a}lez}, P.~G., {et~al.} 2024, \mnras, 531, 355

\bibitem[{{Ucci} {et~al.}(2023){Ucci}, {Dayal}, {Hutter}, {Kobayashi}, {Gottl{\"o}ber}, {Yepes}, {Hunt}, {Legrand}, \& {Tortora}}]{2023MNRAS.518.3557U}
{Ucci}, G., {Dayal}, P., {Hutter}, A., {et~al.} 2023, \mnras, 518, 3557

\bibitem[{{Vanzella} {et~al.}(2022){Vanzella}, {Castellano}, {Bergamini}, {Meneghetti}, {Zanella}, {Calura}, {Caminha}, {Rosati}, {Cupani}, {Me{\v{s}}tri{\'c}}, {Brammer}, {Tozzi}, {Mercurio}, {Grillo}, {Sani}, {Cristiani}, {Nonino}, {Merlin}, \& {Pignataro}}]{Vanzella-Sunburst2022}
{Vanzella}, E., {Castellano}, M., {Bergamini}, P., {et~al.} 2022, \aap, 659, A2

\bibitem[{{Vanzella} {et~al.}(2023{\natexlab{a}}){Vanzella}, {Claeyssens}, {Welch}, {Adamo}, {Coe}, {Diego}, {Mahler}, {Khullar}, {Kokorev}, {Oguri}, {Ravindranath}, {Furtak}, {Hsiao}, {Abdurro'uf}, {Mandelker}, {Brammer}, {Bradley}, {Brada{\v{c}}}, {Conselice}, {Dayal}, {Nonino}, {Andrade-Santos}, {Windhorst}, {Pirzkal}, {Sharon}, {de Mink}, {Fujimoto}, {Zitrin}, {Eldridge}, \& {Norman}}]{Vanzella+23}
{Vanzella}, E., {Claeyssens}, A., {Welch}, B., {et~al.} 2023{\natexlab{a}}, \apj, 945, 53

\bibitem[{{Vanzella} {et~al.}(2023{\natexlab{b}}){Vanzella}, {Loiacono}, {Bergamini}, {Me{\v{s}}tri{\'c}}, {Castellano}, {Rosati}, {Meneghetti}, {Grillo}, {Calura}, {Mignoli}, {Brada{\v{c}}}, {Adamo}, {Rihtar{\v{s}}i{\v{c}}}, {Dickinson}, {Gronke}, {Zanella}, {Annibali}, {Willott}, {Messa}, {Sani}, {Acebron}, {Bolamperti}, {Comastri}, {Gilli}, {Caputi}, {Ricotti}, {Gruppioni}, {Ravindranath}, {Mercurio}, {Strait}, {Martis}, {Pascale}, {Caminha}, {Annunziatella}, \& {Nonino}}]{Vanzella2023}
{Vanzella}, E., {Loiacono}, F., {Bergamini}, P., {et~al.} 2023{\natexlab{b}}, \aap, 678, A173

\bibitem[{{Vestergaard} \& {Osmer}(2009)}]{Vestergaard2009}
{Vestergaard}, M. \& {Osmer}, P.~S. 2009, \apj, 699, 800

\bibitem[{{Wells} {et~al.}(2015){Wells}, {Pel}, {Glasse}, {Wright}, {Aitink-Kroes}, {Azzollini}, {Beard}, {Brandl}, {Gallie}, {Geers}, {Glauser}, {Hastings}, {Henning}, {Jager}, {Justtanont}, {Kruizinga}, {Lahuis}, {Lee}, {Martinez-Delgado}, {Mart{\'\i}nez-Galarza}, {Meijers}, {Morrison}, {M{\"u}ller}, {Nakos}, {O'Sullivan}, {Oudenhuysen}, {Parr-Burman}, {Pauwels}, {Rohloff}, {Schmalzl}, {Sykes}, {Thelen}, {van Dishoeck}, {Vandenbussche}, {Venema}, {Visser}, {Waters}, \& {Wright}}]{Wells+15}
{Wells}, M., {Pel}, J.~W., {Glasse}, A., {et~al.} 2015, \pasp, 127, 646

\bibitem[{{Williams} {et~al.}(2023){Williams}, {Kelly}, {Chen}, {Brammer}, {Zitrin}, {Treu}, {Scarlata}, {Koekemoer}, {Oguri}, {Lin}, {Diego}, {Nonino}, {Hjorth}, {Langeroodi}, {Broadhurst}, {Rogers}, {Perez-Fournon}, {Foley}, {Jha}, {Filippenko}, {Strolger}, {Pierel}, {Poidevin}, \& {Yang}}]{Williams+23}
{Williams}, H., {Kelly}, P.~L., {Chen}, W., {et~al.} 2023, Science, 380, 416

\bibitem[{{Wright} {et~al.}(2023){Wright}, {Rieke}, {Glasse}, {Ressler}, {Garc{\'\i}a Mar{\'\i}n}, {Aguilar}, {Alberts}, {{\'A}lvarez-M{\'a}rquez}, {Argyriou}, {Banks}, {Baudoz}, {Boccaletti}, {Bouchet}, {Bouwman}, {Brandl}, {Breda}, {Bright}, {Cale}, {Colina}, {Cossou}, {Coulais}, {Cracraft}, {De Meester}, {Dicken}, {Engesser}, {Etxaluze}, {Fox}, {Friedman}, {Fu}, {Gasman}, {G{\'a}sp{\'a}r}, {Gastaud}, {Geers}, {Glauser}, {Gordon}, {Greene}, {Greve}, {Grundy}, {G{\"u}del}, {Guillard}, {Haderlein}, {Hashimoto}, {Henning}, {Hines}, {Holler}, {Detre}, {Jahromi}, {James}, {Jones}, {Justtanont}, {Kavanagh}, {Kendrew}, {Klaassen}, {Krause}, {Labiano}, {Lagage}, {Lambros}, {Larson}, {Law}, {Lee}, {Libralato}, {Lorenzo Alverez}, {Meixner}, {Morrison}, {Mueller}, {Murray}, {Mycroft}, {Myers}, {Nayak}, {Naylor}, {Nickson}, {Noriega-Crespo}, {{\"O}stlin}, {O'Sullivan}, {Ottens}, {Patapis}, {Penanen}, {Pietraszkiewicz}, {Ray}, {Regan}, {Roteliuk}, {Royer}, {Samara-Ratna}, {Samuelson}, {Sargent}, {Scheithauer},
  {Schneider}, {Schreiber}, {Shaughnessy}, {Sheehan}, {Shivaei}, {Sloan}, {Tamas}, {Teague}, {Temim}, {Tikkanen}, {Tustain}, {van Dishoeck}, {Vandenbussche}, {Weilert}, {Whitehouse}, \& {Wolff}}]{Wright+23}
{Wright}, G.~S., {Rieke}, G.~H., {Glasse}, A., {et~al.} 2023, \pasp, 135, 048003

\bibitem[{{Wright} {et~al.}(2015){Wright}, {Wright}, {Goodson}, {Rieke}, {Aitink-Kroes}, {Amiaux}, {Aricha-Yanguas}, {Azzollini}, {Banks}, {Barrado-Navascues}, {Belenguer-Davila}, {Bloemmart}, {Bouchet}, {Brandl}, {Colina}, {Detre}, {Diaz-Catala}, {Eccleston}, {Friedman}, {Garc{\'\i}a-Mar{\'\i}n}, {G{\"u}del}, {Glasse}, {Glauser}, {Greene}, {Groezinger}, {Grundy}, {Hastings}, {Henning}, {Hofferbert}, {Hunter}, {Jessen}, {Justtanont}, {Karnik}, {Khorrami}, {Krause}, {Labiano}, {Lagage}, {Langer}, {Lemke}, {Lim}, {Lorenzo-Alvarez}, {Mazy}, {McGowan}, {Meixner}, {Morris}, {Morrison}, {M{\"u}ller}, {rgaard-Nielson}, {Olofsson}, {O'Sullivan}, {Pel}, {Penanen}, {Petach}, {Pye}, {Ray}, {Renotte}, {Renouf}, {Ressler}, {Samara-Ratna}, {Scheithauer}, {Schneider}, {Shaughnessy}, {Stevenson}, {Sukhatme}, {Swinyard}, {Sykes}, {Thatcher}, {Tikkanen}, {van Dishoeck}, {Waelkens}, {Walker}, {Wells}, \& {Zhender}}]{Wright+15}
{Wright}, G.~S., {Wright}, D., {Goodson}, G.~B., {et~al.} 2015, \pasp, 127, 595

\bibitem[{{Xu} {et~al.}(2024){Xu}, {Ouchi}, {Yajima}, {Fukushima}, {Harikane}, {Isobe}, {Nakajima}, {Nakane}, {Ono}, {Umeda}, {Yanagisawa}, \& {Zhang}}]{Xu2024}
{Xu}, Y., {Ouchi}, M., {Yajima}, H., {et~al.} 2024, arXiv e-prints, arXiv:2404.16963

\bibitem[{{Zavala} {et~al.}(2024){Zavala}, {Castellano}, {Akins}, {Bakx}, {Burgarella}, {Casey}, {Ch{\'a}vez Ortiz}, {Dickinson}, {Finkelstein}, {Mitsuhashi}, {Nakajima}, {P{\'e}rez-Gonz{\'a}lez}, {Arrabal Haro}, {Buat}, {Backhaus}, {Calabr{\`o}}, {Cleri}, {Fern{\'a}ndez-Arenas}, {Fontana}, {Franco}, {Giavalisco}, {Grogin}, {Hathi}, {Hirschmann}, {Ikeda}, {Jung}, {Kartaltepe}, {Koekemoer}, {Larson}, {McKinney}, {Papovich}, {Saito}, {Santini}, {Terlevich}, {Terlevich}, {Treu}, \& {Yung}}]{Zavala+2024}
{Zavala}, J.~A., {Castellano}, M., {Akins}, H.~B., {et~al.} 2024, arXiv e-prints, arXiv:2403.10491

\bibitem[{{Ziparo} {et~al.}(2023){Ziparo}, {Ferrara}, {Sommovigo}, \& {Kohandel}}]{Ziparo23}
{Ziparo}, F., {Ferrara}, A., {Sommovigo}, L., \& {Kohandel}, M. 2023, \mnras, 520, 2445

\end{thebibliography}

\end{document}